\documentclass[a4paper,12pt]{article}
\usepackage{jcappub} 

\arxivnumber{2507.05207}

\usepackage{bm}
\usepackage{newtxmath,newtxtext}

\usepackage{graphicx}

\usepackage{textcomp}
\usepackage{float}
\usepackage{booktabs}
\usepackage{dcolumn}
\usepackage{ragged2e}
\usepackage{hyperref}

\hypersetup{colorlinks=true,linkcolor=red,urlcolor=blue,	citecolor=blue}

\usepackage{xcolor}
\usepackage{epsfig}
\usepackage{caption}
\usepackage{appendix}
\usepackage{subcaption}

\usepackage{epsfig}
\usepackage{caption}

\usepackage{float}

\usepackage{multirow}
\usepackage{tcolorbox}

\newcommand{\bq}{\begin{equation}}
	\newcommand{\eq}{\end{equation}}
\newcommand{\bqn}{\begin{eqnarray}}
	\newcommand{\eqn}{\end{eqnarray}}
\newcommand{\nb}{\nonumber}
\newcommand{\lb}{\label}
\newcommand{\pp}{\partial}

\newcommand{\ct}{\ensuremath{c_{\rm T}^{2}}}

\newcommand{\ti}[1]{\ensuremath{\tilde{#1}}}
\newcommand{\nbb}{\ensuremath{ \nabla }}
\newcommand{\m}{\ensuremath{{\mu \nu}}}

\newcommand{\ma}{\ensuremath{m_{1}}}

\newcommand{\eos}{\ensuremath{\omega_{\rm eff}}}

	\title{Interacting Scalar Fields as Dark Energy and Dark Matter in Einstein scalar Gauss Bonnet Gravity}

	\author[1]{Saddam Hussain,}
	\emailAdd{saddamh@zjut.edu.cn}
	
	\author[2]{Simran Arora}
	\emailAdd{arora.simran@yukawa.kyoto-u.ac.jp}
	
	\author[3]{Yamuna Rana}
	\emailAdd{Yamuna$\_$Rana1@baylor.edu}
	
	\author[4]{Benjamin Rose}
	\emailAdd{Ben$\_$Rose@baylor.edu}
	
	\author[5]{Anzhong Wang}
	\emailAdd{anzhong$\_$wang@baylor.edu}

	\affiliation[1]{Institute for Theoretical Physics and Cosmology, Zhejiang University of Technology, Hangzhou 310023, China}
	\affiliation[2]{Center for Gravitational Physics and Quantum Information, Yukawa Institute for Theoretical Physics, Kyoto University, 606-8502, Kyoto, Japan.}
	\affiliation[3,5]{ GCAP-CASPER, Department of Physics and Astronomy, Baylor University, Waco, TX 76798-7316, USA}
	\affiliation[4]{ Department of Physics and Astronomy, Baylor University, Waco, TX 76798-7316, USA}
	
	\abstract{A Gauss-Bonnet (GB) coupled scalar field $\phi$, responsible for the late-time cosmic acceleration and interacting with a coherent scalar field $\psi$ through an interaction potential $W(\phi,\psi)$, is considered from the point of view of particle physics for two different models. The non-minimal coupling between the GB curvature term and the field $\phi$ leads to a time-dependent speed of gravitational waves (GWs), which is fixed to unity in order to be consistent with current GW observations, rendering the GB coupling function model-independent. We investigate the dynamical stability of the system by formulating it as an autonomous system, and provide a detailed discussion on the choice of initial conditions required to obtain stable background evolution of the models. We constrain the model parameters using various sets of observational data, including both early- and late-time probes. We incorporate the improved Dark Energy Survey (DES) 5-year Type Ia supernova sample (DES-SN5YR), referred to as DES-Dovekie, which exhibits substantially lower tension with the Pantheon+ supernova sample. We find that both models are physically viable and closely follow the $\Lambda$CDM trend for the Pantheon+ and DES samples. However, upon including the Roman mock data, a significant departure is observed at higher redshifts, yielding statistically strong preference over the flat $\Lambda$CDM model.}

\begin{document}

	\date{\today}
	\maketitle
	\flushbottom
	
	\section{Introduction}
	
	A large body of observational evidence points toward the existence of mysterious, exotic components that dominate approximately $96\%$ of the Universe's total energy budget \cite{SupernovaCosmologyProject:1998vns,SupernovaSearchTeam:1998fmf,WMAP:2003elm,Sherwin:2011gv,Wright:2007vr,DES:2016qvw,DES:2021esc,SDSS:2005xqv}. Roughly $70\%$ of this budget is attributed to a component known as dark energy, which drives the accelerated expansion of the Universe. The remaining $26\%$ consists of a pressureless form of matter, often referred to as dark matter, which behaves like dust and plays a crucial role in the formation of the cosmic structures observed today \cite{Primack:1997av,DelPopolo:2007dna,Diao:2023tor,Mina:2020eik,Blumenthal:1984bp}. The widely accepted cosmological model that successfully describes a broad range of observations is built upon a cosmological constant $\Lambda$, responsible for the effects of dark energy, and a non-relativistic, electromagnetically neutral, gravitating matter component known as cold dark matter (CDM). However, despite its observational success, the model suffers from several theoretical challenges \cite{Copeland:2006wr,Weinberg:1988cp,Rugh:2000ji,Padmanabhan:2002ji,Carroll:1991mt,Bengochea:2019daa,Kohri:2016lsj,Lopez-Corredoira:2017rqn}. Recent observational results, particularly those from Baryon Acoustic Oscillation (BAO) measurements by the Dark Energy Spectroscopic Instrument (DESI) \cite{DESI:2025zgx, DESI:2024mwx, DESI:2019jxc, Moon:2023jgl}, pose significant challenges to the cosmological constant paradigm and point toward the possibility of a dynamical form of dark energy \cite{Gao:2024ily,Sousa-Neto:2025gpj,Hussain:2025nqy,Myrzakulov:2025jpk,Hussain:2024qrd,Hussain:2024jdt,Chaudhary:2025yvz,Roy:2024kni,Wu:2025wyk,Gialamas:2024lyw,Scherer:2025esj,Paliathanasis:2025cuc,Adil:2023ara}. Various alternative models have been proposed in this context, including potential-driven scalar fields known as quintessence \cite{Peebles:2002gy,Hussain:2023kwk,Roy:2022fif,Setare:2008sf,Cai:2025mas,Dutta:2009yb}, kinetically driven fields such as $k$-essence \cite{Armendariz-Picon:2000nqq,Armendariz-Picon:2000ulo,Chiba:1999ka,Armendariz-Picon:2005oog,Arkani-Hamed:2003pdi,Scherrer:2004au,Chatterjee:2021ijw,Hussain:2022osn,Bhattacharya:2022wzu}, as well as tachyonic and ghost fields \cite{Bagla:2002yn,Khoeini-Moghaddam:2018znw,Liu:2020bmp,Cai:2010uf,Hussain:2022dhp}. Other promising directions include modified gravity theories \cite{Wang:2017brl,Shankaranarayanan:2022wbx,Sotiriou:2008rp,Nojiri:2006ri,Yang:2010hw,Anagnostopoulos:2021ydo,Sokoliuk:2023ccw} and scalar-tensor frameworks \cite{Kobayashi:2019hrl,new_horn,Kainulainen:2004vk,Saratov:2012ni,Langlois:2018dxi}. These models include the potential dependent scalar fields known as quintessential field \cite{Peebles:2002gy,Hussain:2023kwk,Roy:2022fif,Setare:2008sf,Cai:2025mas,Dutta:2009yb}, kinetically driven fields: $k$-essence field \cite{Armendariz-Picon:2000nqq,Armendariz-Picon:2000ulo,Chiba:1999ka,Armendariz-Picon:2005oog,Arkani-Hamed:2003pdi,Scherrer:2004au,Chatterjee:2021ijw,Hussain:2022osn,Bhattacharya:2022wzu}, tachyonic fields, ghost fields \cite{Bagla:2002yn,Khoeini-Moghaddam:2018znw,Liu:2020bmp,Cai:2010uf,Hussain:2022dhp}, modified gravity models \cite{Wang:2017brl,Shankaranarayanan:2022wbx,Sotiriou:2008rp,Nojiri:2006ri,Yang:2010hw,Anagnostopoulos:2021ydo,Sokoliuk:2023ccw} and scalar-tensor theories \cite{Kobayashi:2019hrl,new_horn,Kainulainen:2004vk,Saratov:2012ni,Langlois:2018dxi,Odintsov:2025kyw,Elizalde:2023rds}. 
	
	Besides the nature of the dark energy equation of state, a major crisis has emerged concerning the measurement of the Hubble constant $H_0$. High-redshift measurements from the Planck observations yield $H_0 = 67.4 \pm 0.5$ km/s/Mpc, while direct, model-independent measurements by the SH0ES collaboration report a significantly higher value of $H_0 = 73.04 \pm 1.04$ km/s/Mpc, resulting in a $\sim 5\sigma$ tension between the two determinations \cite{Planck:2018vyg,Riess:2020fzl,Brout_2022}. Numerous theoretical efforts have been made to reconcile this discrepancy by modifying both early- and late-time cosmological physics. However, most attempts have struggled to fully resolve the tension \cite{Schoneberg:2021qvd,DiValentino:2020naf,Bernal:2016gxb}. Among these proposals, interacting dark matter–dark energy models have attracted considerable attention, as they can alleviate the Hubble tension to some extent \cite{Wang:2024vmw,Benisty:2024lmj,Pan:2023mie,DiValentino:2021izs} and simultaneously address some of the theoretical limitations of the $\Lambda$CDM framework \cite{Huey:2004qv,Cai:2004dk}. Such interactions between the dark sectors are commonly modeled by introducing modifications to the continuity equations, thereby altering the energy density evolution of both dark matter and dark energy. 
	Although interacting dark energy models can, in some cases, yield moderately 	higher values of $H_0$, the tension between the SH0ES and Planck measurements often persists \cite{Montani:2024pou,Pan:2023mie}. Therefore, the primary aim
	of the present study is not to fully resolve the Hubble tension, but rather to develop an alternative framework motivated by particle physics principles and to test its viability using current observational data.
	
	In this work, we adopt a scalar field description for dark matter, represented by a coherent scalar field \(\psi\) with a quadratic potential, which effectively mimics pressureless cold dark matter \cite{Turner:1983he}. Although the fluid description of dark matter—characterized by a vanishing equation of state—successfully reproduces observational trends at both high and low redshifts, a scalar field approach offers a more fundamental perspective. From a particle physics standpoint, dark matter is expected to be a weakly interacting particle, distinct from Standard Model constituents. Consequently, dark matter may originate from the quantization of a scalar field, possibly as a relic of the inflaton field from the early universe. Numerous studies have investigated scalar field dark matter models, demonstrating their capacity to reproduce key features of cold dark matter, such as dark halo structures, galaxy rotation curves, late-time cosmic evolution, and large-scale structure formation \cite{Hwang:1996xd,Magana:2012ph,Urena-Lopez:2019kud,Gutierrez-Luna:2021tmq,Matos:2008ag,Solis-Lopez:2019lvz,Peebles:1998qn,Matos:2000jx,Matos:2000ng,Matos:2000ss,matos2010accelerated,Robles:2018fur,Li:2013nal,Magana:2012xe,Aboubrahim:2024spa,Foidl:2022bpn,Suarez:2013iw,Suarez:2011yf,Poulot:2024sex,Lora:2011yc,Dev:2016hxv}.
	
	The dark energy component is modeled by a slowly evolving quintessence field $\phi$ with an exponential potential. A non-gravitational interaction between the two fields is introduced at the Lagrangian level via the interaction term $W(\phi, \psi)$, providing a fundamental framework to study their coupling. This interaction leads to modified equations of motion and facilitates energy exchange between the fields.
	
	{ The scalar field $\phi$, responsible for late-time cosmic acceleration, is also allowed to couple non-minimally to the curvature invariant known as the Gauss-Bonnet (GB) term. Such a coupling is well motivated by string/M-theory \cite{Gross:1986mw,Bento:1995qc,Ferrara:1996hh,Antoniadis:1997eg}, and has been shown to prevent both past and future cosmological singularities \cite{Nojiri:2005vv,Tsujikawa:2006ph,DeFelice:2006pg,Minamitsuji:2024twp,Tsujikawa:2022aar,DeFelice:2009rw,DeFelice:2006pg,Gross:1986mw,Bento:1995qc,Ferrara:1996hh}. This coupling affects the propagation speed $c_T$ of gravitational waves, which is tightly constrained by observations: $|\ct - 1| < 10^{-16}$ \cite{LIGOScientific:2017zic,Odintsov:2019clh,Ezquiaga:2017ekz,TerenteDiaz:2023iqk,Fier:2025}. In general, the behavior of \(\ct\) depends on the form of the GB coupling function \footnote{In principle the coupling function $f$ could depend on both $\phi$ and $\psi$. 
			This would require the specification of a particular coupling model, thereby increasing the parameter space and making the dynamics more complicated. Therefore,  in the present work we mainly restrict the GB coupling to depend only on $\phi$.}.   However, since the deviation from the speed of light is experimentally negligible and possibly undetectable in the late evolution of the universe,   we adopt the approximation \(\ct \simeq 1\). This allows the time derivative of the coupling function to scale with the scale factor, i.e., \(\dot{f} \propto a\), simplifying the model and reducing the parameter space. This approach renders the analysis relatively model-independent \cite{Hussain:2024yee}.}
	
	{In \cite{Hussain:2024yee}, we studied the interacting dark energy and dark matter  for both of the two cases, $\ct = 1$ and $\ct \not= 1$, and found that the final results do not sensitively depend on the precise values of $\ct$, as long as the bound $|\ct - 1| < 10^{-16}$ is satisfied, which is quite different from the considerations of gravitational waves \cite{Fier:2025}. However, in  \cite{Hussain:2024yee} we modeled the dark matter by a  pressure-less perfect fluid. So, in this paper we shall generalize our previous studies to the case where the dark matter is also described by a scalar field and interacting with the dark energy component from the point of view of particle physics. } This approach is more fundamental and closely aligned with particle physics principles. Although the interaction is introduced phenomenologically at the action level, the framework yields consistent field equations and constrains the allowed forms of interaction, preventing arbitrary choices. In contrast, fluid-based models typically introduce interactions by postulating a non-zero term in the continuity equation of the dark sector, i.e., $\nabla_\mu T^{\mu \nu}_{\rm DE} = -Q^\nu = - \nabla_\mu T^{\mu \nu}_{\rm DM}$, where $Q^\nu$ can be arbitrarily chosen without a guiding Lagrangian formulation. This often leads to ambiguities and makes constructing a corresponding Lagrangian a challenging, if not infeasible, task.\\
	Moreover, deriving the interacting equations of motion directly from an action provides a significant advantage when analyzing cosmological perturbations. Without a well-defined interaction Lagrangian, fluid models frequently encounter stability issues that must be resolved in an ad hoc manner \cite{Jackson:2009mz, Kase:2019veo,Tamanini:2015iia}. Such instabilities do not arise in the present framework, where the interaction structure is consistently derived from first principles.

	
	The dynamics of the combined system are analyzed using dynamical system techniques, considering two types of interaction potentials: exponential and power-law forms. We numerically investigate the stability of the system and constrain the model parameters using current observational data, including model-independent Hubble parameter measurements, Type Ia Supernovae observations (Pantheon+ and DES5YR), DESI Baryon Acoustic Oscillation (DRII), Planck CMB data, and Roman Space Telescope mock data.
	
	The paper is organized as follows. In Sec.~\ref{sec:theoretical_framework}, we present the mathematical framework. The background equations of motion and relevant cosmological quantities are derived in a flat FLRW metric in Sec.~\ref{sec:background}. A fluid-equivalent description of the dark matter scalar field is provided in Sec.~\ref{sec:fluid_picture}, and the dynamics of the interacting models are explored in Sec.~\ref{sec:dyn_system}. The observational data sets are summarized in Sec.~\ref{sec:data_presentation}, and the results are discussed in Sec.~\ref{sec:result}. We conclude with a summary of our findings in Sec.~\ref{sec:conclusion}.
	

	\section{Coupled Einstein-scalar-Gauss-Bonnet gravity}
	\label{sec:theoretical_framework}	
	
	In the framework of {\em Einstein-scalar-Gauss-Bonnet (EsGB) gravity}, we consider interacting models of dark energy and dark matter, represented by two scalar fields, $\phi$ and $\psi$, respectively. These fields are coupled via an interaction potential. Initially, both scalar fields are non-minimally coupled to the curvature invariant Gauss-Bonnet (GB) term $\mathcal{G}$ through a general function \(f(\phi,\psi)\), while the remaining components of the Universe are assumed to interact only through gravity. The action for the considered system is \cite{Zhang20a,Cicoli:2023opf}
	\begin{multline}
		S_{\text{EsGB}}= \int dx^4\sqrt{-g}\Big[ \frac{R}{2\kappa^2}   + f(\phi, \psi){\cal G} +
		{\cal L}_{\phi}\left(\phi, \nbb_{\mu} \phi \right)+  {\cal L}_{\psi}\left(\psi, \nbb_{\mu} \psi \right)  +  {\cal L}_{\rm int}\left(\phi, \psi\right)\Big] + S_{\rm M} \ ,
		\label{eq2.1}
	\end{multline}
	where $\kappa^2  \equiv 8\pi G/c^4 = 1/M_{\rm pl}^2$, and $M_{\rm pl} = 2.44\times 10^{18}$ (in natural units where ($c=\hbar = 1$)) \cite{Tomberg:2021ajh}, with $G$ being the Newtonian constant, $c$ the speed of light in vacuum, and $g \;[\equiv \text{det}(g_{\mu\nu})]$ and \(R\) is the Ricci scalar. $S_{\rm M}$ represents the action for the remaining Standard Model particles, such as radiation and barotropic fluids. These background components interact only gravitationally; consequently, their energy-momentum tensors are conserved independently. The GB term is expressed as
	\begin{equation}
		{\cal G} \equiv R^2 -4R_{\mu\nu}R^{\mu\nu} +R_{\mu\nu\rho\sigma}R^{\mu\nu\rho\sigma}\ .
	\end{equation}
	In addition to the GB coupling, the fields $\phi$ and $\psi$ also interact directly through the interaction Lagrangian term $\mathcal{L}_{\rm int}$, where 
	\bqn
	{\cal L}_{\phi}\left(\phi, \nbb_{\mu} \phi \right) &=& - \frac{1}{2}\left(\nbb_{\mu} \phi \nbb^{\mu} \phi \right) - V(\phi),\nb\\
	{\cal L}_{\psi}\left(\psi, \nbb_{\mu} \psi \right) &=& - \frac{1}{2}\left(\nbb_{\mu} \psi \nbb^{\mu} \psi \right) - U(\psi),\nb\\
	\mathcal{L}_{\rm int}\left(\phi, \psi\right) &=&   - W(\phi, \psi).
	\label{dark_matter_lag}
	\eqn
	Here $V(\phi)$ and $U(\psi)$ denote the potentials for the scalar fields, and $W(\phi, \psi)$ represents the interaction potential between them. $\nbb_{\mu}$ denotes the covariant derivative with respect to the metric $g_{\mu\nu}$. The energy-momentum tensors corresponding to each component are given by: 
	\begin{eqnarray}
		\lb{eq2.4}
		T^{GB}_{\mu\nu}&\equiv&2\left(\nbb_\mu \nbb_\nu f\right)R-2g_{\mu\nu}\left(\nbb_\rho \nbb^\rho f\right)R 
		-4\left(\nbb^\rho \nbb_\nu f\right)R_{\mu\rho} 
		- 4\left(\nbb^\rho \nbb_\mu f\right)R_{\nu\rho}+4\left(\nbb^\rho \nbb_\rho f\right)R_{\mu\nu} \nonumber\\
		&&
		+ 4 g_{\mu\nu}\left(\nbb^\rho \nbb^\sigma f\right)R_{\rho\sigma} -4\left(\nbb^\rho \nbb^\sigma f\right)R_{\mu\rho\nu\sigma}\ ,\\
		T_{\m}^{\phi} &=& \nbb_{\mu}\phi \nbb_{\nu} \phi - \frac{1}{2}g_{\m} \big[\nbb^{\alpha}\phi\nbb_{\alpha}\phi + 2V(\phi, \psi)\big]\ ,\\	
		T_{\m}^{\psi} &=& \nbb_{\mu}\psi \nbb_{\nu} \psi - \frac{1}{2}g_{\m} \big[\nbb^{\alpha}\psi\nbb_{\alpha}\psi + 2U(\psi)\big]\ ,\\	
		T_{\m}^{M} &=& \rho_{M} \ u_{\mu} u_{\nu} + P_{M}(u_{\mu} u_{\nu} + g_{\m}) \ ,
	\end{eqnarray}
	where $V(\phi, \psi) \equiv V(\phi) + W(\phi, \psi)$, and the stress tensor is defined as: 
	\begin{equation}
		T_{\m}  \equiv \frac{-2}{\sqrt{-g}} \frac{\delta S}{\delta g^{\m}}\ .
	\end{equation}
	Here $u^{\mu}$ denotes the four-velocity of the fluid, satisfying the normalization condition \(u^{\mu}u_{\mu} = -1\). $\rho_{M}$ represents the energy density of the background fluid, and \(P_{\rm M}\) denotes its corresponding pressure. The subscript $M$ is used to collectively represent both radiation and pressureless baryonic fluids. By varying the action with respect to the field variables $\phi$ and $\psi$, we obtain the corresponding field equations as follows:
	\begin{eqnarray}
		\nbb^{2}\phi & = & - \frac{\partial V(\phi,\psi)}{\partial \phi} +  \frac{\partial f}{\partial \phi}{\cal G} \ ,\\
		\nbb^{2}\psi & = & - \frac{\partial V(\phi,\psi)}{\partial \psi} + \frac{\partial f}{\partial \psi}{\cal G} \ ,
	\end{eqnarray}
	where $\nbb^{2} \equiv \nbb_{\mu} \nbb^{\mu}$. 
	
	\section{The background equations}
	\label{sec:background}
	
	In the flat FLRW metric \(ds^2 = -dt^2 + a(t)^2 d\vec{x}^2\), where \(a\) is the scale factor, the field equations can be expressed as:  
	\begin{eqnarray}
		\ddot{\phi}  + 3 H \dot{\phi} + \frac{\pp V_{\rm int}(\phi,\psi)}{\pp \phi}  - \mathcal{G} f_{,\phi} & = & 0\ ,\\
		\ddot{\psi} + 3 H \dot{\psi} + \frac{\pp V_{\rm int}\phi,\psi)}{\pp \psi}  - \mathcal{G} f_{,\psi} & = &  0\, .
	\end{eqnarray} 
	The GB invariant term becomes:
	\begin{equation}
		\mathcal{G} = 24 H^2(\dot{H} + H^2) \ .
	\end{equation}
	The field equations  can be written as: 
	\begin{equation}
		\dfrac{1}{\kappa^2}G_{\m } = T_{\m}^{\phi} + T_{\m}^{\psi} + T_{\m}^{\rm GB} + T_{\m} ^{\rm M}\ .
	\end{equation}
	The energy density and pressure of each component can be defined by comparing its energy-momentum tensor with the standard perfect fluid form, $T_{\m} =  \rho \ u_{\mu} u_{\nu} + P(u_{\mu} u_{\nu} + g_{\m}) $, which yield
	\begin{eqnarray}
		\rho_{\phi} & = &  \frac{1}{2} \dot{\phi}^2 + V(\phi) \ ,\\
		P_{\phi} & = &   \frac{1}{2} \dot{\phi}^2 - V(\phi)  \ ,\\
		\rho_{\psi}  & =  & \frac{1}{2} \dot{\psi}^2 + U(\psi) +  W(\phi, \psi)  \label{psi_energy} \ ,\\
		P_{\psi} & = &   \frac{1}{2} \dot{\psi}^2 - U(\psi) - W(\phi, \psi) \ .
	\end{eqnarray}
	Here we incorporate the interaction potential term into the $\psi$ field. It is not necessary to include the interaction term only with the dark energy or dark matter, since so far the nature of dark sector is unknown and the available observation can only probe the total contribution of energy densities. Therefore, the total energy density remains unchanged after including the interaction term in any of the two components. The first Friedmann equation reads:
	\begin{equation}
		3 H^2/\kappa^2 = \rho_{\phi} + \rho_{\psi} - 24 \dot{f}(\psi, \phi) H^3 + \rho_{M}\, , 
	\end{equation}
	and the Raychaudhari equation becomes:
		\begin{multline}
			\dot{H} = 	-\frac{\kappa^2 }{2 \left(96 H^4 \kappa ^2 f_{, \psi }^2+96 H^4 \kappa ^2 f_{, \phi }^2+8 H \kappa ^2 \dot{\psi } f_{, \psi }+8 H \kappa ^2 \dot{\phi } f_{, \phi }+1\right)} \Bigg[ -32 H^3 \dot{\psi } f_{, \psi }-32 H^3 \dot{\phi } f_{, \phi } \\ +\dot{\psi }^2 \left(8 H^2 f_{, \psi \psi }+1\right)+\dot{\phi }^2 \left(8 H^2 f_{, \phi \phi }+1\right)+8 H^2 \left(24 H^4 f_{, \psi }^2+24 H^4 f_{, \phi }^2-f_{, \psi } V_{, \psi }-f_{, \phi } V_{, \phi }\right)\\ + \sum_{M = \{r,b\}}(1+ \omega_{M}) \rho_{M}\Bigg]\ .
			\label{hdot_general}
		\end{multline}
	Here, a subscript preceded by a comma denotes a partial derivative with respect to that quantity. The parameter \(\omega_M = {P_{M}}/{\rho_{M}}\) defines the equation of state (EoS) for the background fluid. For radiation, \(\omega_{r} = 1/3\), while for baryons, \(\omega_{m} = 0\). So far, we have considered a generalized case where both scalar fields are non-minimally coupled to the GB term so that $f= f(\phi, \psi)$. However, in the remainder of this paper, we will focus exclusively on the scenario where the GB coupling function \(f\) depends only on the field \(\phi\), that is
	\begin{equation}
		f = f(\phi). 
	\end{equation} 
	The field equations can be further cast into energy-momentum conservation equations, which makes it convenient to identify the interaction term governing the energy exchange between the two scalar fields. After rearrangement, the continuity equations become:
	\begin{eqnarray}
		\dot{\rho}_{\phi} + 3 H (\rho_{\phi} + P_{\phi}) &= & -Q +  24 H^2 \dot{\phi} (\dot{H} + H^2) f_{,\phi}, ~~~~~~~ \label{rhophi_eqn} \\
		\dot{\rho}_{\psi} + 3 H (\rho_{\psi} + P_{\psi}) &= &  Q  \label{rhopsi_eqn},\\
		\dot{\rho}_{M} +  3 H (\rho_{M} + P_{M}) & = & 0\ .
	\end{eqnarray}
	Here, \(Q \equiv \dot{\phi} W_{,\phi}\) represents the interaction term responsible for energy transfer between the two scalar fields. If \(Q > 0\), energy flows from the field \(\phi\) to the field \(\psi\). These equations are equivalent to the conservation equations presented in ~\cite{Tsujikawa:2006ph}. The background fluids are individually conserved. In principle, one could consider couplings between the scalar fields and these background fluids; however, current observations indicate that their contribution relative to the dark sector is negligible. Therefore, neglecting any coupling between these components does not significantly affect large-scale structure formation or the late-time acceleration epoch.
	
	We further transform Eq.~\eqref{rhophi_eqn} by redefining its pressure term as:
	\begin{equation}
		\dot{\rho}_{\phi} + 3 H (\rho_{\phi} + P_{\rm eff} ) = -Q\ ,
	\end{equation}  
	where the effective pressure of the field becomes: 
	\begin{equation}
		P_{\rm eff} = P_{\phi} - 8 H \dot{\phi} (\dot{H} + H^2) f_{,\phi} \ . 
	\end{equation}
	With this transformation, the conservation equation aligns with the $\psi$ field equation; however, it modifies the corresponding field’s pressure, resulting in a modification of the individual equation of state (EoS) as:
	\begin{eqnarray}
		\omega_{\phi} & = &\frac{P_{\rm eff}}{\rho_{\phi}} \ \label{field_eos},\\
		\omega_{\psi }  & = & \frac{P_{\psi}}{\rho_{\psi}} = \frac{1/2 \dot{\psi}^2 - U - W}{1/2 \dot{\psi}^2 + U +W}	.
	\end{eqnarray}
	Nevertheless, the system's effective equation of state remains unchanged and is given by:
	\begin{equation}
		\omega_{\rm eff}  = - \frac{2 \dot{H}}{3 H^2} -1\, .
	\end{equation}
	This quantity plays a significant role in characterizing the accelerating or non-accelerating nature of different epochs in the Universe. For instance, \(-1 \leq \eos < -0.3\) corresponds to the accelerating regime, \(\eos < -1\) denotes the phantom phase, and \(\eos \simeq 0\) identifies the matter-dominated epoch. Additionally, the tensor propagation speed, which depends on \(f\), is given by \cite{Tsujikawa:2006ph,Odintsov:2019clh,Fier:2025}:
	\begin{equation}
		\ct =  \frac{1 + 8 \kappa^2 \ddot{f}}{1 + 8 \kappa^2 H \dot{f}}\ .
		\label{gw_speed}
	\end{equation}
	Recent observations have shown that the speed of gravitational waves is approximately equal to the speed of light \cite{LIGOScientific:2017zic}, imposing the bound:
	\begin{equation}
		|\ct-1 | \le 5 \times 10^{-16}, \label{gw_bound}
	\end{equation}
	in natural units where \(c = 1\) \cite{Ezquiaga:2017ekz}. Additionally, the propagation speed at the perturbation level must be positive to avoid ghost instabilities \cite{Fier:2025}. Therefore, to incorporate these constraints, throughout the rest of this study, we impose the tensor speed constraint as:
	\begin{equation}
		\ct =1 \implies \dot{f} \propto a \ .
	\end{equation}
	Hence, this constraint simplifies the mathematical complexity, making the analysis model-independent. Consequently, the time derivative of the GB coupling function becomes:
	\begin{equation}
		\dot{f}(\phi) = \ma a, \quad \ddot{f} = \ma \dot{a}\ , \label{f_const}
	\end{equation} 
	where \(\ma\) is the GB mass dimensional constant. 
	
	To determine the dynamics of the composite system, it is necessary to specify the forms of the potentials. We choose an exponential potential for the scalar field $\phi$, so that it can act as a dark energy candidate during the late-time epoch. For the field $\psi$, we select a quadratic potential, and for the interaction potential, we assume a quadratic dependence on $\psi$ multiplied by a generalized function of $\phi$:
	\bqn
	&& V(\phi) = V_0\exp\left(\frac{\lambda\kappa^2\phi^2}{2}\right), \quad 	U(\psi) = \frac{1}{2} m^2 \psi^2,\nb\\ 
	&& W(\phi, \psi) = \frac{1}{2} \xi(\phi) \psi^2 \ .
	\label{potentials}
	\eqn
	Here $V_0, \lambda, m$ are free parameter. In particular, the parameter \(m\) represents the mass of the $\psi$ field, which is several orders of magnitude greater than the Hubble parameter \((H)\) to ensure that $\psi$ behaves as a pressureless non-relativistic fluid. 

	\subsection{A short note on scalar field dark matter}
	
	Before deriving the  dynamical systems, we first demonstrate that the coherent oscillations of a scalar field with a quadratic potential can effectively mimic a pressureless, non-relativistic fluid. This equivalence arises when the oscillation frequency of the scalar field is much greater than the expansion rate of the Universe \cite{Turner:1983he}. To illustrate this, let us consider the dark matter field \(\psi\) minimally coupled to gravity, neglecting all interactions. The equation of motion then becomes:
	\begin{equation}
		\ddot{\psi} + 3 H \psi + \frac{\pp U}{\pp \psi} = 0 \ . \label{field_psi}
	\end{equation}
	Refs.~ \cite{Ratra:1990me, Saha:2024irh} have shown that an oscillating scalar field can mimic a pressureless fluid. Therefore, we assume the solution for \(\psi\) to be an oscillatory function:
	\begin{equation}
		\psi(t) = c_1(t) \sin\alpha (t) + c_2(t) \cos\alpha (t) \ , \label{psi_solu}
	\end{equation}
	where, \(c_{1,2}(t)\) and \(\alpha(t)\) are time-dependent functions, with the assumption that the rate of change of the trigonometric argument \(\alpha(t)\) is much greater than the rate of change of the amplitude functions \(c_{1,2}(t)\). Substituting Eq.~\eqref{psi_solu} into Eq.~\eqref{field_psi} and using the quadratic potential from Eq.~\eqref{potentials}, we obtain	
	\bqn
	&& \cos\alpha \bigg(2 \dot{\alpha} \dot{c}_{1} +  \ddot{c}_{2} +  c_1 \ddot{\alpha}  - c_2 \dot{\alpha}^2 +  3 H \dot{c}_{2} + 3 H c_{1} \dot{\alpha}
	 + m^2 c_{2} \bigg) + \sin\alpha \bigg(\ddot{c}_{1} - 2\dot{\alpha} \dot{c}_{2} - c_1 \dot{\alpha}^2  - c_2 \ddot{\alpha} \nb\\
	&& ~~~~~~~~~~~ + 3 H \dot{c}_{1}   - 3 H c_2 \dot{\alpha} + m^2 c_1\bigg) =0.
	\eqn
	Since the trigonometric functions are linearly independent, their coefficients must vanish independently. As a result, we obtain differential equations corresponding to different orders of \(\mathcal{O}(m)\). The equation containing the \(\mathcal{O}(m^2)\) term is:
	\begin{equation}
		\dot{\alpha}^2 - m^2 = 0\ .
	\end{equation}
	This yields \(\alpha = m(t - t_0)\), where \(t_0\) represents the initial time and can be set to zero. Substituting this relation into the above expression, we obtain the terms of order \(\mathcal{O}(m)\); for example, the equations corresponding to \(\mathcal{O}(m^1)\) are:
	\begin{equation}
		2 \dot{c}_{1} + 3 H c_{1} = 0, \quad 	2 \dot{c}_{2} + 3 H c_{2} = 0\ , 
		\label{mass_1_term}	
	\end{equation}
	while the $\mathcal{O}(m^{0})$ terms are: 
	\begin{equation}
		\ddot{c}_{1} + 3 H \dot{c}_{1}=0, \quad  \ddot{c}_{2} + 3 H \dot{c}_{2}=0\ .
	\end{equation}
	Since this equation does not contain any mass terms, it can be discarded. Solving Eq.~\eqref{mass_1_term}, we find that the amplitude functions become:
	\begin{equation}
		c_{1,2}(t) = {\ti{c}_{1,2}} \ {a^{-3/2}} \ . \label{c_solu}
	\end{equation}	
	Here, \(\tilde{c}_{1,2}\) are dimensionless integration constants, and the amplitude of the \(\psi\) field evolves with the scale factor. In contrast, the trigonometric argument \(\alpha\) varies with cosmic time, implying that \(\alpha\) changes much more rapidly than the amplitudes \(c_{1,2}\) when \(m \gg H\). Since the field \(\psi\) oscillates rapidly, it is essential to compute its time-averaged behavior over one oscillation period, \(T = 2\pi/m\). The field \(\psi\) can be expressed as the product of a rapidly oscillating component and a slowly varying amplitude. Therefore, the time averaging is applied specifically to the oscillatory part in order to extract its effective contribution over longer timescales. As a result, the averaged energy density becomes:
	\begin{equation}
		\langle \rho_{\psi} \rangle = \left\langle \frac{1}{2} \dot{\psi}^2 + U \right \rangle\ . 
	\end{equation}
	Using Eqs.~\eqref{psi_solu}, ~\eqref{mass_1_term} and ~\eqref{c_solu}, we find that the average density becomes: 
	\bqn
	\langle \rho_{\psi} \rangle &=& \frac{9}{16} (c_{1}^2 + c_{2}^2) H^2 + \frac{1}{2} m^2 (c_{1}^2 + c_{2}^2) 
= \frac{1}{2} m^2 (\ti{c}_{1}^2 + \ti{c}_{2}^2) a^{-3} \left(1 + \frac{9 H^2}{8 m^2}\right)\ .
	\eqn
	For \(m \gg H\), the averaged energy density scales as \(a^{-3}\), which is identical in form to that of a pressureless cold dark matter fluid.

	\section{Fluid approximation of $\psi$ field}
	\label{sec:fluid_picture}
	
	In the previous section, we considered the $\psi$ field to be minimally coupled to the metric, and thus no energy transfer occurred between the fields $\phi$ and $\psi$. We demonstrated that the field $\psi$ undergoes rapid oscillations and behaves as a pressureless fluid when its mass is significantly larger than the Hubble parameter. Previous studies have shown that numerically resolving such rapid oscillations becomes computationally expensive \cite{Urena-Lopez:2015gur}. To address this, various numerical techniques have been implemented within the CLASS code framework \cite{Urena-Lopez:2023ngt}.
	
	In this section, we take an alternative approach by recasting the field equation for $\psi$ into an approximate fluid description using the WKB approximation. We follow the formalism developed in ~\cite{Bertolami:2012xn, Poulot:2024sex, vandeBruck:2022xbk}. To derive the fluid-approximated equations, we use the time-averaged energy density and pressure defined in Eq.~\eqref{psi_energy} as:
	\begin{equation}
		\langle \rho_{\psi} \rangle + \langle P_{\psi} \rangle =  \langle \dot{\psi}^2 \rangle \ .
	\end{equation} 
	The field equation corresponding to the $\psi$ field is 
	\begin{equation}
		\ddot{\psi} + 3 H \dot{\psi} + \frac{\pp U}{\pp \psi} + \frac{\pp W}{\pp \psi} = 0 \ .
	\end{equation}
	Multiplying this equation by the time-averaged value of \(\dot{\psi}\), the field equation becomes:
	\begin{equation}
		\left(\dot{\psi} \ddot{\psi} + \dot{\psi} U_{,\psi}+ \dot{\psi} W_{,\psi} + \dot{\phi} W_{,\phi}\right) + 3 H \dot{\psi}^2 - \dot{\phi} W_{, \phi} = 0 \ .
	\end{equation}
	Here, we have added and subtracted \(\dot{\phi} W_{,\phi}\), and by taking the time average of the entire equation, we obtain:
	\begin{equation}
		\frac{d \langle \rho_{\psi} \rangle}{d t} + 3 H \langle \rho_{\psi} \rangle = \dot{\phi} \left\langle\frac{\pp W}{\pp \phi} \right\rangle\ .
	\end{equation}
	Here, we have assumed that the dark matter mass satisfies \(m \gg H\). To obtain an analytical solution to this equation, it is necessary to specify the form of the interaction potential \(W(\psi, \phi)\) as that given by Eq.(\ref{potentials}).
	For this interaction term, the continuity equation becomes \cite{Bertolami:2012xn}: 
	\begin{equation}
		\dot{\langle {\rho}_{\psi} \rangle } + 3 H \langle \rho_{\psi} \rangle = \dot{\phi} \frac{\pp \ln(m^2 + \xi(\phi))}{\pp \phi} \langle \rho_{\psi} \rangle \ . \label{dar_matter_mass}
	\end{equation}
	Here, we have used \(\langle P_{\psi} \rangle = 0 = \frac{1}{2} \langle \dot{\psi}^2 \rangle - \left\langle U + W \right\rangle\), and \(\langle \rho_{\psi} \rangle = \langle \dot{\psi}^2 \rangle\). Since the \(\phi\) field evolves slowly, we do not apply any time averaging to \(\phi\) or its derivatives. Integrating the above equation, we obtain:
	\begin{equation}
		\rho_{\psi} = \rho_{\psi0} e^{-3 N} c_0 (m^2  + \xi(\phi)) \ . \label{dark_matter_equation}
	\end{equation}
	Here, we have omitted the angle brackets, although \(\rho_{\psi}\) still denotes the averaged value. We define \(N = \ln a = -\ln(1 + z)\),\footnote{where \(z\) is the redshift.} and \(\rho_{\psi0}\) and \(c_0\) are integration constants, with \(c_0\) being absorbable into the mass term. From the solution, we observe that the averaged energy density of the field scales as \(a^{-3}\), with a slight correction arising from the interaction term \(\xi(\phi)\). The solution for \(\rho_{\psi}\) imposes the condition that the quantity \((m^2 + \xi(\phi))\) must remain non-negative, since \(\rho_{\psi}\) increases as redshift increases (\(N < 0\)). From an observational perspective, this suggests that the magnitude of the interaction has to be extremely small at late times, as numerous model-independent analyses confirm that deviations from CDM are minimal.
	
	On the other hand, considering the quintessence continuity equation \eqref{rhophi_eqn} along with the solution for \(\rho_{\psi}\), the equation becomes:
	\begin{equation}
		\dot{\rho}_{\phi} + 3 H (\rho_{\phi} + P_{\phi}) = -\dot{\phi} \left\langle\frac{\pp W}{\pp \phi} \right\rangle\ +  24 H^2 \dot{\phi} (\dot{H} + H^2) f_{,\phi}\ ,
	\end{equation}
	which further simplifies to 
	\begin{equation}
		\dot{\rho}_{\phi} + 3 H (\rho_{\phi} + P_{\rm eff}) = - \dot{\phi} \frac{\pp \ln(m^2 + \xi(\phi))}{\pp \phi} \times \rho_{\psi}  \ .
	\end{equation}
	Thus, in the past, if the coefficient of \(\rho_{\psi}\) remains positive, the dark energy field continuously loses energy to the matter field, thereby enhancing the matter energy density.
	
	\section{Dynamics for approximated system}
	\label{sec:dyn_system}
	
	In this section, we elaborate on determining the dynamics of the system using the field equations and Friedmann equation for suitable choices of potentials, subject to the constraint \(\ct = 1\). The relevant equations are as follows:
	\begin{align}
		&	3 H^2/\kappa^2 = \rho_{\phi} + \rho_{\psi} - 24 \dot{f}(\phi) H^3 + \rho_{M}\, , \\ 
		& \dot{H} = \frac{\kappa ^2 \left(-\dot{\psi }^2-\dot{\phi }^2 -  (1+\omega_M)\rho_{M}\right)}{16 \ma a H \kappa ^2+2}\ ,\\ 
		&	\ddot{\phi}  + 3 H \dot{\phi} + \frac{\pp V(\phi,\psi)}{\pp \phi}  - \mathcal{G} f_{,\phi} = 0 \ ,\\
		&\rho_{\psi} = \rho_{\psi0} e^{-3 N} c_0 (m^2  + \xi(\phi)) \ , \\
		&	\dot{\rho}_{\phi} + 3 H (\rho_{\phi} + P_{\rm eff}) = - \dot{\phi} \frac{\pp \ln(m^2 + \xi(\phi))}{\pp \phi} \times \rho_{\psi} \ , \\
		& \dot{\rho}_{M} + 3 H (1+ w_M)\rho_{M} = 0 \ .
	\end{align}
	We use this set of equations within the dynamical systems framework to analyze the qualitative behavior of the system. To evaluate the system, we adopt the following form of the potential for the quintessence field \(\phi\), as given in \cite{Lee:2004vm}:
	\begin{equation}
		V(\phi) = V_0 \exp\left(\frac{\lambda \kappa^2 \phi^2}{2}\right), 
	\end{equation}
	and choose the exponential form of interaction potential as Model I:
	\begin{equation}
		\rm{Model \ I:}	 \ \xi(\phi) = M \left(\frac{b_c + V(\phi)/V_0}{1+ b_c} \right)^{n_c},   \text{with} \ b_c + 1>0 \ ,
	\end{equation}
	where $b_c, V_0, \lambda, n_c, M$ are the constant parameters and $M$ has the dimensions of mass squared. Here we shall consider only the \(b_c = 0\) case \cite{Lee:2006za,Lee:2004vm}. 
	
	{
		As we introduce the interaction term between the dark sectors
		at the Lagrangian level through the operator
		$\frac{1}{2}\xi(\phi)\psi^2$, which becomes an 
		contribution to the effective dark matter mass, Eq.~\eqref{dar_matter_mass},
		\begin{equation}
			m_{\rm eff}^2 = m_\psi^2 + \xi(\phi).
		\end{equation}
		Such a structure is common in scalar--tensor theories and interacting dark
		energy models, where the dark matter mass depends on a light scalar degree of
		freedom.
		
		In the absence of a fundamental microscopic description of the dark sector,
		an effective field theory (EFT) approach suggests that $\xi(\phi)$ should be a
		smooth function of $\phi$, expandable as
		\begin{equation}
			\xi(\phi) = \sum_n c_n \left( \frac{\phi}{M} \right)^n ,
		\end{equation}
		with $M$ denoting the cutoff scale \cite{Lee:2004vm}. The power-law interaction studied in later section as Model II, corresponds to the leading-order truncation of this expansion and
		thus represents a minimal EFT realization.
		
		Exponential couplings of the form $\xi(\phi ) \propto \exp(\lambda \phi)$
		naturally arise in scalar--tensor and dilaton-like theories, as well as in
		string-inspired compactifications, where scalar fields determine effective
		masses and couplings in the Einstein frame \cite{Mignemi:1992nt,Kanti:1995vq,Damour:1994zq,Ghosh:2025pbn,Costa:2025kwt,Brax:2025ahm,He:2011qn,Carson:2020ter}. Similar forms have been widely
		considered in coupled quintessence and interacting dark sector models \cite{Amendola:1999er}.
		
		The exponential and power-law choices therefore represent two theoretically
		motivated and complementary realizations of scalar-mediated dark
		matter--dark energy interactions, allowing us to explore distinct dynamical
		behaviors within a consistent effective framework.
	}
	
	\subsection{Model I: \boldmath \(\xi(\phi) = M\exp\left(\frac{\lambda \kappa^2 \phi^2}{2}\right)^{n_c}\)} 
	
	The dynamics of the system can be obtained by reducing the second-order differential equations to a system of first-order ordinary differential equations through the introduction of the following dimensionless dynamical variables:
		\bqn
		&& x = \dfrac{\kappa\dot{\phi}}{\sqrt{6} H}, \;\; y= \dfrac{\kappa\sqrt{V}}{\sqrt{3} H}, \;\;  z  = \kappa \phi, \;\; \Omega_{\psi} = \frac{\kappa^2 \rho_{\psi}}{3 H^2}, \nb\\
		&& \Omega_{\rm GB} =  8\kappa^2 m_{1} a H, \;\;  \Omega_{r} = \frac{\kappa^2 \rho_{r}}{3 H^2}, \;\; \Omega_{b} = \frac{\kappa^2 \rho_{b}}{3 H^2}. \label{dyn_variable}
		\eqn
		In terms of these variables, the Friedmann equation becomes: 
		\begin{equation}
			1 = \Omega_{\phi} + \Omega_{\psi} - \Omega_{\rm GB} + \Omega_{r} + \Omega_{b}\ , 
		\end{equation}
		where the field fractional density is given by 
		\begin{equation}
			\Omega_{\phi} = x^2 + y^2 \ ,
		\end{equation}
		and the Hubble derivative is given by:
		\begin{equation}
			\frac{\dot{H}}{H^2} = \frac{-3}{2} \left(\dfrac{\Omega_{\psi} + 2 x^2 + 4/3 \Omega_{r} + \Omega_{b}}{\Omega_{\rm GB} + 1}\right)\ .
			\label{H_deriv}
		\end{equation}	
		We now construct the autonomous system for the interacting fields. The autonomous system of equations is given by:
		\begin{eqnarray}
			x' & = & -3x - \frac{3 y^2 \lambda z}{\sqrt{6}} - \frac{3}{2} \frac{\xi(z)}{(m^2 + \xi(z))} \frac{n_c \lambda z \Omega_{\psi} }{\sqrt{6}}
			  + \frac{\Omega_{\rm GB}}{2x} \left(\frac{\dot{H}}{H^2} +1\right) - x \frac{\dot{H}}{H^2} \ , \label{x_prime}\\
			y' & = & \frac{1}{2}\lambda y z \sqrt{6} x - y \frac{\dot{H}}{H^2}\ , \label{y_prime}\\
			z' & = & \sqrt{6} x \ , \label{z_prime}\\
			\Omega_{\rm GB}' & =& \Omega_{\rm GB} \left(1 + \frac{\dot{H}}{H^2}\right) \ , \label{gb_prime}\\
			\Omega_{r}' & = & - 4 \Omega_{r} - 2 \Omega_{r} \frac{\dot{H}}{H^2} \ , \label{omega_r_prime}\\
			\Omega_{b}' & = & -3 \Omega_{b} - 2 \Omega_{b} \frac{\dot{H}}{H^2} \ . \label{b_prime}
		\end{eqnarray}
		Here, \(()' \equiv \frac{d}{dN}\), where \(dN = H\,dt\), and \(\xi(z) = M \exp\left(\frac{\lambda z^2 n_c}{2}\right)\). To close the system, we use the relation derived in the previous section, \((m^2 + \xi(\phi)) \langle \psi^2 \rangle = \langle \rho_{\psi} \rangle\). The dynamics of the \(\psi\) field can then be obtained using the Hubble constraint as:
		\begin{equation}
			\Omega_{\psi} = 1 - \Omega_{\phi} + \Omega_{\rm GB} - \Omega_{r}-\Omega_{b} \ .
			\label{psi_evo_eq}
		\end{equation}
		The fractional energy densities corresponding to the fields and fluids must lie within the range \([0,1]\), whereas \(\Omega_{\rm GB}\) can take any value in the range \([-1,1]\). In this work, our approach is entirely focused on numerically determining the dynamics of the system and constraining the model parameter space against observational data. Therefore, we assess the stability of the system based on the numerical evolution of the dynamical variables.
		
\begin{figure}[t]
			\centering
			\includegraphics[scale=0.79]{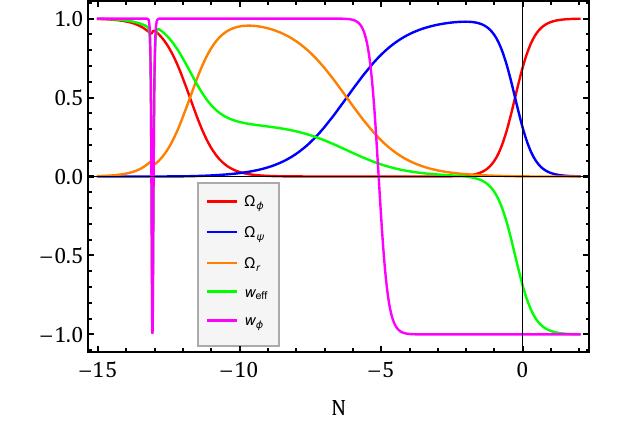}
			\caption{The evolution of the minimally coupled quintessence field for the exponential potential with the initial conditions \(\lambda = 4.0, x_0 = 10^{-6}, y_0 = 0.8327, z_0 = 10^{-8}, \Omega_{\psi 0} = 0.306\). }
			\label{fig:quint_min_evo}
		\end{figure}

		First, we present the dynamics of the minimally coupled quintessence field by switching off both the interaction and the GB coupling terms. The evolution is shown in Fig.~\ref{fig:quint_min_evo} as a function of $N$. The plot includes relevant cosmological parameters such as the fractional energy densities of the quintessence field \(\Omega_{\phi}\), the fluid \(\Omega_{\psi}\), and radiation \(\Omega_{\rm r}\). To characterize the different epochs of the universe, we also plot the effective equation of state \(w_{\rm eff}\) alongside the field’s equation of state \(w_{\phi}\).
		
		The plot reveals that the energy density of the \(\phi\) field dominates the current epoch, while the \(\psi\) field energy density is subdominant, with \(\Omega_{\psi_0} = 0.306\) for the selected initial conditions. In the future, i.e., \(N \gg 1\), \(\Omega_{\phi}\) saturates to 1, indicating the stable nature of the model under the considered potentials. In the intermediate past, the matter energy density dominates and its corresponding equation of state (EoS) nearly vanishes. For \(N < -5\), the radiation energy density dominates, and the effective EoS is approximately \(1/3\). In the asymptotic past, the quintessence field dominates again, and the effective EoS approaches 1, representing stiff matter. 
		
		We observe that around \(N = -14\), the radiation density becomes subdominant, and the quintessence energy density begins to dominate due to its kinetic term \(\dot{\phi} \sim x\). However, the effect of field domination in the past can be regulated by tuning the initial condition on the parameter \(x\), thereby producing a proper radiation-dominated phase that satisfies necessary observational constraints. 
		
		Hence, the system exhibits a stable de Sitter phase in the asymptotic future and generates different epochs in the asymptotic past characterized by saddle behavior.\footnote{The term ``saddle'' refers to a region in phase space where trajectories initially approach a fixed point--representing a certain cosmological regime, for instance, a fixed point with matter fractional energy density dominance \(\Omega_{\psi} \sim 1\)--but subsequently move away, indicating repulsive behavior. In this context, the radiation phase becomes unstable while the matter-dominated phase exhibits saddle characteristics. Interested readers may consult refs.~\cite{Das:2019ixt,Shahalam:2017rit} for a comprehensive understanding of phase space analysis and observational implications.}

\begin{figure*}
			\centering
			\includegraphics[scale=0.9]{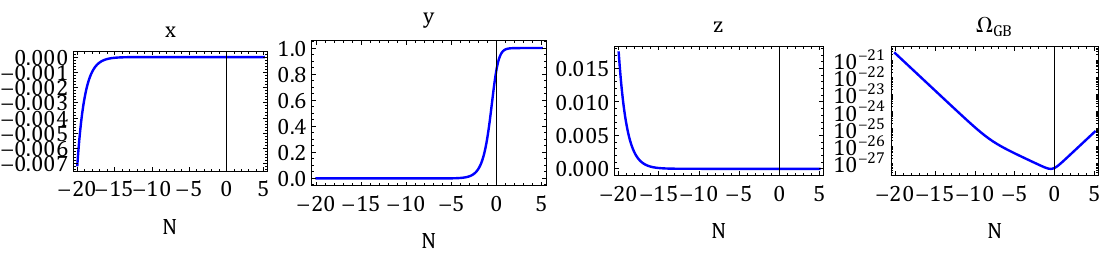}
			\includegraphics[scale=0.6]{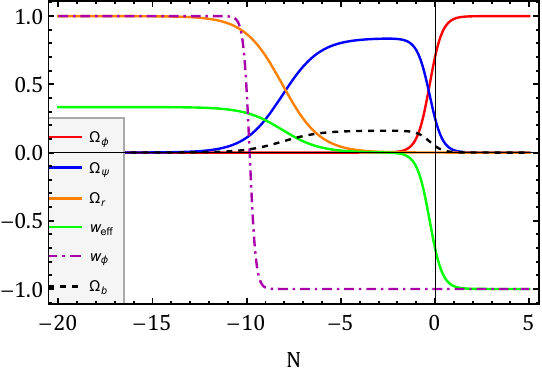}
			\includegraphics[scale=0.6]{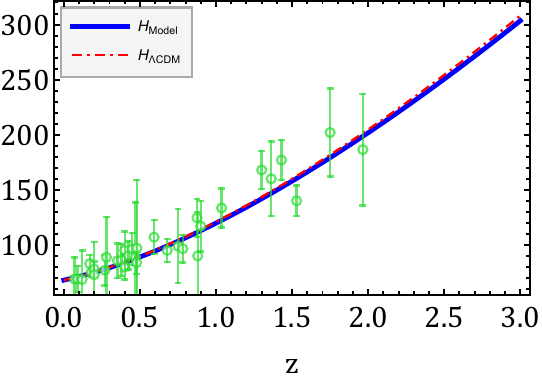}
			\caption{The evolution of the cosmological parameters for the Model I. In the figure $\ti{\Omega}_{A}(N) = \frac{\kappa^2 \rho_{A}(N)}{3H_0^2}$, where $\ti{\Omega}_{\phi}(0) = 0.70$ and $ H_0 = 68.0$. The initial conditions corresponding to the dynamical variables are \(x_0 = 10^{-18}, z_0 = 10^{-16}, \Omega_{b_0}h^2 = 0.0223\) and the parameters value are $\lambda = -2.0, n_c = 0.3, m=10^{-2}, M = 1.0, m_1 = 10^{-30}, N_{\rm eff}=3.046$. }
			\label{fig:evo_model1_numeric}
		\end{figure*}

		We now turn our attention to the interacting scenario, where the cosmological solutions are obtained by numerically evolving the autonomous system of Eqs.~\eqref{x_prime}--\eqref{b_prime}. {A detailed stability analysis and numerical evolution of Model I are presented in Appendix \ref{appen:stabilityofmodel}. We find that the current model produces an attractor stable phase for $0<|\lambda|, |n_c|<10$. However, for opposite signs of the parameters, the model yields consistent solutions only within a narrow region, namely $0<n_c<1$ and $-10<\lambda<0$ or $-10< n_c <0$ and $0 < \lambda<1$. We also observe that certain parameter combinations can produce significantly different evolutionary behaviors, such as an extended radiation-dominated phase or enhanced baryonic peaks. These scenarios are physically unacceptable, as they may drive other density components to negative values. Therefore, meaningful constraints on the parameter space must be obtained through confrontation with observational data. Before proceeding to the observational analysis, we outline our strategy for testing the model against cosmological data sets.
			
			In Appendix \ref{appen:stabilityofmodel}, the initial conditions were set deep in the radiation epoch. However, certain parameters, such as $y_i$ or $z_i$ (where the subscript denotes the value at the initial time), cannot be uniquely specified. Their values influence the scalar-field energy density profile, and therefore a two-dimensional shooting method would be required to ensure that, within the allowed range of $y_i$ and $z_i$, the field energy density matches the desired value at the current epoch and simultaneously satisfies the Hubble constraint relation. This matching procedure is computationally expensive and can require many hours of numerical computation.
			
			To avoid the shooting method, we instead sample the dynamical variables at the present epoch. From the dynamical evolution shown in Fig.~\ref{fig:dyn_evo_model1_lam_nc} and related figures, one observes that although $y$ is initially set to a small value, it dynamically evolves and approaches a finite value at late times. Therefore, the present value of \(y\) essentially determines the current dark energy density. In contrast, \(x\) remains extremely small throughout the evolution and exhibits only mild variation. Since the present baryon density is already constrained observationally, and the scalar-field density is defined as $\Omega_\phi = x^2 + y^2$, we treat $\Omega_{\phi,0}$ as a free parameter and fix the value of (x(0)), from which $y(0)$ can be determined.
			
			In the present analysis, we fix the radiation density using the Planck value $N_{\rm eff} = 3.046$, from which ${\Omega}_{r,0}$ is determined. The baryon density is constrained by varying $\Omega_{b,0}h^2$, where $h \equiv H_0/100$. The Gauss–Bonnet density is constrained by sampling its parameter $m_1$, since from Eq.~\eqref{dyn_variable}, $\Omega_{\rm GB,0} = 8 m_1 H_0$, with $a(0)=1$. The Hubble constant is constrained by sampling $H_0$.
			
			The remaining variable $z$ can be fixed similarly to $x$. From its dynamical evolution, $z_i$ may start from an arbitrary value in the early epoch, but as the system approaches the late-time attractor, $z$ becomes very small and remains in the low-value regime. Hence, for the parameter ranges of $\lambda$ and $n_c$ that lead to stable attractor solutions, the trajectories of $z$ are insensitive to its initial value. Therefore, fixing $z(0)$ does not reduce the dynamical freedom of the system.
			
			Sampling the dynamical variables at the present epoch for an attractor system significantly reduces computational cost, as no shooting method is required. This increases the computational efficiency by several orders of magnitude without compromising the physical consistency of the model.
			
			Before implementing this structure in the Python pipeline, we numerically demonstrate that for an attractor system, imposing initial conditions at $N=0$ reproduces the same past evolution when extrapolated backward. The evolution is shown in Fig.~\ref{fig:evo_model1_numeric}, where the dynamical variables follow nearly identical trajectories and attain comparable magnitudes when extrapolated to $N=-20$.}

		Comparing Figs. \ref{fig:quint_min_evo} and \ref{fig:evo_model1_numeric} we can see clearly that once the interaction between the dark sector is introduced,
		the evolution of the universe has a very desirable behavior not only in the current epoch but also in the radiation- and matter-dominated epochs. 
		Without such interactions, Fig. \ref{fig:quint_min_evo} shows that the universe was dominated by a stiff fluid when $N < 13$. In addition, with the interaction, a long period of matter-dominated epoch exists, as shown explicitly in  Fig. \ref{fig:evo_model1_numeric}, which provides an effective mechanism for the formation of large-scale structure, sharply in contrast to the case without interactions between the two dark components  \cite{TerenteDiaz:2023iqk}. 
		
		Additionally, we compute the sound horizon distance, a crucial standard ruler in cosmology that constrains baryon density and, indirectly, the GB coupling parameter as well as \(z_0\) and \(x_0\). Using constraints on these parameters, we evaluate the sound horizon at the drag epoch \(r_d\) and the recombination epoch \(r_s\), corresponding approximately to redshifts \(\tilde{z} \sim 1059.93\) and \(1089.92\), respectively. The Planck collaboration reports values for \(\Lambda\)CDM as \(r_d = 147.21\) Mpc and \(r_s = 144.60\) Mpc for \(H_0 = 67.66\) km/s/Mpc \cite{Planck:2018vyg}. The sound horizons at both epochs are calculated as \cite{DESI:2024mwx} \footnote{In our current model, the dark sector interacts with photons and baryons only gravitationally, so we have $\dot\rho_i + 3H(\rho_i + P_i) = 0, (i = \gamma, b)$.
			Then, for super-Hubble modes it can be shown  that $\delta\rho_{\gamma}/(\rho_{\gamma} + P_{\gamma}) = \delta\rho_{b}/\rho_{b}$ \cite{Lesgourgues:2013qba}. Since $\delta P_{\gamma} \gg \delta P_{b}$, we find that the speed of the photon-baryon plasm can be written as $c_s^2 \equiv (\delta P_{\gamma} + \delta P_{b})/(\delta\rho_{\gamma} +\delta\rho_{b}) \simeq 1/[3(1 + 3\rho_b/(4\rho_{\gamma})]$. Then,  the radius of the sound horizon of the photon-baryon plasm defined by
			$r_s \equiv \int_{z_s}^{\infty}{c_s(z)/H(z) dz}$ can be written in the form of Eq.(\ref{sound_distance}), where $z_s$ denotes the redshift at the recombination time. The same argument will lead to the expression for $r_d$.}: 
		\begin{equation}
			r_{d,s} =\int_{ z_{d,s}}^{\infty} \dfrac{3 \times 10^{5} d\ti{z}}{H \sqrt{3\left(1 + \frac{3 \Omega_{b_0}h^2}{4 \Omega_{\gamma_0}h^2(1+\ti{z})}\right)}} \ ,
			\label{sound_distance}
		\end{equation}
		where \(\Omega_{\gamma_0} h^2\) denotes the photon density parameter and \(h \equiv H_0/100\), with the value \(2.472 \times 10^{-5}\) \cite{Chen:2018dbv}. 
		For the current model, with the initial conditions specified in Fig.~\ref{fig:evo_model1_numeric}, {the sound horizon distances at the two epochs are
			\begin{equation}
				r_d = 148.06 \ \mathrm{Mpc}, \quad r_s = 145.34 \ \mathrm{Mpc},
			\end{equation}
			for \(H_0 = 68.0\) km/s/Mpc, 
			demonstrating that the current interacting model produces sound horizon distances similar to those of \(\Lambda\)CDM when the Hubble parameter is chosen close to the Planck value.}	
		
		\subsection{Model II: \boldmath \(\xi(\phi) = M (\kappa \phi)^{n_c}\)}
		
		For this interacting model, the autonomous equations from Eq.~\eqref{y_prime} to Eq.~\eqref{b_prime} remain unchanged. The only modification occurs in the equation for \(x'\), which is given by
		\bqn 
		x' &=&  -3x - \frac{3 y^2 \lambda z}{\sqrt{6}} - \frac{3}{2 \sqrt{6}} \frac{M n_c (z)^{n_c - 1 } \Omega_{\psi}}{(m^2 + M z^{n_c})} \nb\\
		&& + \frac{\Omega_{\rm GB}}{2x} \left(\frac{\dot{H}}{H^2} +1\right) - x \frac{\dot{H}}{H^2} \ . \label{x_prime_mod2}
		\eqn 
		
		{We determine the model's dynamics and stability conditions in Appendix \ref{appen:model2_stability}. The dynamics show that for comparable values of $-3 < \lambda < 10$ and $n_c \in [-30,40]$, the model does not produce a late-time accelerating solution. Consequently, $\Omega_{\psi}$ remains the dominant component of the universe, and the corresponding effective EoS remains zero, exhibiting a dust-like behavior. Since observational analyses indicate the existence of an accelerated expansion of the universe, this parameter range, although leading to a stable system, will not be able to confront the observational data.
			
			Nevertheless, the interacting model also features a stable late-time de-Sitter phase when the potential parameter becomes of the order of $<10^{-11}$. In this range, the interaction parameter can take any value $n_c \in [-3,3]$. From the dynamical analysis, we find that no significant change is visible in the density evolution (see Figs. \ref{fig:dyn_evo_model2_neglam_nc} and \ref{fig:dyn_evo_model2_neglam_negnc}); hence, the model remains insensitive to parameter variations. Although no significant change is visible, we shall still vary these model parameters while carrying out the observational analysis. However, we fix the value of ($z(0) = 10^{4}$), as the dynamical evolution shows that the variation in $z$ is negligible across the redshift. We follow a similar procedure for this model while confronting it with observational data. }

		\begin{figure*}
			\centering
			\includegraphics[scale=0.9]{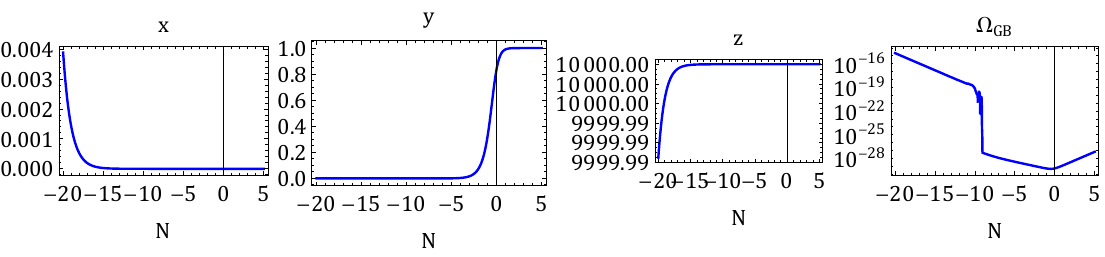}
			\includegraphics[scale=0.6]{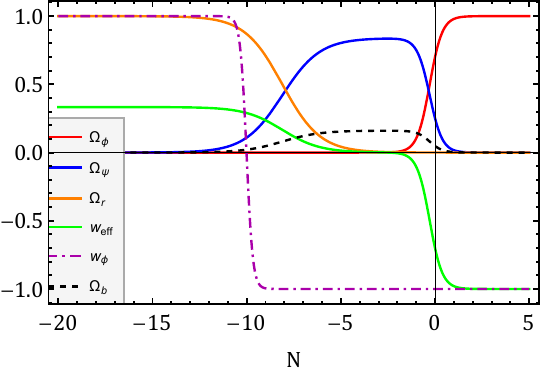}
			\includegraphics[scale=0.6]{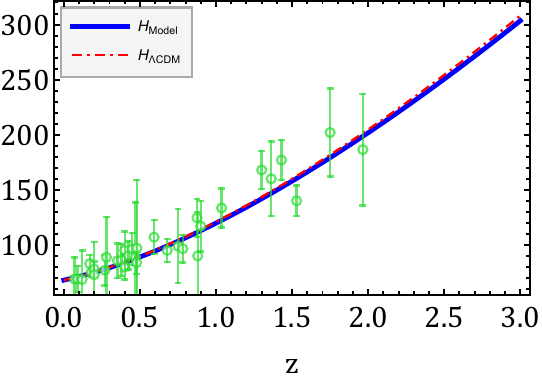}
			\caption{The evolution of the cosmological parameters for the Model II. In the figure $\ti{\Omega}_{A}(N) = \frac{\kappa^2 \rho_{A}(N)}{3H_0^2}$, where $\ti{\Omega}_{\phi}(0) = 0.70$ and $ H_0 = 68.0$. The initial conditions corresponding to the dynamical variables are \(x_0 = 10^{-13}, z_0 = 10^{4}, \Omega_{b_0}h^2 = 0.0223\) and the parameters value are $\lambda = -10^{-18}, n_c = -2.3, m=1, M = 1.0, m_1 = 10^{-31}, N_{\rm eff}=3.046$. }
			\label{fig:evo_model2_numeric}
		\end{figure*}
		
		{Before proceeding to the next section, we wish to show numerically that, for the considered parameter range where the model exhibits an accelerating epoch, setting the initial conditions at a particular order of magnitude produces dynamics similar to those shown in Fig. \ref{fig:dyn_evo_model2_neglam_negnc}, and yields three distinct phases. The evolution of the cosmological parameters is shown in Fig.~\ref{fig:evo_model2_numeric}. Based on this numerical evolution, we will use a similar order of initial conditions for $x_0$ and $z_0$ to carry out the observational analysis.
			
			Within the range $\lambda < -10^{-16}$ and $n_c < -1.5$, the model exhibits a stable de-Sitter solution at late times, along with unstable matter- and radiation-dominated phases at early epochs. We find that for positive values of $\lambda$ of the same order, the model becomes unstable irrespective of the choice of $n_c$.

			We calculate the sound horizon at two different epochs for \(H_0=68.0\) km/s/Mpc,
			\begin{equation}
				r_d = 148.06 \ \mathrm{Mpc}, \quad r_s = 145.34 \ \mathrm{Mpc},
			\end{equation}
			demonstrating that the current model exhibits behavior similar to that of \(\Lambda\)CDM.}

		\section{Data Analysis}
		\label{sec:data_presentation}
		In this section, we outline the observational data sets used to constrain the model parameters via the MCMC method.
		
		\begin{itemize}
			\item \textbf{CC Data:} This data set contains 32 model-independent observational points corresponding to the Hubble parameter. We refer to this as Cosmic Chronometers (CC) \cite{Vagnozzi:2020dfn,Jimenez:2001gg,Moresco:2015cya}. {We use the covariance matrix constructed in \cite{Moresco:2012jh,Moresco:2015cya,Moresco:2016mzx} to evaluate the likelihood.\footnote{The Python code to construct the covariance matrix for the $15$ highly correlated samples is available at \url{https://gitlab.com/mmoresco/CCcovariance}}}
			
			\item \textbf{PP Data:} This data set corresponds to the Pantheon+ compilation, which consists of 1550 spectroscopically confirmed Type Ia Supernovae \cite{Brout_2022}. The catalog contains 1770 data samples, from which we use the observational column corresponding to the non-SH0ES calibrated apparent magnitude \(m_{\rm obs}\). {We apply a redshift cut $z > 0.01$, neglecting the nearby samples that contain larger error bars. Additionally, we incorporate the Cepheid-calibrated samples; as a result, the total number of samples reduces to 1657. To compute the likelihood, we analytically marginalize over $M_B$ following the prescription in \cite{Goliath:2001af}, as implemented in the \texttt{Cobaya} repository\footnote{Likelihood estimation code: \url{https://github.com/CobayaSampler/cobaya/blob/master/cobaya/likelihoods/sn/pantheonplus.py}.}.} We refer to this set as ``PP''.

			\item \textbf{DES Data:} This data set also includes Type Ia Supernovae, but from a different observation -- the Dark Energy Survey (DES-SN5YR) -- which includes 1829 distinct SNe \cite{DES:2024jxu,DES:2024hip}. It consists of 194 nearby SNe samples with redshift \(z < 0.1\) and 1635 DES SNe samples. {However, the initially released data set showed significant tension with the Pantheon+ sample, and no correlation matrix was provided. After the release of this manuscript, we were notified that the earlier version of the sample has been replaced with an improved data set, now referred to as DES-Dovekie. This updated sample consists of 197 SNe Ia at low redshift and 1623 DES likely SNe Ia. The sample is approximately 20\% more robust than the earlier release and exhibits only slight tension with the Pantheon+ sample. The improved data set also shows a weak preference for dynamical dark energy over a cosmological constant when analyzed with $w_0 w_a$CDM and $\Lambda$CDM models \cite{DES:2025sig}. The new sample, along with the full covariance matrix and updated likelihood estimation, is available in the GitHub repository.\footnote{ The updated  \href{https://github.com/des-science/DES-SN5YR/tree/main/4_DISTANCES_COVMAT}{DES-Dovekie sample} and likelihood calculation can be found here-- \url{https://github.com/des-science/DES-SN5YR/blob/main/5_COSMOLOGY/Dovekie_cosmosis_likelihood.py}.} } 
			
			We compute the likelihood using the data column corresponding to the distance modulus \(\mu\), along with the full covariance matrix. The distance modulus is defined as:
			\begin{equation}
				\mu \equiv m_{B}-M_{B} = 5\log(D_L/\text{Mpc}) + 25 \  ,
			\end{equation}
			where, \(m\) denotes the apparent magnitude of the supernova and \(D_L\) is the luminosity distance: 	
			\begin{equation}
				D_L({z}) = c(1+{z}) \int_0^{{z}} \frac{dz'}{H(z')} \ ,
			\end{equation} 
			assuming a flat FLRW metric, and \(c\) is the speed of light in km/s. The model parameters are constrained by minimizing the chi-square ($\chi^2$) likelihood, defined as:
			\begin{equation}
				-2 \ln (\mathcal{L}) = \chi^2 = \rm \Delta D_{i} \mathcal{C}^{-1}_{ij} \Delta D_j\ ,
			\end{equation}
			where $\rm	\Delta D = \mu_{\rm Obs} - \mu_{\rm Model}  .$

			\item \textbf{DES BAO:} This set includes samples of Baryon Acoustic Oscillations (BAO) from the Dark Energy Spectroscopic Instrument (DESI) Release II \cite{DESI:2025zgx}, which is an enhanced version of DR1 \cite{DESI:2024mwx,DESI:2019jxc,Moon:2023jgl}. The observables correspond to the ratios \(\{D_M/r_d, D_H/r_d, D_V/r_d\}\), where \(D_M\) is the comoving angular diameter distance, \(D_H\) the Hubble distance, \(D_V\) the spherically averaged distance, and \(r_d\) the sound horizon at the drag epoch \cite{eBOSS:2020yzd,DESI:2024mwx}.

			\item \textbf{PLA Data:} This data set is derived by compressing the full Planck data set \cite{Planck:2018vyg} as reported in \cite{Chen:2018dbv}. The observables correspond to distance priors -- the acoustic scale \(l_A\) and shift parameter \(R\). These priors effectively encode information about the CMB temperature power spectrum. The acoustic scale characterizes the angular scale of the sound horizon, influencing peak spacing, while the shift parameter affects the line-of-sight direction, impacting peak heights. The shift parameter is defined as \cite{Elgaroy:2007bv}: 
			\begin{equation}
				R(z_*) \equiv \frac{D_A(z_*) \sqrt{\Omega_{m_0} H_0^2}}{c}\ ,
			\end{equation}
			where \(z_* = 1089.92\) is the redshift at photon decoupling \cite{Planck:2018vyg}, and \(D_A\) is the angular diameter distance for a flat geometry given by:
			\begin{equation}
				D_A = c \int_{0}^{z} \frac{dz}{H(z)} \ .
			\end{equation}
			Here, \(\Omega_{m_0}\) denotes the total matter density at present. For the minimally coupled case, where both baryonic and dark matter are pressureless and individually conserved, baryon density does not explicitly appear in the Hubble parameter \(H = H_0 \sqrt{\Omega_\Lambda + \Omega_{m_0} (1+z)^3 + \Omega_r (1+z)^4}\). However, in the current interacting scenario, the total matter density is \(\Omega_{m_0} = \Omega_{\psi_0} + \Omega_{b_0}\), since the non-gravitational coupling only modifies the dark matter profile while baryons remain pressureless with \(\Omega_b \propto (1+z)^3\). Therefore, care must be taken when applying such models.\footnote{Many previous studies have ignored baryon contributions when analyzing dynamical systems or constraining parameters with late-time data, which might be acceptable for late-time analysis but leads to inaccuracies for early-time observations.} The acoustic scale is defined as:
			\begin{equation}
				l_A = \frac{\pi D_A(z_*)}{r_s(z_*)}
			\end{equation}
			where \(r_s\) is the comoving sound horizon at the photon decoupling epoch. Along with these, the compressed CMB data includes \(\Omega_{b_0} h^2\). The data set includes the correlation matrix between \(\{R, l_A, \Omega_{b_0} h^2\}\), whose inverse covariance matrix is reported in the appendix of \cite{Chen:2018dbv}. The chi-square is calculated as:
			\begin{equation}
				\chi^2_{\rm PLA} = \sum \rm (D_i^{ obs}- D_i^{th}) C_{ij}^{-1} (D_j^{ obs}- D_j^{th}) \ .
			\end{equation}
			{In the current study, we use the compressed dataset, which has been further improved in \cite{Arendse:2019hev} by including an additional observable. The sample is extended to a four-parameter likelihood, $\{100,\Omega_b h^2, 100,\theta_{*}, R, \Omega_{\rm dm} h^2\}$, along with its correlation matrix. This extension has been shown to be efficient in constraining models beyond $\Lambda$CDM. The current likelihood replaces the acoustic scale with the angular scale defined as $\theta_{*} = r_s(z_*)/D_A(z_*)$ and introduces an additional likelihood parameter corresponding to the dark matter density $\Omega_{\rm dm}$. Additionally, we compute the redshift at recombination and at the photon–baryon decoupling epoch using the Hu–Sugiyama fitting formula \cite{Hu:1995en}
				\begin{equation}
					\begin{aligned}
						& z_{\mathrm{d}}=1345 \frac{\left.\left(\Omega_{\mathrm{m}} h^2\right)^{0.251}\left[1+b_1\left(\Omega_{\mathrm{b}} h^2\right)^{b_2}\right)\right]}{1+0.659\left(\Omega_{\mathrm{m}} h^2\right)^{0.828}}\ , \\
						& b_1=0.313\left(\Omega_{\mathrm{m}} h^2\right)^{-0.419}\left[1+0.607\left(\Omega_{\mathrm{m}} h^2\right)^{0.674}\right] \ , \\
						& b_2=0.238\left(\Omega_{\mathrm{m}} h^2\right)^{0.223} \ .
					\end{aligned}
				\end{equation}
				and 
				\begin{equation}
					\begin{aligned}
						& z_*=1047\left[1+0.00124\left(\Omega_{\mathrm{b}} h^2\right)^{-0.738}\right]\left[1+g_1\left(\Omega_{\mathrm{m}} h^2\right)^{g_2}\right] \\
						& g_1=0.0783\left(\Omega_{\mathrm{b}} h^2\right)^{-0.238}\left[1+39.5\left(\Omega_{\mathrm{b}} h^2\right)^{0.763}\right]^{-1} \\
						& g_2=0.56\left[1+21.1\left(\Omega_{\mathrm{b}} h^2\right)^{1.81}\right] .
					\end{aligned}
				\end{equation}
				It should be noted that $\Omega_m$ represents the total matter density including the current density of baryon and dark matter $\Omega_{m} = \Omega_{b} + \Omega_{\psi}$.\footnote{Reader can find the python code to estimate the likelihood in the GitHub repository \url{https://github.com/sleonardokap/cosmological_data/tree/Main-execution-files/BAO\%20data}} 
			}
			
			\item \textbf{Roman Data:} This dataset comprises 20,824 data points gathered from the forthcoming Roman Space Telescope, which will provide future insights into cosmological parameters. A crucial element of the mission is the High Latitude Wide-Area Survey, which includes both imaging and spectroscopic elements. The intended High Latitude Time-Domain Survey by Roman enables the detection and light-curve monitoring of thousands of Type Ia supernovae at redshifts up to $z \approx 3$. Currently, we can produce this dataset by utilizing the simulation features of the SNANA software \cite{Kessler:2009yy} alongside the PIPPIN pipeline manager, based on the assumption of a $\Lambda$CDM cosmological model. The specifics related to the Roman simulations were acquired from discussions concerning a manuscript that is being prepared by the Roman Supernova Project Infrastructure Team. For further information regarding this dataset, we refer readers  to the references \cite{Kessler:2025eib,Hussain:2024yee,Hounsell:2023xds}. 
			
		\end{itemize}
		For the above data sets, we compute the joint likelihood corresponding to two combinations of data: 
		(i) CC + DESBAO + PLANCK + PP, and 
		(ii) CC + DESBAO + PLANCK + DES, (iii) CC+ DESBAO + PLANCK + ROMAN, as
		\begin{equation}
			-2 \ln \mathcal{L}_{\rm tot} = \chi^2_{\rm tot}.
		\end{equation}
		{For both Supernova data sets, PP and DES, no direct constraint on $H_0$ can be obtained, as the data are degenerate with the absolute magnitude $M_B$. Similarly, the DESI BAO data are insufficient to constrain $H_0$, since they are degenerate with the sound horizon scale $r_d$. Hence, we combine these data sets with an external measurement, such as the Hubble data, which acts as an anchor. However, due to the complexity of the model, relying only on late-time data sets becomes infeasible for probing the consistency of the model across all redshifts and cannot adequately constrain other sensitive parameters, such as the GB coupling parameter or the baryon matter density. Therefore, including the Planck data provides a stronger anchor that not only constrains $H_0$ but also places reliable bounds on the model parameters and reduces their degeneracies.}
		
		We estimate the likelihood by implementing the model in Python and use the publicly available affine-invariant Markov Chain Monte Carlo (MCMC) ensemble sampler \texttt{emcee} \cite{Foreman-Mackey:2012any} to obtain the posterior distributions of the model parameters. The resulting samples are then analyzed using \texttt{GetDist} \cite{Lewis:2019xzd} to obtain marginalized 1D and 2D posterior distributions.
		
		Finally, we compare the statistical preference of the current model relative to the flat \(\Lambda\)CDM model using information criteria such as the Akaike Information Criterion (AIC) and the Bayesian Information Criterion (BIC) \cite{Akaike:1974vps,bic_criterion,Trotta:2008qt}:
		\begin{align}
			\mathrm{AIC} &= -2 \ln \mathcal{L}_{\rm max} + 2k, \\
			\mathrm{BIC} &= -2 \ln \mathcal{L}_{\rm max} + k \ln N,
		\end{align}
		where \(k\) is the number of model parameters, \(N\) is the total number of observational data points, and \(\mathcal{L}_{\rm max}\) is the maximum likelihood.
		
		\section{Results}
		\label{sec:result}
		
		{A detailed strategy to constrain both models has been laid out in the previous section. Therefore, we constrain the following parameters of both models using the above-mentioned data sets: \(\{\Omega_{\phi}, H_0, \lambda, n_c, \Omega_{b}h^2, m_1\}\). We adopt uniform priors on these parameters within the intervals listed in Tab.~\ref{tab:priors}, while fixing the remaining parameters.}
		
		\begin{table}[t]
			\centering
			\begin{tabular}{|lrr|}
				\hline
				\hline
				\multicolumn{2}{c}{Model I}&  Model II \\
				\hline
				Parameters & \multicolumn{2}{c}{Range}  \\
				\hline
				$\Omega_{\phi}$ & [0.5, 1.0] & [0.5,1.0]\\
				$H_0$  & [30,100] & [30,100]\\
				
				$\lambda$ & $[0,4.0]$, & $[-10^{-22}, -10^{-14}]$\\
				$n_c$ & $[-1.5,5.5]$ & $[-4.0,3.0]$\\
				$ \Omega_{b}h^2$ & $[10^{-5}, 0.1]$ & $[10^{-5}, 0.1]$\\
				$10^{m_1}$ & $[-50, -19]$ & $[-39,-20]$\\
				$z_0$ & $10^{-16}$ & $10^{4}$ \\
				$x_0$ & $10^{-18}$ & $10^{-13}$\\
				$m$ & $10^{-2}$ & 1 \\
				$M$ & 1 & 1 \\
				$N_{\rm eff} $ & $3.046$ & $3.046$ \\
				\hline
				\hline
			\end{tabular}
			\caption{The priors range of the model parameters and values of the fixed parameters. }
			\label{tab:priors}
		\end{table}
		
\begin{figure}
			\centering
			\includegraphics[scale=0.6]{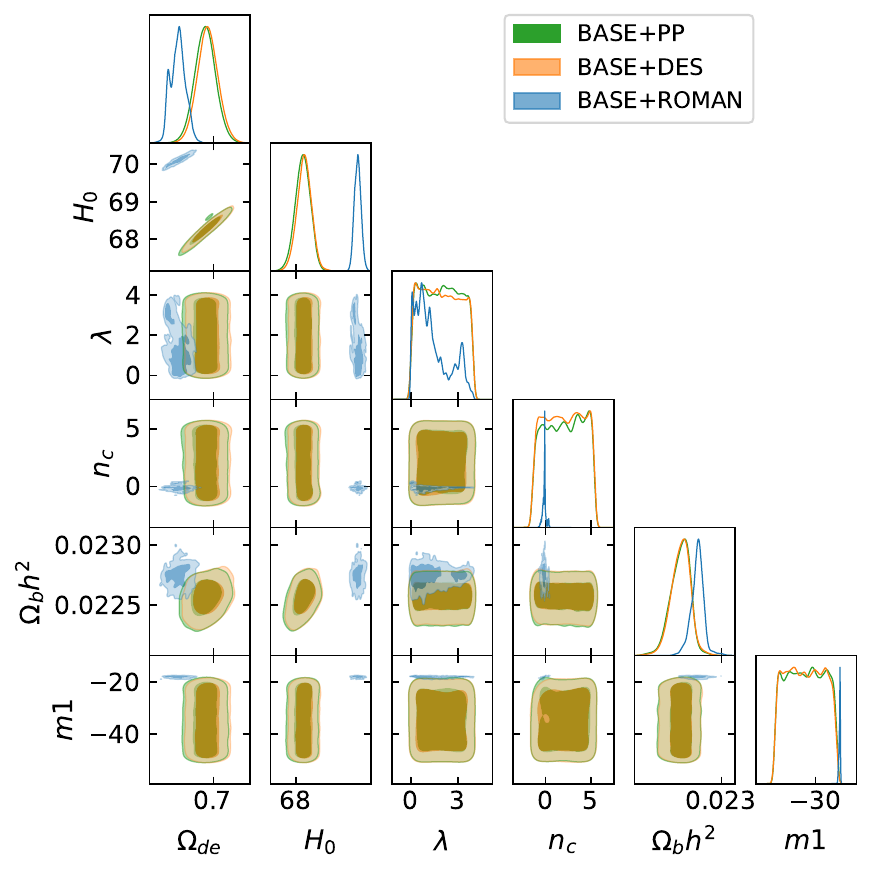}
			\caption{1$\sigma$, 2$\sigma$ distribution of parameter estimation of model I. }
			\label{fig:model1_corner_plots}
		\end{figure}

\begin{figure}
			\centering
			\includegraphics[scale=0.6]{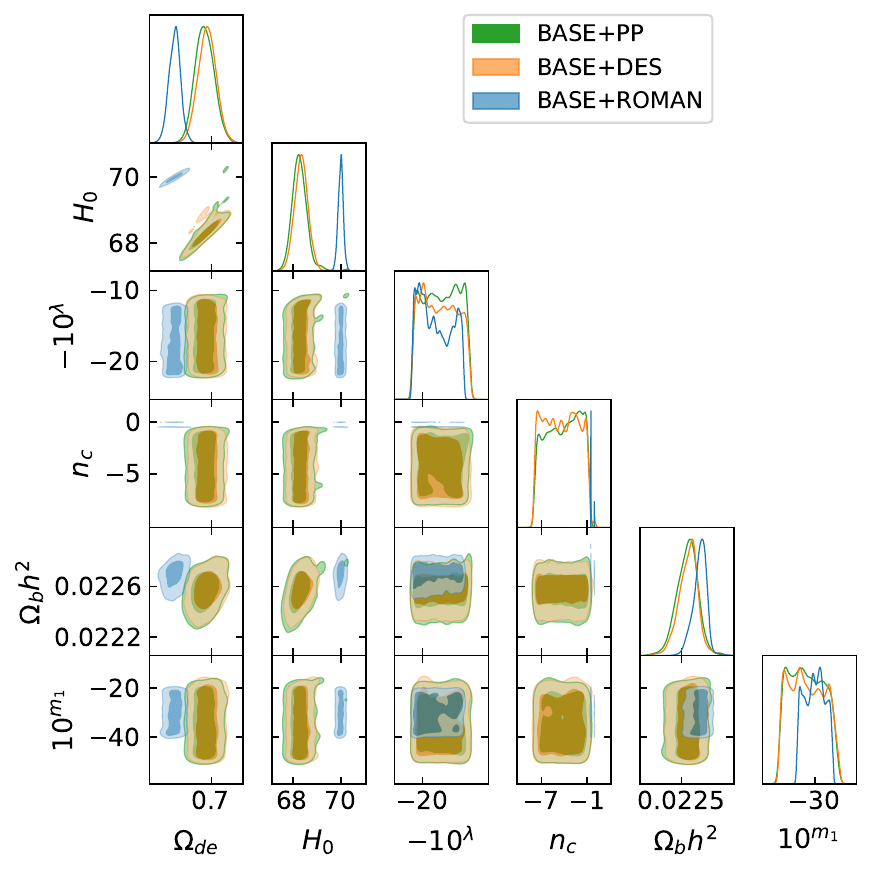}
			\caption{1$\sigma$, 2$\sigma$ distribution of parameter estimation of the Model II.}
			\label{fig:model2_corner_plots}	
		\end{figure}

		{The best-fit values of the model parameters at the 68\% confidence level are presented in Tab.~\ref{tab:model_best_fit} for all three combinations of data sets. We select the combination CC + DESI BAO + PLANCK (BASE) as the baseline data set, common to both models, and then include the supernova data separately for comparison with the flat \(\Lambda\)CDM model. The marginalized posterior distributions for all models obtained from the MCMC analysis are shown in Figs.~\ref{fig:model1_corner_plots} and \ref{fig:model2_corner_plots}. We do not show the corner plot for flat $\Lambda$CDM and only report its best-fit values in Tab.~\ref{tab:model_best_fit}.}

		\begin{table}[t]
			\centering
			\begin{tabular} { l  c c c}
				\noalign{\vskip 3pt}\hline\noalign{\vskip 1.5pt}\hline\noalign{\vskip 5pt}
				\multicolumn{1}{c}{\bf Model I } &  \multicolumn{1}{c}{\bf BASE+PP} &  \multicolumn{1}{c}{\bf BASE+DES} & \multicolumn{1}{c}{\bf BASE+Roman}\\
				
				\noalign{\vskip 3pt}\cline{2-4}\noalign{\vskip 3pt}
				
				Parameters &  68\% limits &  68\% limits & 68\% limits\\
				\hline
				{\boldmath$\Omega_{\phi}  $} & $0.6973\pm 0.0034         $ & $0.6981\pm 0.0034          $ & $ 0.6884^{+0.0023}_{-0.0036}$\\
				
				{\boldmath$H_0            $} & $68.23\pm 0.26              $ & $68.29\pm 0.25             $& $70.10 ^{+0.10}_{-0.12}$\\
				
				{\boldmath$ \lambda    $} & $2.0 \pm 1.1 $ & $1.9 \pm 1.2$ & $1.4^{+2.0}_{-1.4}$ \\
				
				{\boldmath$n_c    $} & $2.1 \pm 2.0 $ & $2.1 \pm 2.0$ & $-0.12^{+0.12}_{-0.17}$\\
				
				{\boldmath$\Omega_{b}h^2  $} & $0.02256^{+0.00012}_{-0.000084}        $ & $0.02257^{+0.00011}_{-0.000087}    $ & $0.022744^{+0.000070}_{-0.000054}$\\
				
				{\boldmath$10^{m_1}           $} & $-34.9\pm 8.8 $ &                   $-34.7 \pm 8.7$  & $-18.14^{+0.36}_{-0.22}$        \\
				\hline
				\hline
				\multicolumn{4}{c}{\textbf{Model II}}\\
				\hline
				\hline
				{\boldmath$\Omega_{\phi}  $} & $0.6974 \pm 0.0037$ & $0.6984 \pm 0.0036$ & $0.6857 \pm 0.0023$\\
				
				{\boldmath$H_0            $} & $68.30 ^{+0.25}_{-0.34} $ & $68.37 \pm 0.33$ & $69.98 \pm 0.11$\\
				
				{\boldmath$ -10^{\lambda }$} & $-16.4 \pm 3.2 $ & $-16.7^{+3.1}_{-4.9}$ & $-17.4^{+5.1}_{-4.4}$ \\
				
				{\boldmath$n_c    $} & $-4.1^{+3.2}_{-2.1} $ & $-4.3 \pm 2.1$ & $-3.70^{+0.088}_{-0.10} $\\
				
				{\boldmath$\Omega_{b}h^2  $} & $0.02256^{+0.00012}_{-0.00010}        $ & $0.02257^{+0.00011}_{-0.00009}    $ & $0.022699^{+0.000083}_{-0.000053} $\\
				
				{\boldmath$10^{m_1} $} & $-34.4^{+9.6}_{-14.0} $ &                   $-34.3 \pm9.4$  & $-30.1 \pm 5.6$   \\
				\hline
				\hline
				\multicolumn{4}{c}{\boldmath \bfseries$\Lambda$CDM}\\
				\hline
				\hline
				{\boldmath$\Omega_{\phi}  $} & $0.6982\pm 0.0036        $ & $0.6989\pm 0.0036          $ & $0.7192\pm 0.0017$\\
				
				{\boldmath$H_0            $} & $68.67\pm 0.28               $ & $68.73\pm 0.28            $ & $71.314\pm 0.099$\\

				{\boldmath$\Omega_{b}h^2  $} & $0.02259^{+0.00013}_{-0.00011}         $ & $0.02260^{+0.00012}_{-0.00011}       $ &$0.02420^{+0.00011}_{-0.000083}$\\

				\hline
				\hline
			\end{tabular}
			\caption{The best fit value of the model parameters at $68\%$ level.}
			\label{tab:model_best_fit}
		\end{table}
		
\begin{figure}
			\centering
			\includegraphics[scale=0.45]{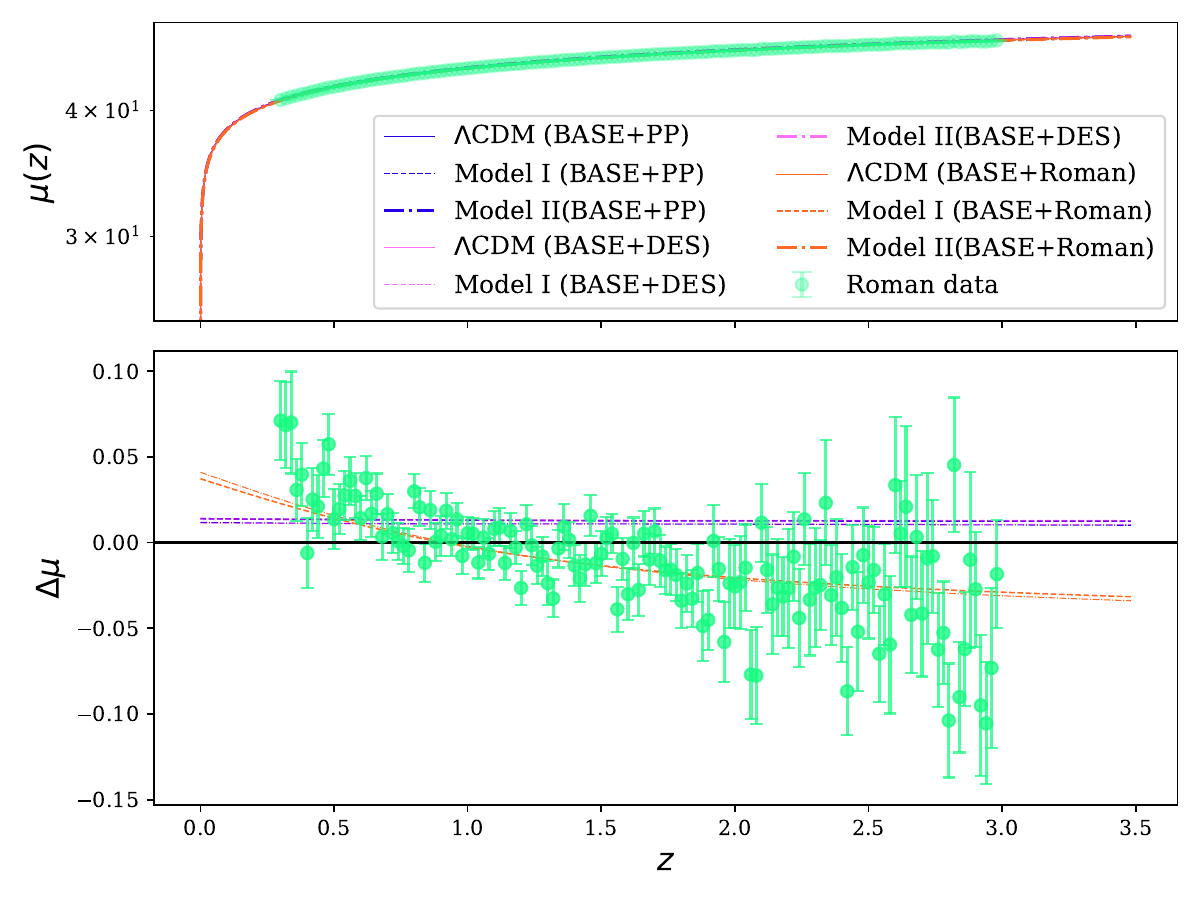}
			\caption{
				Distance modulus \(\mu\) for both interacting models and the \(\Lambda\)CDM model, plotted using the best-fit parameters from Tab.~\ref{tab:model_best_fit}. The curves are shown against the Roman mock binned observational data. The residuals are defined as $\Delta\mu \equiv \mu_{\Lambda \mathrm{CDM}} - \mu_{\mathrm{Obs}}$ and $\Delta\mu_{\mathrm{Model}} \equiv \mu_{\mathrm{Model}} - \mu_{\Lambda \mathrm{CDM}}$.
			}
			\label{fig:roman_distance_modulus}
		\end{figure}

		{Model I yields a Hubble parameter value around $68.26$ km/s/Mpc for the PP and DES data sets, whereas a value of $70.10$ km/s/Mpc is obtained for the Roman data. These values are slightly lower than those obtained for the flat $\Lambda$CDM model. From the corner plot in Fig.~\ref{fig:model1_corner_plots}, it can be seen that the potential, interaction, and GB parameters become insensitive and remain unconstrained for the PP and DES samples, whereas a tight constraint is observed for the Roman data set. This clearly indicates that the future Roman survey will be able to distinguish between dynamical dark energy models and $\Lambda$CDM. The 2D distribution plots show that the value of $H_0$ for the Roman simulated data is in tension with the rest of the data at the $>2\sigma$ level, whereas the other parameter values overlap within the $2\sigma$ region for this model.
			
			From the dynamical analysis of Model II, we obtain the range of the potential parameter for which the system produces a stable accelerating solution during the late-time epoch; consequently, we set the prior range of the potential parameter accordingly. We find that the considered model yields values similar to Model I for the dark energy density, Hubble parameter, baryon density, and GB parameter. The marginalized parameter distribution plot in Fig.~\ref{fig:model2_corner_plots} shows that the flat 2D distributions signify that $\lambda$ and \(n_c\) remain unconstrained, and their variation does not significantly alter the system's dynamics. The data also indicate that a GB coupling parameter below a certain limit (in this case $m_1 < -25.0$) makes the model dynamically robust due to its small magnitude. In contrast to the previous model, even for the Roman data set these parameters remain unconstrained, although they yield an $H_0$ value close to $70.0$ km/s/Mpc.
			
			Both models exhibit only a minute deviation from $\Lambda$CDM. We therefore compare the model preference statistically based on the benchmarks reported in Tab.~\ref{tab:stat}. We report the Akaike Information Criterion (AIC), Bayesian Information Criterion (BIC), and the reduced chi-squared $\chi_{\rm rd} \equiv \chi^2_{\min}/\nu$, where $\nu = N_{\rm obs} - k$ denotes the number of degrees of freedom, with $N_{\rm obs}$ being the total number of observational data points and $k$ the number of free parameters. Values of $\chi_{\rm rd} \gtrless 1$ indicate poor (over) fitting, while $\chi_{\rm rd} \sim 1$ suggests a good fit.
			
			We also compare the AIC and BIC values of the current models with those of the $\Lambda$CDM model. The difference $\Delta \mathrm{IC} = \mathrm{IC}_{\rm Model} - \mathrm{IC}_{\Lambda \mathrm{CDM}}$ serves as a measure of relative model support, where $\Delta \mathrm{IC} < 2$ indicates substantial support for the model, values between 2 and 7 indicate weak support, and $\Delta \mathrm{IC} > 10$ suggests no support.
			
			The statistical analysis indicates that both Model I and Model II show a weak preference over $\Lambda$CDM according to the AIC, but no support based on the BIC criterion for the PP and DES data sets. This apparent penalization arises primarily due to the inclusion of additional free parameters in the sampling procedure, which are explicitly penalized by information criteria such as BIC. If these additional parameters are fixed, the corresponding information criteria are significantly reduced and tend to align more closely with those of the standard $\Lambda$CDM model. Therefore, despite the comparatively higher BIC values, the models cannot be regarded as moderately unsupported relative to $\Lambda$CDM.
			
			For the Roman mock samples, however, both models show strong statistical support over $\Lambda$CDM, suggesting that future high-precision surveys may provide meaningful evidence for physics beyond the standard cosmological model.
			
			A visual representation is shown in Fig.~\ref{fig:roman_distance_modulus}, where the distance modulus $\mu$ is plotted for all models using their respective best-fit values against the Roman mock samples. The plot shows that both models exhibit only a minute deviation for the PP and DES data sets, while for Roman a significant departure is noticeable at higher redshift $z>1$. The evolution of the density parameters for both models is similar to that shown in Figs.~\ref{fig:evo_model1_numeric} and \ref{fig:evo_model2_numeric}; hence, we do not explicitly present the evolution corresponding to the best-fit values.
			
			One of the underlying reasons the models do not show a significant deviation from the baseline model and do not exhibit phantom behavior is that the GB coupling function naturally evolves to very small values at low redshifts when constrained by observational data.}
		As a result, the influence of the GB term becomes negligible in the present epoch, and the theory approaches standard GR. At higher redshifts, although the coupling strength increases (as do interaction terms), allowing for non-trivial dynamics, the parameter space required to maintain stability suppresses the GB term enough to prevent it from pushing the EoS into the phantom regime.\\		
		This balancing act is critical. In previous GB or interacting scalar field (fluid) models, instabilities often arose even without entering the phantom regime \cite{Valiviita:2008iv,Jackson:2009mz,Wang:2016lxa}. To mitigate such issues, we carefully constrain the GB parameters to ensure that the system remains free from ghost or gradient instabilities while still matching cosmological observations. In particular, the model respects the gravitational wave speed constraint $c_T^2 = 1$, ensuring compatibility with the stringent bounds.\\
		Therefore, the absence of phantom behavior in our results is a \textit{consequence of both theoretical consistency and observational viability}, rather than a deficiency of the model.

		\begin{table}[t]
			\centering
			\begin{tabular}{|l | c | c | r| r|r|}
				\hline
				Model & AIC & BIC  & $\chi^2/\nu $ & $ \rm \Delta AIC$ & $\rm \Delta BIC$\\
				\hline
				I- D1 & 1741.66 & 1774.29 & 1.021 & 6.94 & 23.25 \\
				I- D2 & 1691.71& 1724.89 & 0.905 & 7.33 &  23.92\\
				RDP & 20979.84 & 21027.52 & 1.005 & $-416.41$ & $-392.56$ \\
				\hline
				II- D1 & 1741.44 & 1774.07 & 1.021&  6.72& 23.03\\
				II- D2 & 1691.33& 1724.51 & 0.904 & 6.95& 23.54\\
				RDP &  20981.22 & 21028.90 & 1.005 & $-315.03$ & $-391.18$\\
				\hline
				$\Lambda$CDM- D1 & 1734.72 & 1751.04 & 1.019 & 0 & 0\\
				$\Lambda$CDM- D2 & 1684.38 & 1700.97 & 0.902 & 0 & 0 \\
				RDP & 21396.25 & 21420.08 & 1.025 & 0 &0 \\
				\hline
				\hline
			\end{tabular}
			\caption{The statistical comparison of the models, where D1-- BASE+PP, D2--BASE+DES and RDP--BASE+ROMAN. }
			\label{tab:stat}
		\end{table}

		\section{Conclusions}
		\label{sec:conclusion}
		
		In this work, within the framework of Einstein scalar Gauss-Bonnet gravity,  we explore the non-gravitational interaction between dark matter and dark energy from an action principle by considering exponential and power-law types of interaction from the point of view of particle physics. We choose scalar-field models for both dark matter and dark energy, with quadratic and exponential potentials, respectively. The scalar field with the quadratic potential behaves as cold dark matter around the minima of the potential when its mass is greater than the Hubble parameter. The quintessence scalar field with an exponential potential produces the late-time cosmic acceleration. The quintessence field is also coupled to the Gauss-Bonnet  term via an arbitrary function \(f(\phi)\), whose form is constrained from gravitational wave observations, making the analysis model-independent.
		
		We constructed the autonomous system of equations to check the stability of the system {for two different interactions between the two dark components, and showed that both interacting models produce a stable de-Sitter solution in the later evolution of the universe.} { Additionally, we provide a detailed discussion on the choice of initial conditions corresponding to the field variables, which is applicable to general dynamical systems. Such a systematic discussion on the selection of initial conditions, particularly in studies aligned with observational constraints and aimed at ensuring numerical stability without explicitly deriving the critical points, has not been thoroughly addressed in earlier works.
			
			Both models successfully reproduce a $\Lambda$CDM-type behavior and constrain the GB coupling parameter to lie below certain thresholds, $\Omega_{\rm GB} \lesssim 10^{-25}$ at low redshift, remaining suppressed throughout the cosmic evolution (see Figs.~\ref{fig:evo_model1_numeric} and \ref{fig:evo_model2_numeric}).}  We further test the models using two combinations of data sets, incorporating both late-time and early-time cosmological observations. Our analysis shows that both models are consistent with current data and represent physically viable alternatives to the standard \(\Lambda\)CDM paradigm. Since our analysis includes high- and low-redshift observations, particularly those from the Planck observations, the Gauss–Bonnet  coupling function is tightly constrained. Consequently, its contribution becomes negligible at low redshifts, effectively recovering General Relativity, while it may become significant in the asymptotic past due to its growth at higher redshifts in both models. Throughout the cosmic evolution, the interacting models closely mimic \(\Lambda\)CDM, with the scalar field's equation of state remaining in the accelerating regime, ensuring ghost-free dynamics. For all data sets used to constrain the model parameters, the inferred Hubble constant \(H_0\) remains close to \(69.0\) km/s/Mpc, slightly higher than the Planck estimate, but still in tension with the SH0ES measurement.
		
		As shown explicitly in Figs.   \ref{fig:evo_model1_numeric} and \ref{fig:evo_model2_numeric}, when  interactions between the two dark components are introduced, 
		the models fit not only with current observational data well but also produce two well-separated and desirable epochs, the matter- and radiation-dominated epochs. This is consistent with our previous studies presented in \cite{Hussain:2024yee},  but sharply in contrast to the case without interactions \cite{TerenteDiaz:2023iqk}.  {In Fig.~\ref{fig:roman_distance_modulus}, the comparison of the two models with $\Lambda$CDM is further investigated using the distance modulus $\mu$ against the Roman mock data. It can be seen that {although both models mimic $\Lambda$CDM-type behavior for the PP and DES samples, they show a significant departure at higher redshifts when the Roman sample is included. This suggests that future Supernova surveys will be capable of probing and distinguishing alternative dark energy models from the standard $\Lambda$CDM scenario.
				
				Upon performing the statistical comparison through the evaluation of the AIC and BIC, we find that both models provide an excellent fit to the PP and DES data, as the reduced chi-squared values are very close to unity. However, the AIC indicates only weak preference, while the BIC shows no support for the current models over $\Lambda$CDM. In contrast, both information criteria show strong support for the present models over the baseline model when the Roman mock sample is included.}}

		Other important tests of the current models include the stability analysis of the interaction, the consistency of the models with current observations in terms of the slip parameter, lensing function and growth factor \cite{Ishak:2019aay}, and the possibilities of alleviating the Hubble tension and other anomalies found in current cosmological observations. Most of these issues are closely related to linear perturbations of the theory, and currently, following \cite{Kobayashi:2019hrl,Bellini:2014fua}, we are working on it, and expect to report our results on these important issues soon in another occasion, as such studies are clearly out of the  scope of the current paper. 
		
		Finally, we would like to note that tests of the theory with the Solar System and gravitational wave observations had already initiated. In particular, the solar system tests lead to the following constraint \cite{Amendola:2007ni}
		\begin{equation}
			\left|\alpha f_{,\phi}(\phi_0)\right| \lesssim 1.6 \times 10^{14}\; (\text{km})^2\,, 
		\end{equation}
		where $\phi_0$ is the current value of the scalar field $\phi$. However,  the observations of low-mass x-ray binary and gravitational waves imposed much severer  constraint \cite{Yagi:2012gp,Lyu:2022gdr} 
		\begin{equation}
			\left|\alpha f_{,\phi}(\phi_0)\right| \lesssim 1.18\; (\text{km})^2.
		\end{equation}
		It had been also shown that the theory is stable in the FLRW background  \cite{Fier:2025huc}.
		However, all these studies were carried out under the assumptions that the GB scalar field $\phi$ does not interact with other matter fields, except for the gravitational interaction. Once such interaction is included, the equations of perturbations will be changed. As a result, the stability conditions shall be also changed
		\cite{Kobayashi:2019hrl}. As mentioned above, we have been investigating these issues and hope to report such studies soon in another occasion.

		\begin{acknowledgments}
			\noindent	S.H. acknowledges the support of National Natural Science Foundation of China under Grants No. W2433018 and No.
			11675143, and the National Key Reserach and development Program of China under Grant No. 2020YFC2201503. SA acknowledges the Japan Society for the Promotion of Science (JSPS) for providing a postdoctoral fellowship during 2024-2026 (JSPS ID No.: P24318). This work of SA is supported by the JSPS KAKENHI grant (Number: 24KF0229). Y.R. acknowledges the support from Baylor University through the graduate program of Department of Physics and Astronomy. 
			B.R. is partially supported as a member of the Roman Supernova Project Infrastructure Team under NASA contract 80NSSC24M0023.  A.W. is partly supported by the US NSF grant, PHY-2308845.
			This work was completed, in part, with resources provided by the University of Chicago's Research Computing Center.
		\end{acknowledgments}
		
		\appendix
		
		\section{Stability analysis of the interacting models \label{appen:stabilityofmodel}}
		
		\subsection{Model I: \boldmath \(\xi(\phi) = M\exp\left(\frac{\lambda \kappa^2 \phi^2}{2}\right)^{n_c}\)}
		
		In this Appendix, we determine the stability of the interacting system at the background level. In the previous sections, the observational analysis was performed using the autonomous system written in terms of compact phase-space variables. However, due to the additional degrees of freedom, the dimensionality of the phase space exceeds three. Therefore, to carry out the stability analysis, we solve the autonomous system of equations numerically by fixing the initial
		conditions deep in the radiation epoch (i.e., $N \simeq -20$) and evolving the system toward the future epoch ($N > 0$). By varying the model parameters, we examine whether all dynamical variables remain well-behaved at late times. If any variable diverges abruptly during the evolution, the system is considered unstable.
		
		Conventionally, the stability analysis of an autonomous system is performed
		through the following procedure \cite{Bahamonde:2017ize,Das:2019ixt,Alho:2020cdg,Hussain:2024qrd}:
		\begin{itemize}
			\item The critical points of the autonomous system are obtained by setting
			the right-hand side of the coupled differential equations to zero, i.e.,
			$\vec{x}' = 0$.
			
			\item Some critical points may depend explicitly on the model parameters. These points can be physically classified by evaluating the corresponding energy density parameters and the effective equation of state. For example, if at a given critical point the scalar field density satisfies $\Omega_\phi > 0.5$ and the effective equation of state obeys $w_{\rm eff} < -0.3$, the point represents a dark energy dominated phase. By imposing such physical conditions, viable ranges of the model parameters
			can be identified.
			
			\item The stability of a critical point is then analyzed by linearizing
			the autonomous system around that point. Expanding to first order,
			the system can be written in matrix form, and the corresponding eigenvalues are determined. For a three-dimensional phase space, the Jacobian matrix is $3 \times 3$, yielding three eigenvalues $(E_1, E_2, E_3)$. If the real parts
			of all eigenvalues are negative (positive), the critical point is stable
			(unstable). If the eigenvalues have mixed signs, the point is a saddle. If one eigenvalue has vanishing real part while the others are negative,
			the linear analysis becomes inconclusive, and more advanced techniques,
			such as the center manifold method, are required.
			
			Instead of employing such analytical techniques, one estimates stability numerically by evolving the system in the vicinity
			of the critical points while varying model parameters or initial conditions.
			If the point is stable, small deviations decay and the dynamical variables
			converge toward the corresponding critical point, confirming the stability
			of the system.
		\end{itemize}
		
		For the present model, as stated earlier, instead of determining the critical points analytically, we demonstrate the stability of the system numerically by evolving it from the deep past to the future epoch while varying the model parameters. This procedure also allows us to explore correlations among the parameters. Accordingly, we solve the autonomous system of equations
		Eqs.~(\ref{x_prime})--\eqref{b_prime} in the range
		$N = -20$ to $N = 5$.
		
		Before evolving the system, it is important to specify consistent initial conditions in the early universe. In many studies involving complex dynamical systems, the discussion of initial conditions is either absent or incomplete. Therefore, we explicitly outline the procedure adopted here to ensure that the evolution closely reproduces the standard cosmological epochs.
		
		The primary dynamical variables of the system are
		$(x, y, z, \Omega_{\rm GB}, \Omega_r, \Omega_b)$,
		while the dark matter density is determined from the Friedmann constraint
		relation Eq.~\eqref{psi_evo_eq}. Owing to this constraint, the density
		parameters of all physical components lie within the interval $[0,1]$.
		We set the initial conditions deep in the radiation-dominated epoch,
		where radiation is the dominant component and the remaining contributions
		are subdominant.
		
		To ensure a realistic cosmological evolution from radiation domination to late-time acceleration, the initial conditions cannot be chosen arbitrarily.
		Instead, we use observational constraints on present-day density parameters
		and extrapolate them backward. For example, Planck observations constrain
		the present radiation density through the effective number of relativistic
		species $N_{\rm eff}$, according to
		\begin{equation}
			\Omega_{r0} h^2 = \Omega_{\gamma 0} h^2
			\left(1 + 0.227107\, N_{\rm eff}\right) \, .
		\end{equation}
		Early universe observations (CMB) give
		$N_{\rm eff} = 3.046$,
		$\Omega_{\gamma 0} h^2 = 2.472 \times 10^{-5}$,
		with $h \equiv H_0/100$.
		From the Friedmann constraint Eq.~\eqref{psi_evo_eq}, the sum of all density
		components must equal unity at any epoch. Therefore, deep in the radiation era we set $\Omega_r(N=-20) = 1 - \Omega_{r0}$, since all other components are extremely small at that time but not zero, hence $\Omega_r(N=-20) \ne 1$. Similarly, baryons and cold dark matter are negligible at $N=-20$, but they are not exactly zero. For the minimally coupled baryonic component, the density
		parameter evolves as
		\begin{equation}
			\Omega_{b}(N) = \frac{\kappa^2 \rho_b(N)}{3 H(N)^2} = \frac{\kappa^2 \rho_{b0} e^{-3 N} }{3 H_0^2 }\dfrac{H_0^2}{H^2} = \frac{\ti{\Omega}_{b0} e^{-3 N} H_0^2}{ H^2} \ ,
		\end{equation}
		where we have used the continuity equation for a minimally coupled fluid. Observational constraints on $\tilde{\Omega}_{b0} h^2$ allow us to determine its present value and hence extrapolate its early-time value. 
		
		A key step is determining the initial value of the Hubble parameter.
		Since radiation dominates at $N=-20$, the Hubble function can be safely
		approximated as
		\begin{equation}
			H(N) = H_0 \sqrt{\Omega_{r0} e^{-4 N}} \ .
		\end{equation}
		Including baryonic or CDM contributions produces only negligible corrections
		at such early times. This extrapolation procedure is valid only for
		minimally coupled matter components. If radiation were interacting with
		dark matter, for example, its energy-momentum tensor would not be separately conserved, and this simple backward extrapolation would not apply. 
		
		In the present model, the $\psi$-matter sector is non-minimally coupled
		to the scalar field $\phi$, so its density cannot be independently fixed.
		However, the Friedmann constraint relation automatically determines it,
		and therefore no additional initial condition is required.
		
		Another relevant variable is $\Omega_{\rm GB} = 8 m_1 e^{N} H$ (see Eq.~\eqref{dyn_variable}), where $\kappa = 1$ and $N=\ln a$.
		Since this expression contains a single free parameter $m_1$,
		its initial value can be directly specified once $m_1$ is chosen within
		its observationally allowed range.
		
		The auxiliary variable $z(N)$ does not enter explicitly in the Friedmann
		constraint and can be chosen such that the radiation, matter, and
		late-time accelerating epochs are clearly realized.
		A shooting method, similar to the one implemented in the CLASS code,
		or a controlled trial-and-error procedure may be adopted to determine
		a suitable value.
		
		The remaining variables correspond to the scalar field $\phi$.
		Its initial energy density can be safely assumed to be very small
		(e.g., $\lesssim 10^{-13}$, depending on the potential),
		ensuring that it remains subdominant during the radiation and matter
		epochs. As the system evolves dynamically, if the chosen potential
		admits an accelerating attractor solution for the selected parameter
		values, the scalar field eventually dominates at late times.
		In that regime, its dynamical variables approach constant values,
		signaling the presence of a stable attractor solution,
		provided that the other dynamical variables also asymptotically
		converge. Only in such a case can the full system be considered
		asymptotically stable at the background level.

\begin{figure*}[t]
			\centering
			\includegraphics[scale=0.7]{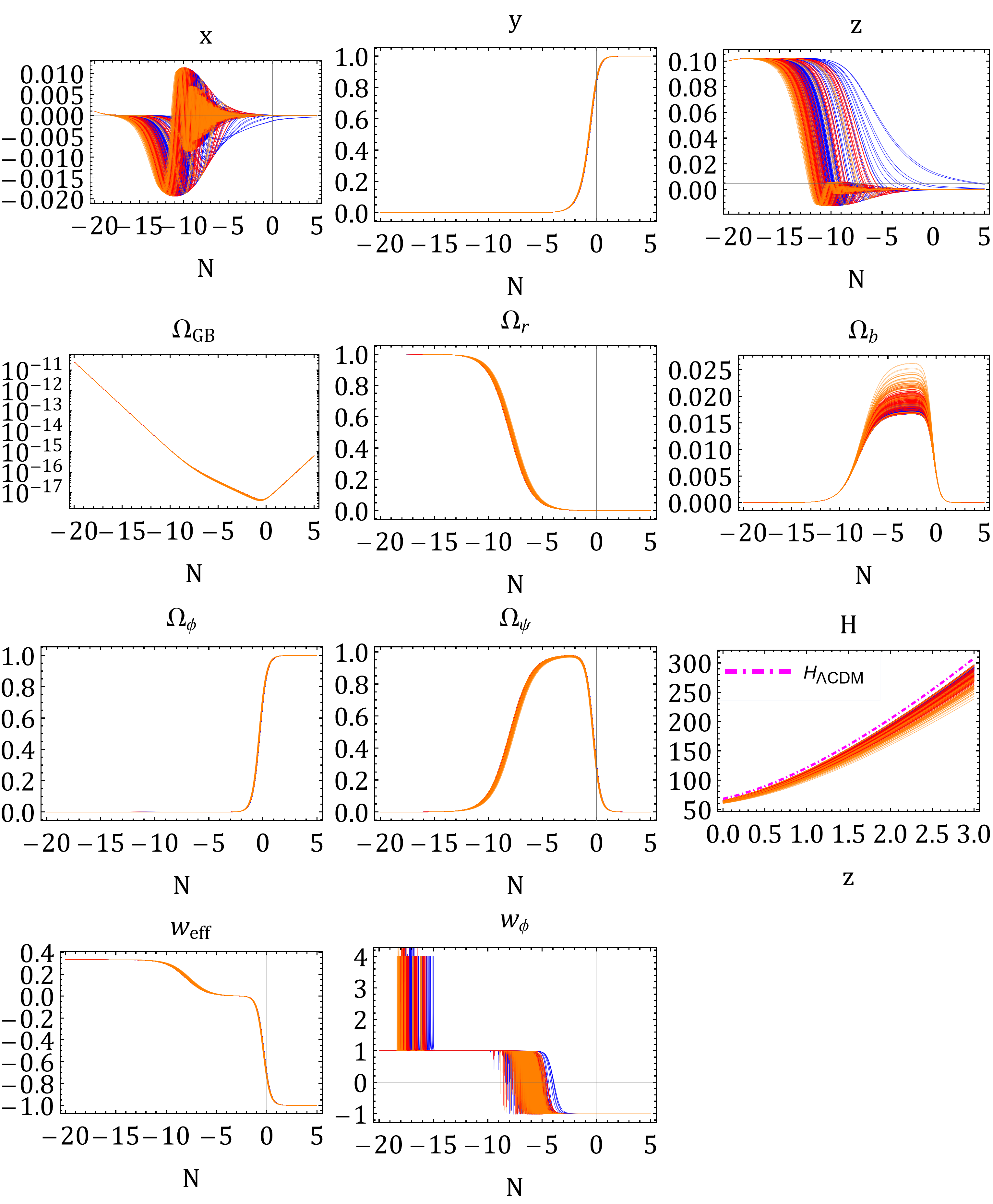}
			\caption{The evolution of the dynamical variables and the density parameters of Model I is estimated by fixing the initial conditions at $N=-20$, where $x(-20) = 10^{-3}, y(-20) = 10^{-16}, z(-20) = 0.1, m_1 = 10^{-20}, \Omega_{\rm GB} = 2.5 \times 10^{-11}, \Omega_{b}(-20) = 1.0989 \times 10^{-7}, H(-20) = 1.522 \times 10^{17}$, for $H_0 = 68 \ {\rm km/s/Mpc}, m = 10^{-2}, M=1$. Here, we randomly generate points for $\lambda$ in the range $0 \le \lambda < 16$, and for each $\lambda$, we determine the plot for $1 \le n_c \le 4$, $4 < n_c \le 8$, and $8 < n_c < 12$, which are shown in blue, red, and orange curves, respectively. The Hubble evolution is compared with the $\Lambda$CDM model $H(z) = 68.0 \sqrt{(1-0.31) + 0.31 (1+z)^3}$.}
			\label{fig:dyn_evo_model1_lam_nc}
		\end{figure*}

		{$\bullet$ \bf Impact of ($+\lambda$, $+n_c$):} To demonstrate the stability of the system, we illustrate the impact of the
		potential parameter $\lambda$ and the interaction parameter $n_c$ in
		Fig.~\ref{fig:dyn_evo_model1_lam_nc}. We vary $\lambda$ within the range $0$ to $16$, and for each value of $\lambda$, we consider three intervals for $n_c$, namely $n_c \in [1,4]$, $[4,8]$, and $[8,12]$. 
		
		From the figure, it is evident that the variable $x$, initially set to
		$\sim 10^{-3}$, exhibits only a very small variation throughout the evolution and eventually stabilizes near zero at late times. The variable $y$, which is initially close to zero, grows around $N \sim 0$ and saturates to unity in the future epoch, signaling the onset of scalar-field domination. For this evolution, the initial condition for $z$ is chosen as $z \sim 0.1$. It decays during the evolution and asymptotically approaches zero for all considered parameter combinations. We have verified that choosing different initial values of $z$ does not alter the qualitative behavior or the stability
		of the system.
		
		The initial condition for $\Omega_{\rm GB}$ is set by taking
		$m_1 < 10^{-20}$. We find that larger values do not yield physically viable solutions and typically lead to instability. Consequently,
		$\Omega_{\rm GB}$ remains extremely small throughout the evolution and
		decreases further at low redshift. Its small magnitude is primarily due
		to the stringent tensor speed constraint imposed through
		Eq.~\eqref{gw_bound}. In earlier studies, it has been shown that a sizable Gauss–Bonnet contribution can drive the system into a phantom regime ($w_{\rm eff} < -1$) even for a non-phantom scalar field
		\cite{Tsujikawa:2006ph}. However, those analyses were based on specific
		choices of the Gauss–Bonnet coupling function and did not incorporate
		gravitational wave constraints. In the present, model-independent scenario, the tight GW bound forces $\Omega_{\rm GB}$ to remain negligible, rendering the modified gravity effects practically indistinguishable from standard GR at the background level.
		
		Examining the density evolution, we observe that the $\psi$ component
		dominates during the intermediate redshift regime, while the scalar field
		$\phi$ begins to dominate near the present epoch and eventually saturates
		to unity in the far future. The effective equation of state evolves from
		$w_{\rm eff} \simeq 1/3$ in the radiation era to $w_{\rm eff} \simeq 0$
		during the matter-dominated phase (where $\Omega_\psi$ is dominant),
		and finally approaches $w_{\rm eff} \to -1$, indicating a stable de Sitter phase.
		
		The scalar field equation of state, defined in Eq.~\eqref{field_eos},
		\begin{equation}
			w_\phi = \frac{P_{\rm eff}}{\rho_\phi} = 
			\frac{x^2 - y^2
				-\frac{\Omega_{\rm GB}}{3}
				\left(\frac{-3(1+w_{\rm eff})}{2} + 1\right)}
			{x^2 + y^2} \, , \label{field_eos_eff_dyn}
		\end{equation}
		initially behaves as a stiff fluid ($w_\phi \simeq 1$). For certain
		parameter combinations, rapid oscillations appear in the range
		$N \sim -18$ to $-15$, and may also persist into the matter era.
		Nevertheless, these oscillations do not significantly affect the
		background dynamics and the field equation of state eventually
		approaches $w_\phi \to -1$ at late times. Although negligible at the
		background level, such oscillatory behavior could potentially leave
		imprints on structure formation, which warrants further investigation.
		
		Throughout the entire evolution, neither the effective equation of state  nor the scalar field equation of state crosses the phantom divide. We also compare the Hubble evolution with that of $\Lambda$CDM using the Planck-estimated value $H_0 = 68.0\,{\rm km\,s^{-1}\,Mpc^{-1}}$. The background expansion closely mimics the standard model, with mild deviations depending on the specific combination of $(\lambda, n_c)$. Therefore, observational constraints are required to determine the allowed parameter space.

\begin{figure*}[t]
			\centering
			\includegraphics[scale=0.6]{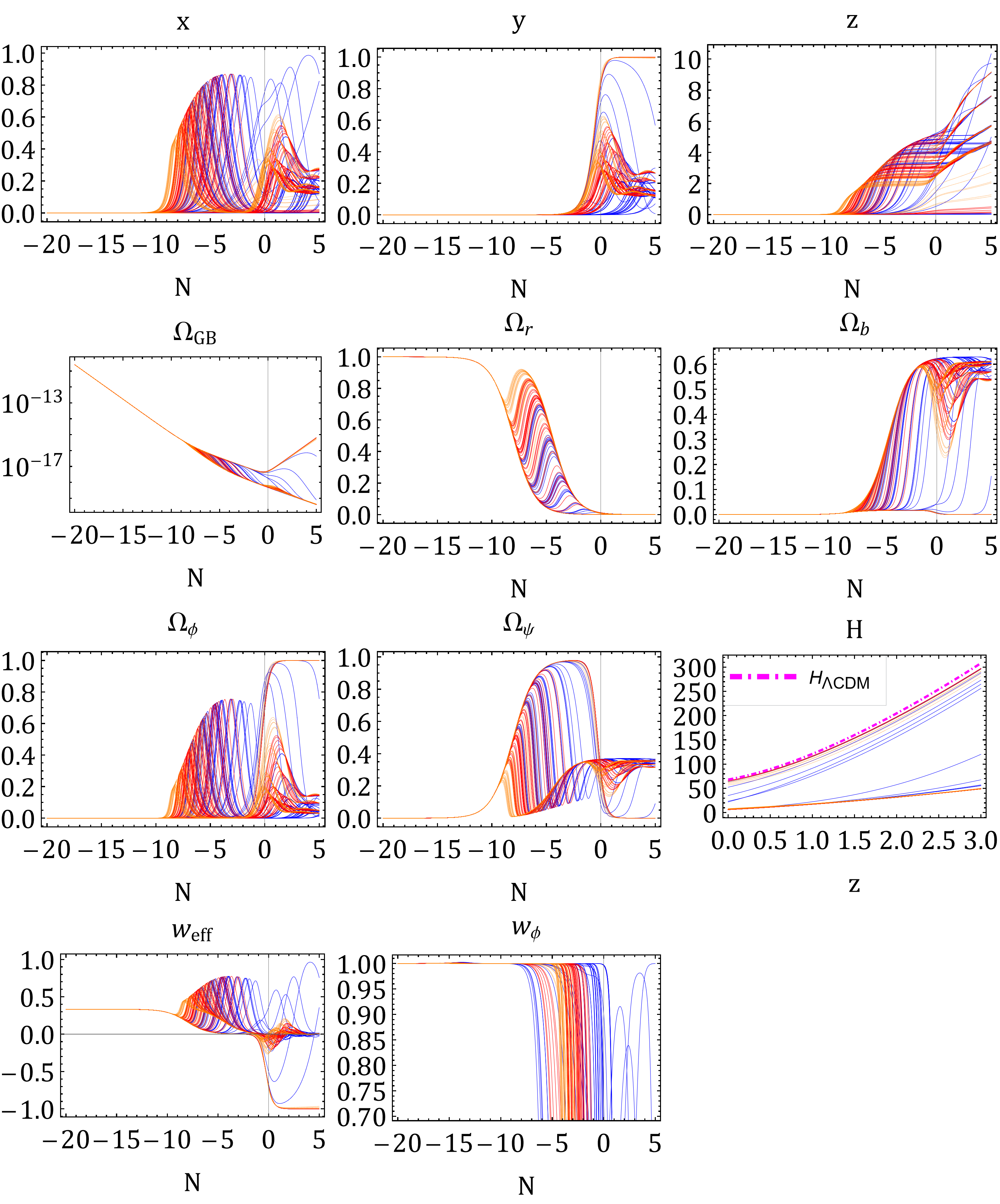}
			\caption{The evolution of the dynamical variables and the density parameters of Model I is estimated by fixing the initial conditions at $N=-20$, where $x(-20) = 10^{-3}, y(-20) = 10^{-16}, z(-20) = 0.001, m_1 = 10^{-20}, \Omega_{\rm GB} = 2.5 \times 10^{-11}, \Omega_{b}(-20) = 1.0989 \times 10^{-7}, H(-20) = 1.522 \times 10^{17}$, for $H_0 = 68 \ {\rm km/s/Mpc}, m = 10^{-2}, M=1$. Here, we randomly generate points for $\lambda$ in the range $-3.0 \le \lambda < -0.1$, and for each $\lambda$, we determine the plot for $1 \le n_c \le 4$, $4 < n_c \le 8$, and $8 < n_c < 12$, which are shown in blue, red, and orange curves, respectively. The Hubble evolution is compared with the $\Lambda$CDM model $H(z) = 68.0 \sqrt{(1-0.31) + 0.31 (1+z)^3}$.}
			
			\label{fig:dyn_evo_model1_neglam_nc}
		\end{figure*}

		{$\bullet$ \bf ($-\lambda$, $+n_c$):}  We determine the evolution for negative values of $\lambda$ \(\in (-3, -0.1)\) with $n_c \in [1,4],\ [4,8],\ [8,12]$, as shown in Fig. \ref{fig:dyn_evo_model1_neglam_nc}. We observe that, for certain blue curves corresponding to lower values of $n_c$, the Hubble evolution closely follows that of the standard model. In some cases, a few higher values of $n_c$ also mimic the standard cosmological evolution. However, in several instances, $\Omega_b$ dominates in the early epoch, which is not physically viable. Therefore, only specific ranges of negative $\lambda$ yield a consistent cosmological sequence, evolving from a radiation-dominated era to a stable de Sitter attractor phase.

\begin{figure*}[t]
			\centering
			\includegraphics[scale=0.6]{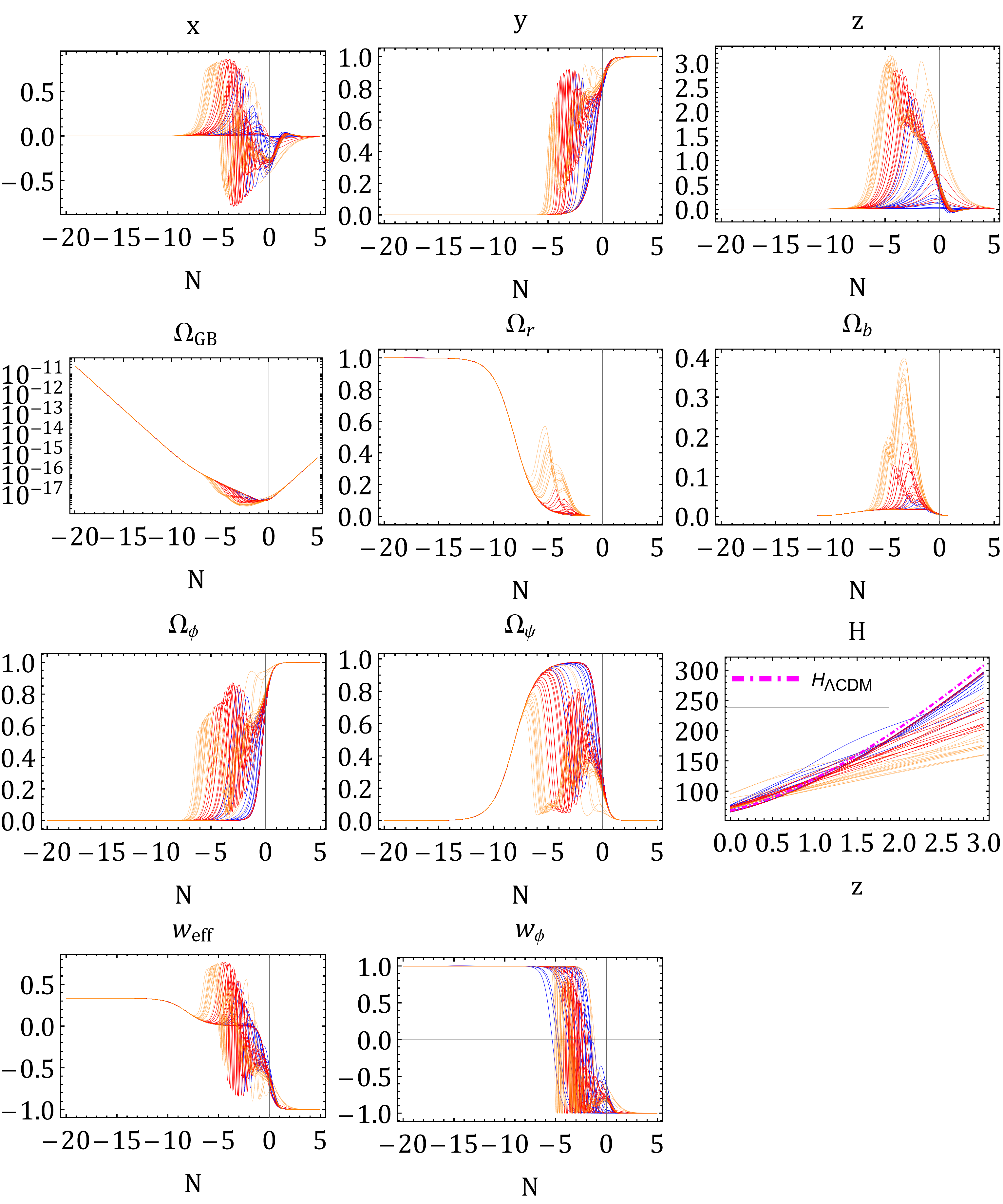}
			\caption{The evolution of the dynamical variables and the density parameters of Model I is estimated by fixing the initial conditions at $N=-20$, where $x(-20) = 10^{-3}, y(-20) = 10^{-16}, z(-20) = 0.001, m_1 = 10^{-20}, \Omega_{\rm GB} = 2.5 \times 10^{-11}, \Omega_{b}(-20) = 1.0989 \times 10^{-7}, H(-20) = 1.522 \times 10^{17}$, for $H_0 = 68 \ {\rm km/s/Mpc}, m = 10^{-2}, M=1$. Here, we randomly generate points for $\lambda$ in the range $0 < \lambda < 3.0$, and for each $\lambda$, we determine the plot for $-1 \le n_c \le -0.1$, $-2 < n_c \le -1$, and $-3 < n_c < -2$, which are shown in blue, red, and orange curves, respectively. The Hubble evolution is compared with the $\Lambda$CDM model $H(z) = 68.0 \sqrt{(1-0.31) + 0.31 (1+z)^3}$.}
			
			\label{fig:dyn_evo_model1_lam_negnc}
		\end{figure*}

		{$\bullet$ \bf ($+\lambda$, $-n_c$):} We determine the evolution for positive values of $\lambda$ \(\in (0, 3)\) with $n_c \in [-1,-0.1],\ [-2,-1],\ [-3,-2]$, as shown in Fig. \ref{fig:dyn_evo_model1_lam_negnc}. We observe that for nearly all parameter combinations, the dynamical variables saturate to constant values at late times, indicating the presence of a stable accelerating attractor solution. The overall dynamics is qualitatively similar to the previously discussed case.
		
		However, for certain parameter choices, the baryon density becomes excessively large during the evolution, rendering those scenarios observationally non-viable. In all cases, the Hubble evolution remains very close to that of $\Lambda$CDM, while still exhibiting small deviations. Therefore, although this parameter range presents an interesting dynamical behavior compared to the first case, only a restricted region of the parameter space can be considered physically viable, and a detailed observational analysis would likely disfavour most of this range statistically.

\begin{figure*}[t]
			\centering
			\includegraphics[scale=0.6]{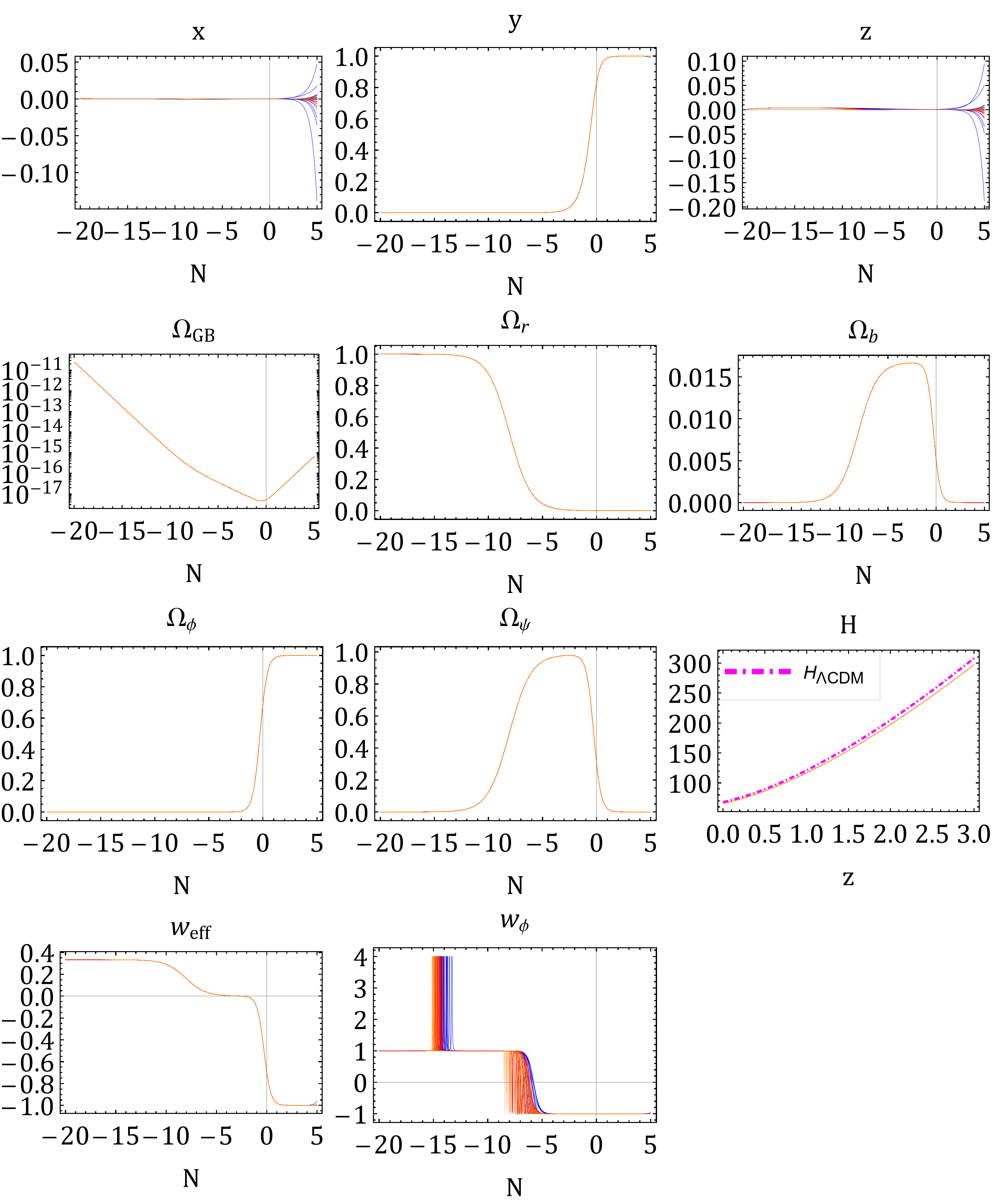}
			\caption{The evolution of the dynamical variables and the density parameters of Model I is estimated by fixing the initial conditions at $N=-20$, where $x(-20) = 10^{-3}, y(-20) = 10^{-16}, z(-20) = 0.001, m_1 = 10^{-20}, \Omega_{\rm GB} = 2.5 \times 10^{-11}, \Omega_{b}(-20) = 1.0989 \times 10^{-7}, H(-20) = 1.522 \times 10^{17}$, for $H_0 = 68 \ {\rm km/s/Mpc}, m = 10^{-2}, M=1$. Here, we randomly generate points for $\lambda$ in the range $-3 < \lambda < 0$, and for each $\lambda$, we determine the plot for $-1 \le n_c \le -0.1$, $-2 < n_c \le -1$, and $-3 < n_c < -2$, which are shown in blue, red, and orange curves, respectively. The Hubble evolution is compared with the $\Lambda$CDM model $H(z) = 68.0 \sqrt{(1-0.31) + 0.31 (1+z)^3}$.}
			
			\label{fig:dyn_evo_model1_neglam_negnc}
		\end{figure*}

		{$\bullet$ \bf $(-\lambda, -n_c)$:} We determine the evolution for negative values of $\lambda$ \(\in (-3,0)\) with $n_c \in [-1,-0.1],\ [-2,-1],\ [-3,-2]$, as shown in Fig. \ref{fig:dyn_evo_model1_neglam_negnc}. We observe that the overall dynamics is qualitatively similar to the $(+\lambda, +n_c)$ case. However, for certain blue curves corresponding to specific parameter choices, the variables $x$ and $z$ exhibit divergence at late times, indicating instability in those regions of the parameter space.
		
		Nevertheless, for the majority of parameter combinations, the system evolves toward a late-time stable attractor solution, with \(H(z)\) remaining very close to that of $\Lambda$CDM.
		
		It is worth noting that varying the dark matter mass parameter $m$ or the interaction parameter $M$ does not significantly alter the stability or the qualitative evolution of the system.

		\subsection{Model II: \boldmath \(\xi(\phi) = M (\kappa \phi)^{n_c}\)}
		\label{appen:model2_stability}

		We now determine the dynamics of Model II for different combinations of $\lambda$ and \(n_c\). As described earlier, we set the initial conditions deep in the radiation epoch and evolve the system toward the future. Unlike the previous case, for the power-law type interaction, the system becomes highly sensitive to the magnitude of the potential parameter $\lambda$.
		
\begin{figure*}[t]
			\centering
			\includegraphics[scale=0.6]{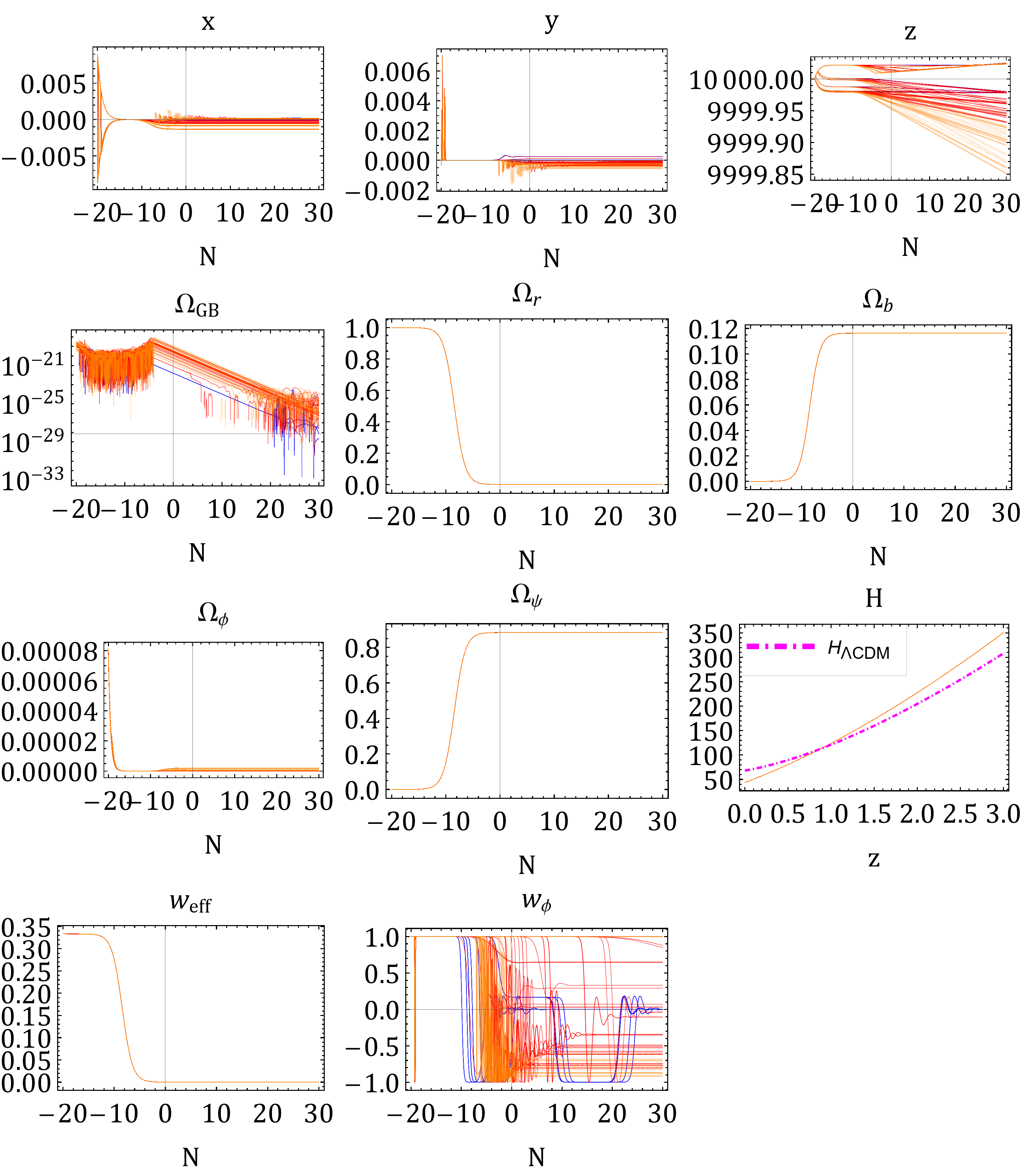}
			\caption{The evolution of the dynamical variables and the density parameters of Model II is estimated by fixing the initial conditions at $N=-20$, where $x(-20) = 0.009, y(-20) = 10^{-16}, z(-20) = 10^4, m_1 = 10^{-28}, \Omega_{b}(-20) = 1.0989 \times 10^{-6}, H(-20) = 1.522 \times 10^{17}$, for $H_0 = 68 \ {\rm km/s/Mpc}, m = 1, M=1$. Here, we randomly generate points for $\lambda$ in the range $-3 < \lambda < 10$, and for each $\lambda$, we determine the plot for $-30 \le n_c \le -0.1$, $0.1 < n_c \le 15$, and $15 < n_c < 40$, which are shown in blue, red, and orange curves, respectively. The Hubble evolution is compared with the $\Lambda$CDM model $H(z) = 68.0 \sqrt{(1-0.31) + 0.31 (1+z)^3}$.}
			\label{fig:dyn_evo_model2_biglam_nc}
		\end{figure*}

		For $\lambda \in [-3,10]$, we plot the evolution of the dynamical variables and cosmological parameters in Fig. \ref{fig:dyn_evo_model2_biglam_nc}, with $n_c$ spanning the range \(\in [-30,40]\). We find that near the present epoch, the variable $y$ becomes negligible. Consequently, the scalar field $\phi$ density remains suppressed, while $\Omega_{\psi}$ dominates throughout the redshift range $N > -8$. This leads to a prolonged matter-dominated phase in which the effective equation of state $(w_{\rm eff})$ remains close to zero, and the universe does not enter an accelerating regime.
		
		Although the scalar field does not dominate in this scenario, the evolution of \(w_\phi\) indicates that for certain combinations of \((\lambda, n_c)\), transient accelerating behavior may arise even when $y$ remains small. The system stabilizes at late times, as the parameters \((x, y, z, \Omega_{\rm GB})\) approach finite values without any abrupt variation over an extended period. In the evolution of (z), small changes are observed for some parameter choices; however, these variations occur extremely slowly, indicating an asymptotically stable configuration.
		
		Therefore, for larger values of $\lambda$, the system remains effectively matter dominated, and no transition to a late-time accelerating phase occurs. The Hubble evolution significantly deviates from that of $\Lambda$CDM, with substantial discrepancies appearing at both low and high redshifts. Although the system exhibits dynamical stability, the chosen parameter range fails to reproduce the observed late-time acceleration of the universe.

\begin{figure*}[t]
			\centering
			\includegraphics[scale=0.6]{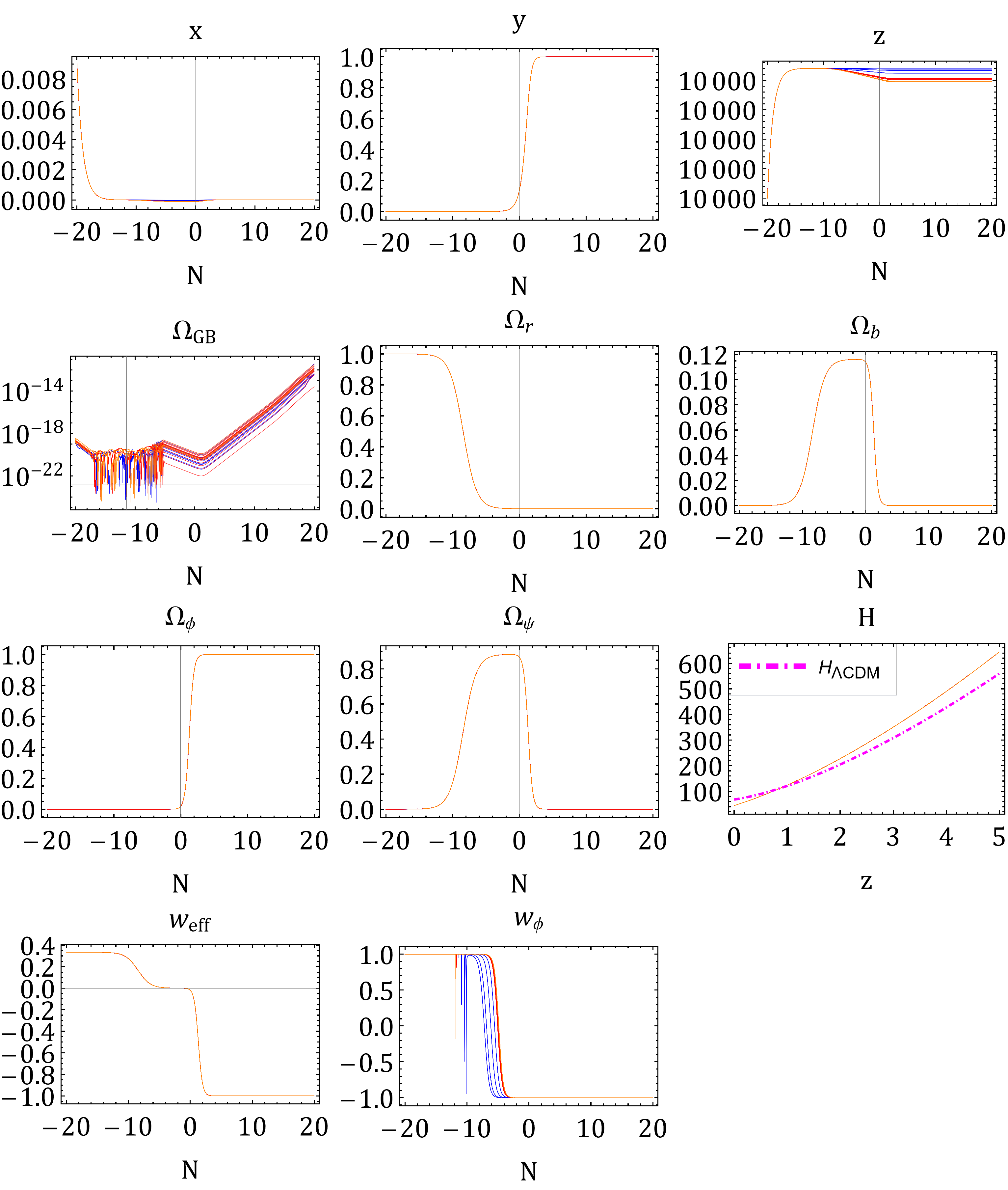}
			\caption{The evolution of the dynamical variables and the density parameters of Model II is estimated by fixing the initial conditions at $N=-20$, where $x(-20) = 0.009$, $y(-20) = 10^{-16}$, $z(-20) = 10^4$, $m_1 = 10^{-28}$, $\Omega_{b}(-20) = 1.0989 \times 10^{-6}$, and $H(-20) = 1.522 \times 10^{17}$, for $H_0 = 68 \ {\rm km/s/Mpc}$, $m = 1$, and $M = 1$. Here, we randomly generate points for $\lambda$ in the range $-10^{-18} < \lambda < -10^{-11}$, and for each $\lambda$, we determine the plots for bins of $0.1 \le n_c \le 3.0$, shown in blue, red, and orange curves. The Hubble evolution is compared with the $\Lambda$CDM model $H(z) = 68.0 \sqrt{(1-0.31) + 0.31 (1+z)^3}$.}
			\label{fig:dyn_evo_model2_neglam_nc}
		\end{figure*}

\begin{figure*}[t]
			\centering
			\includegraphics[scale=0.6]{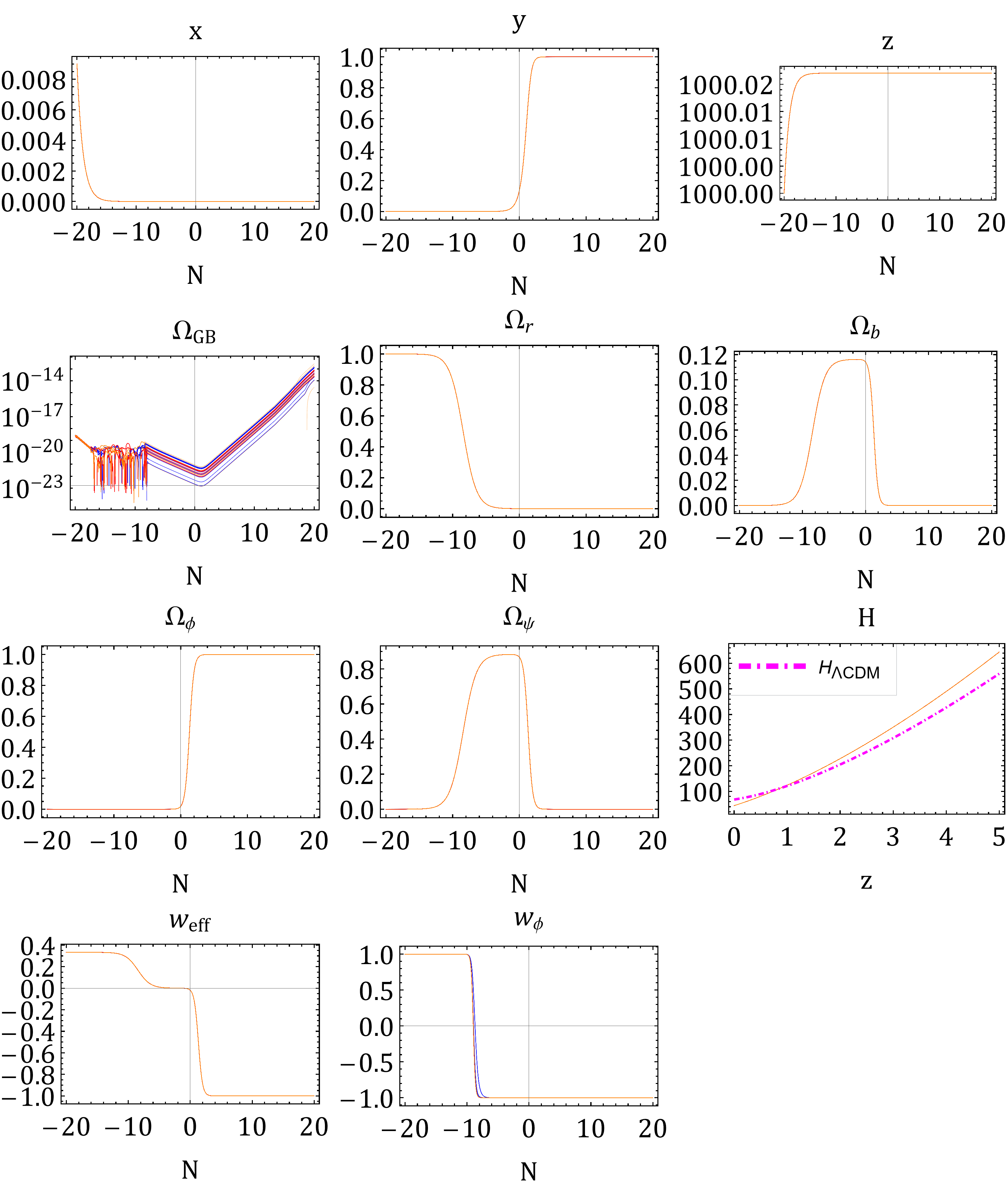}
			\caption{The evolution of the dynamical variables and the density parameters of Model II is estimated by fixing the initial conditions at $N=-20$, where $x(-20) = 0.009$, $y(-20) = 10^{-16}$, $z(-20) = 10^4$, $m_1 = 10^{-28}$, $\Omega_{b}(-20) = 1.0989 \times 10^{-6}$, and $H(-20) = 1.522 \times 10^{17}$, for $H_0 = 68 \ {\rm km/s/Mpc}$, $m = 1$, and $M = 1$. Here, we randomly generate points for $\lambda$ in the range $-10^{-18} < \lambda < -10^{-11}$, and for each $\lambda$, we determine the plots in the bin $-3 \le n_c \le -1$, which are shown in blue, red, and orange curves. The Hubble evolution is compared with the $\Lambda$CDM model $H(z) = 68.0 \sqrt{(1-0.31) + 0.31 (1+z)^3}$.}
			\label{fig:dyn_evo_model2_neglam_negnc}
		\end{figure*}

		In order to explain the current accelerating phase of the universe, we reduce the potential parameter to $\lambda \sim -10^{-15}$ and determine the dynamics for different values of \(n_c\), as shown in Figs. \ref{fig:dyn_evo_model2_neglam_nc} and \ref{fig:dyn_evo_model2_neglam_negnc}. For this model, we choose a higher initial value of \(z\), which remains nearly constant throughout the evolution. $\Omega_{\phi}$ dominates during the current epoch, and the effective EoS \(w_{\rm eff}\) approaches $-1$ in the future epoch.
		
		In this simulation, the value of \(w_{\rm eff}\) at the present epoch ($N=0$) is not close to $-0.5$, which is due to the tuning of the initial condition of \(y\). This can be consistently adjusted while confronting the model with observations. Nevertheless, the system evolves toward a stable attractor de-Sitter phase at late times, where nearly all dynamical variables remain saturated at finite values, indicating the existence of an attractor stable de-Sitter critical point.

		\bibliographystyle{JHEP}
		
		\bibliography{ref}

@article{Mignemi:1992nt,
	author = "Mignemi, S. and Stewart, N. R.",
	title = "{Charged black holes in effective string theory}",
	eprint = "hep-th/9212146",
	archivePrefix = "arXiv",
	reportNumber = "PRINT-93-0038 (PARIS)",
	doi = "10.1103/PhysRevD.47.5259",
	journal = "Phys. Rev. D",
	volume = "47",
	pages = "5259--5269",
	year = "1993"
}

@article{Carson:2020ter,
	author = "Carson, Zack and Yagi, Kent",
	title = "{Probing Einstein-dilaton Gauss-Bonnet Gravity with the inspiral and ringdown of gravitational waves}",
	eprint = "2003.00286",
	archivePrefix = "arXiv",
	primaryClass = "gr-qc",
	doi = "10.1103/PhysRevD.101.104030",
	journal = "Phys. Rev. D",
	volume = "101",
	number = "10",
	pages = "104030",
	year = "2020"
}

@article{Kanti:1995vq,
	author = "Kanti, P. and Mavromatos, N. E. and Rizos, J. and Tamvakis, K. and Winstanley, E.",
	title = "{Dilatonic black holes in higher curvature string gravity}",
	eprint = "hep-th/9511071",
	archivePrefix = "arXiv",
	reportNumber = "CERN-TH-95-297, OUTP-95-43-P, CERN-TH/95-297, OUTP-95-43P",
	doi = "10.1103/PhysRevD.54.5049",
	journal = "Phys. Rev. D",
	volume = "54",
	pages = "5049--5058",
	year = "1996"
}

@article{Yagi:2012gp,
	author = "Yagi, Kent",
	title = "{A New constraint on scalar Gauss-Bonnet gravity and a possible explanation for the excess of the orbital decay rate in a low-mass X-ray binary}",
	eprint = "1204.4524",
	archivePrefix = "arXiv",
	primaryClass = "gr-qc",
	doi = "10.1103/PhysRevD.86.081504",
	journal = "Phys. Rev. D",
	volume = "86",
	pages = "081504",
	year = "2012"
}

@article{Lyu:2022gdr,
	author = "Lyu, Zhenwei and Jiang, Nan and Yagi, Kent",
	title = "{Constraints on Einstein-dilation-Gauss-Bonnet gravity from black hole-neutron star gravitational wave events}",
	eprint = "2201.02543",
	archivePrefix = "arXiv",
	primaryClass = "gr-qc",
	reportNumber = "LIGO-P2100466",
	doi = "10.1103/PhysRevD.105.064001",
	journal = "Phys. Rev. D",
	volume = "105",
	number = "6",
	pages = "064001",
	year = "2022",
	note = "[Erratum: Phys.Rev.D 106, 069901 (2022), Erratum: Phys.Rev.D 106, 069901 (2022)]"
}

@article{Ishak:2019aay,
	author = "Ishak, Mustapha and others",
	title = "{Modified Gravity and Dark Energy models Beyond $w(z)$CDM Testable by LSST}",
	eprint = "1905.09687",
	archivePrefix = "arXiv",
	primaryClass = "astro-ph.CO",
	month = "5",
	year = "2019"
}

@article{Bellini:2014fua,
	author = "Bellini, Emilio and Sawicki, Ignacy",
	title = "{Maximal freedom at minimum cost: linear large-scale structure in general modifications of gravity}",
	eprint = "1404.3713",
	archivePrefix = "arXiv",
	primaryClass = "astro-ph.CO",
	doi = "10.1088/1475-7516/2014/07/050",
	journal = "JCAP",
	volume = "07",
	pages = "050",
	year = "2014"
}

@article{SupernovaCosmologyProject:1998vns,
	author = "Perlmutter, S. and others",
	collaboration = "Supernova Cosmology Project",
	title = "{Measurements of $\Omega$ and $\Lambda$ from 42 high redshift supernovae}",
	eprint = "astro-ph/9812133",
	archivePrefix = "arXiv",
	reportNumber = "LBNL-41801, LBL-41801",
	doi = "10.1086/307221",
	journal = "Astrophys. J.",
	volume = "517",
	pages = "565--586",
	year = "1999"
}

@article{SupernovaSearchTeam:1998fmf,
	author = "Riess, Adam G. and others",
	collaboration = "Supernova Search Team",
	title = "{Observational evidence from supernovae for an accelerating universe and a cosmological constant}",
	eprint = "astro-ph/9805201",
	archivePrefix = "arXiv",
	doi = "10.1086/300499",
	journal = "Astron. J.",
	volume = "116",
	pages = "1009--1038",
	year = "1998"
}

@article{WMAP:2003elm,
	author = "Spergel, D. N. and others",
	collaboration = "WMAP",
	title = "{First year Wilkinson Microwave Anisotropy Probe (WMAP) observations: Determination of cosmological parameters}",
	eprint = "astro-ph/0302209",
	archivePrefix = "arXiv",
	doi = "10.1086/377226",
	journal = "Astrophys. J. Suppl.",
	volume = "148",
	pages = "175--194",
	year = "2003"
}

@article{Sherwin:2011gv,
	author = "Sherwin, Blake D. and others",
	title = "{Evidence for dark energy from the cosmic microwave background alone using the Atacama Cosmology Telescope lensing measurements}",
	eprint = "1105.0419",
	archivePrefix = "arXiv",
	primaryClass = "astro-ph.CO",
	doi = "10.1103/PhysRevLett.107.021302",
	journal = "Phys. Rev. Lett.",
	volume = "107",
	pages = "021302",
	year = "2011"
}

@article{Wright:2007vr,
	author = "Wright, Edward L.",
	title = "{Constraints on Dark Energy from Supernovae, Gamma Ray Bursts, Acoustic Oscillations, Nucleosynthesis and Large Scale Structure and the Hubble constant}",
	eprint = "astro-ph/0701584",
	archivePrefix = "arXiv",
	doi = "10.1086/519274",
	journal = "Astrophys. J.",
	volume = "664",
	pages = "633--639",
	year = "2007"
}

@article{DES:2024jxu,
	author = "Abbott, T. M. C. and others",
	collaboration = "DES",
	title = "{The Dark Energy Survey: Cosmology Results with {\ensuremath{\sim}}1500 New High-redshift Type Ia Supernovae Using the Full 5 yr Data Set}",
	eprint = "2401.02929",
	archivePrefix = "arXiv",
	primaryClass = "astro-ph.CO",
	reportNumber = "FERMILAB-PUB-23-0821-PPD, DES-2023-805",
	doi = "10.3847/2041-8213/ad6f9f",
	journal = "Astrophys. J. Lett.",
	volume = "973",
	number = "1",
	pages = "L14",
	year = "2024"
}

@article{DES:2024hip,
	author = "Vincenzi, M. and others",
	collaboration = "DES",
	title = "{The Dark Energy Survey Supernova Program: Cosmological Analysis and Systematic Uncertainties}",
	eprint = "2401.02945",
	archivePrefix = "arXiv",
	primaryClass = "astro-ph.CO",
	reportNumber = "FERMILAB-PUB-23-693-PPD",
	doi = "10.3847/1538-4357/ad5e6c",
	journal = "Astrophys. J.",
	volume = "975",
	number = "1",
	pages = "86",
	year = "2024"
}

@article{DES:2016qvw,
	author = "Kwan, J. and others",
	collaboration = "DES",
	title = "{Cosmology from large-scale galaxy clustering and galaxy\textendash{}galaxy lensing with Dark Energy Survey Science Verification data}",
	eprint = "1604.07871",
	archivePrefix = "arXiv",
	primaryClass = "astro-ph.CO",
	reportNumber = "FERMILAB-PUB-16-146-A-AE",
	doi = "10.1093/mnras/stw2464",
	journal = "Mon. Not. Roy. Astron. Soc.",
	volume = "464",
	number = "4",
	pages = "4045--4062",
	year = "2017"
}

@article{DES:2021esc,
	author = "Abbott, T. M. C. and others",
	collaboration = "DES",
	title = "{Dark Energy Survey Year 3 results: A 2.7\% measurement of baryon acoustic oscillation distance scale at redshift 0.835}",
	eprint = "2107.04646",
	archivePrefix = "arXiv",
	primaryClass = "astro-ph.CO",
	reportNumber = "FERMILAB-PUB-21-838-PPD, DES-2021-0651",
	doi = "10.1103/PhysRevD.105.043512",
	journal = "Phys. Rev. D",
	volume = "105",
	number = "4",
	pages = "043512",
	year = "2022"
}

@article{SDSS:2005xqv,
	author = "Eisenstein, Daniel J. and others",
	collaboration = "SDSS",
	title = "{Detection of the Baryon Acoustic Peak in the Large-Scale Correlation Function of SDSS Luminous Red Galaxies}",
	eprint = "astro-ph/0501171",
	archivePrefix = "arXiv",
	reportNumber = "FERMILAB-PUB-05-057-A-CD",
	doi = "10.1086/466512",
	journal = "Astrophys. J.",
	volume = "633",
	pages = "560--574",
	year = "2005"
}

@article{Kessler:2025eib,
	author = "Kessler, Richard and Hounsell, Rebekah and Joshi, Bhavin and Rubin, David and Sako, Masao and Chen, Rebecca and Miranda, Vivian and Rose, Benjamin. M.",
	title = "{Cosmology Constraints from Type Ia Supernova Simulations of the Nancy Grace Roman Space Telescope Strategy Recommended by the High Latitude Time Domain Survey Definition Committee}",
	eprint = "2506.04402",
	archivePrefix = "arXiv",
	primaryClass = "astro-ph.CO",
	month = "6",
	year = "2025"
}

@article{Hounsell:2023xds,
	author = "Hounsell, Rebekah and others",
	title = "{Roman CCS White Paper: Measuring Type Ia Supernovae Discovered in the Roman High Latitude Time Domain Survey}",
	eprint = "2307.02670",
	archivePrefix = "arXiv",
	primaryClass = "astro-ph.IM",
	month = "7",
	year = "2023"
}

@article{Kessler:2009yy,
	author = "Kessler, Richard and others",
	title = "{SNANA: A Public Software Package for Supernova Analysis}",
	eprint = "0908.4280",
	archivePrefix = "arXiv",
	primaryClass = "astro-ph.CO",
	reportNumber = "FERMILAB-PUB-09-861-A",
	doi = "10.1086/605984",
	journal = "Publ. Astron. Soc. Pac.",
	volume = "121",
	pages = "1028",
	year = "2009"
}

@inproceedings{Primack:1997av,
	author = "Primack, Joel R.",
	title = "{Dark matter and structure formation}",
	booktitle = "{Midrasha Mathematicae in Jerusalem: Winter School in Dynamical Systems}",
	eprint = "astro-ph/9707285",
	archivePrefix = "arXiv",
	reportNumber = "SCIPP-96-59-REV, SCIPP-96-59",
	month = "7",
	year = "1997"
}

@article{DelPopolo:2007dna,
	author = "Del Popolo, Antonino",
	title = "{Dark matter and structure formation a review}",
	eprint = "0801.1091",
	archivePrefix = "arXiv",
	primaryClass = "astro-ph",
	doi = "10.1134/S1063772907030018",
	journal = "Astron. Rep.",
	volume = "51",
	pages = "169--196",
	year = "2007"
}

@article{Diao:2023tor,
	author = "Diao, Junwen and Wei, Shibiao and Wei, Zherui and Liu, Chang",
	title = "{The impact of the dark matter on galaxy formation}",
	doi = "10.1088/1742-6596/2441/1/012025",
	journal = "J. Phys. Conf. Ser.",
	volume = "2441",
	number = "1",
	pages = "012025",
	year = "2023"
}

@article{Mina:2020eik,
	author = "Mina, Mattia and Mota, David F. and Winther, Hans A.",
	title = "{Solitons in the dark: First approach to non-linear structure formation with fuzzy dark matter}",
	eprint = "2007.04119",
	archivePrefix = "arXiv",
	primaryClass = "astro-ph.CO",
	doi = "10.1051/0004-6361/202038876",
	journal = "Astron. Astrophys.",
	volume = "662",
	pages = "A29",
	year = "2022"
}

@article{Blumenthal:1984bp,
	author = "Blumenthal, George R. and Faber, S. M. and Primack, Joel R. and Rees, Martin J.",
	editor = "Srednicki, M. A.",
	title = "{Formation of Galaxies and Large Scale Structure with Cold Dark Matter}",
	reportNumber = "SLAC-PUB-3307",
	doi = "10.1038/311517a0",
	journal = "Nature",
	volume = "311",
	pages = "517--525",
	year = "1984"
}

@article{Copeland:2006wr,
	author = "Copeland, Edmund J. and Sami, M. and Tsujikawa, Shinji",
	title = "{Dynamics of dark energy}",
	eprint = "hep-th/0603057",
	archivePrefix = "arXiv",
	doi = "10.1142/S021827180600942X",
	journal = "Int. J. Mod. Phys. D",
	volume = "15",
	pages = "1753--1936",
	year = "2006"
}

@article{Weinberg:1988cp,
	author = "Weinberg, Steven",
	editor = "Hsu, Jong-Ping and Fine, D.",
	title = "{The Cosmological Constant Problem}",
	reportNumber = "UTTG-12-88",
	doi = "10.1103/RevModPhys.61.1",
	journal = "Rev. Mod. Phys.",
	volume = "61",
	pages = "1--23",
	year = "1989"
}

@article{Rugh:2000ji,
	author = "Rugh, S. E. and Zinkernagel, H.",
	title = "{The Quantum vacuum and the cosmological constant problem}",
	eprint = "hep-th/0012253",
	archivePrefix = "arXiv",
	doi = "10.1016/S1355-2198(02)00033-3",
	journal = "Stud. Hist. Phil. Sci. B",
	volume = "33",
	pages = "663--705",
	year = "2002"
}

@article{Padmanabhan:2002ji,
	author = "Padmanabhan, T.",
	title = "{Cosmological constant: The Weight of the vacuum}",
	eprint = "hep-th/0212290",
	archivePrefix = "arXiv",
	doi = "10.1016/S0370-1573(03)00120-0",
	journal = "Phys. Rept.",
	volume = "380",
	pages = "235--320",
	year = "2003"
}

@article{Carroll:1991mt,
	author = "Carroll, Sean M. and Press, William H. and Turner, Edwin L.",
	title = "{The Cosmological constant}",
	reportNumber = "CFA-3332",
	doi = "10.1146/annurev.aa.30.090192.002435",
	journal = "Ann. Rev. Astron. Astrophys.",
	volume = "30",
	pages = "499--542",
	year = "1992"
}

@article{Bengochea:2019daa,
	author = "Bengochea, Gabriel R. and Le\'on, Gabriel and Okon, Elias and Sudarsky, Daniel",
	title = "{Can the quantum vacuum fluctuations really solve the cosmological constant problem?}",
	eprint = "1906.05406",
	archivePrefix = "arXiv",
	primaryClass = "gr-qc",
	doi = "10.1140/epjc/s10052-019-7554-1",
	journal = "Eur. Phys. J. C",
	volume = "80",
	number = "1",
	pages = "18",
	year = "2020"
}

@article{Kohri:2016lsj,
	author = "Kohri, Kazunori and Matsui, Hiroki",
	title = "{Cosmological Constant Problem and Renormalized Vacuum Energy Density in Curved Background}",
	eprint = "1612.08818",
	archivePrefix = "arXiv",
	primaryClass = "hep-th",
	reportNumber = "KEK-TH-1949, KEK-COSMO-198",
	doi = "10.1088/1475-7516/2017/06/006",
	journal = "JCAP",
	volume = "06",
	pages = "006",
	year = "2017"
}

@article{Lopez-Corredoira:2017rqn,
	author = "L\'opez-Corredoira, Mart\'\i{}n",
	title = "{Tests and problems of the standard model in Cosmology}",
	eprint = "1701.08720",
	archivePrefix = "arXiv",
	primaryClass = "astro-ph.CO",
	doi = "10.1007/s10701-017-0073-8",
	journal = "Found. Phys.",
	volume = "47",
	number = "6",
	pages = "711--768",
	year = "2017"
}

@article{DESI:2025zgx,
	author = "Abdul Karim, M. and others",
	collaboration = "DESI",
	title = "{DESI DR2 Results II: Measurements of Baryon Acoustic Oscillations and Cosmological Constraints}",
	eprint = "2503.14738",
	archivePrefix = "arXiv",
	primaryClass = "astro-ph.CO",
	reportNumber = "FERMILAB-PUB-25-0169-PPD",
	month = "3",
	year = "2025",
	doi = " ",
	journal = " "
}

@article{DESI:2024mwx,
	author = "Adame, A. G. and others",
	collaboration = "DESI",
	title = "{DESI 2024 VI: cosmological constraints from the measurements of baryon acoustic oscillations}",
	eprint = "2404.03002",
	archivePrefix = "arXiv",
	primaryClass = "astro-ph.CO",
	reportNumber = "FERMILAB-PUB-24-0154-PPD",
	doi = "10.1088/1475-7516/2025/02/021",
	journal = "JCAP",
	volume = "02",
	pages = "021",
	year = "2025"
}

@article{DESI:2019jxc,
	author = "Levi, Michael E. and others",
	collaboration = "DESI",
	title = "{The Dark Energy Spectroscopic Instrument (DESI)}",
	eprint = "1907.10688",
	archivePrefix = "arXiv",
	primaryClass = "astro-ph.IM",
	reportNumber = "FERMILAB-PUB-19-434-AE",
	month = "7",
	year = "2019",
	doi =" ",
	journal =" "
}

@article{Moon:2023jgl,
	author = "Moon, Jeongin and others",
	title = "{First detection of the BAO signal from early DESI data}",
	eprint = "2304.08427",
	archivePrefix = "arXiv",
	primaryClass = "astro-ph.CO",
	reportNumber = "FERMILAB-PUB-23-199-PPD",
	doi = "10.1093/mnras/stad2618",
	journal = "Mon. Not. Roy. Astron. Soc.",
	volume = "525",
	number = "4",
	pages = "5406--5422",
	year = "2023"
}

@article{Gao:2024ily,
	author = "Gao, Qing and Peng, Zhiqian and Gao, Shengqing and Gong, Yungui",
	title = "{On the Evidence of Dynamical Dark Energy}",
	eprint = "2411.16046",
	archivePrefix = "arXiv",
	primaryClass = "astro-ph.CO",
	doi = "10.3390/universe11010010",
	journal = "Universe",
	volume = "11",
	number = "1",
	pages = "10",
	year = "2025"
}

@article{Sousa-Neto:2025gpj,
	author = "Sousa-Neto, Agripino and Bengaly, Carlos and Gonzalez, Javier E. and Alcaniz, Jailson",
	title = "{Evidence for dynamical dark energy from DESI-DR2 and SN data? A symbolic regression analysis}",
	eprint = "2502.10506",
	archivePrefix = "arXiv",
	primaryClass = "astro-ph.CO",
	month = "6",
	year = "2025",
	doi = " "
}

@article{Hussain:2025nqy,
	author = "Hussain, Saddam and Arora, Simran and Wang, Anzhong and Rose, Ben",
	title = "{Probing the Dynamics of Gaussian Dark Energy Equation of State Using DESI BAO}",
	eprint = "2505.09913",
	archivePrefix = "arXiv",
	primaryClass = "astro-ph.CO",
	month = "5",
	year = "2025"
}

@article{Myrzakulov:2025jpk,
	author = "Myrzakulov, Yerlan and Hussain, Saddam and Shahalam, M.",
	title = "{Phase space and Data analyses of a non-minimally coupled scalar field system with decaying dark energy model}",
	eprint = "2506.11755",
	archivePrefix = "arXiv",
	primaryClass = "gr-qc",
	month = "6",
	year = "2025"
}

@article{Hussain:2024qrd,
	author = "Hussain, Saddam and Nelleri, Sarath and Bhattacharya, Kaushik",
	title = "{Comprehensive study of k-essence model: dynamical system analysis and observational constraints from latest Type Ia supernova and BAO observations}",
	eprint = "2406.07179",
	archivePrefix = "arXiv",
	primaryClass = "astro-ph.CO",
	doi = "10.1088/1475-7516/2025/03/025",
	journal = "JCAP",
	volume = "03",
	pages = "025",
	year = "2025"
}

@article{Hussain:2024jdt,
	author = "Hussain, Saddam",
	title = "{Non-adiabatic particle production scenario in algebraically coupled quintessence field with dark matter fluid}",
	eprint = "2403.10215",
	archivePrefix = "arXiv",
	primaryClass = "gr-qc",
	month = "3",
	year = "2024"
}

@article{Chaudhary:2025yvz,
	author = "Chaudhary, Himanshu and Hussain, Saddam",
	title = {{Yano-Schr{\"o}dinger Hyperfluid: Cosmological Implications}},
	eprint = "2503.23115",
	archivePrefix = "arXiv",
	primaryClass = "gr-qc",
	month = "3",
	year = "2025"
}

@article{Roy:2024kni,
	author = "Roy, Nandan",
	title = "{Dynamical dark energy in the light of DESI 2024 data}",
	eprint = "2406.00634",
	archivePrefix = "arXiv",
	primaryClass = "astro-ph.CO",
	doi = "10.1016/j.dark.2025.101912",
	journal = "Phys. Dark Univ.",
	volume = "48",
	pages = "101912",
	year = "2025"
}

@article{Wu:2025wyk,
	author = "Wu, Peng-Ju",
	title = "{Comparison of dark energy models using late-universe observations}",
	eprint = "2504.09054",
	archivePrefix = "arXiv",
	primaryClass = "astro-ph.CO",
	month = "4",
	year = "2025",
	doi = "10.48550/arXiv.2504.09054",
	journal = " "
}

@article{Gialamas:2024lyw,
	author = {Gialamas, Ioannis D. and H{\"u}tsi, Gert and Kannike, Kristjan and Racioppi, Antonio and Raidal, Martti and Vasar, Martin and Veerm{\"a}e, Hardi},
	title = "{Interpreting DESI 2024 BAO: Late-time dynamical dark energy or a local effect?}",
	eprint = "2406.07533",
	archivePrefix = "arXiv",
	primaryClass = "astro-ph.CO",
	doi = "10.1103/PhysRevD.111.043540",
	journal = "Phys. Rev. D",
	volume = "111",
	number = "4",
	pages = "043540",
	year = "2025"
}

@article{Scherer:2025esj,
	author = "Scherer, Mateus and Sabogal, Miguel A. and Nunes, Rafael C. and De Felice, Antonio",
	title = "{Challenging $\Lambda$CDM: 5$\sigma$ Evidence for a Dynamical Dark Energy Late-Time Transition}",
	eprint = "2504.20664",
	archivePrefix = "arXiv",
	primaryClass = "astro-ph.CO",
	month = "4",
	year = "2025",
	doi =" ",
	journal =" "
}

@article{Paliathanasis:2025cuc,
	author = "Paliathanasis, Andronikos",
	title = "{Observational constraints on dark energy models with {\ensuremath{\Lambda}} as an equilibrium point}",
	eprint = "2502.16221",
	archivePrefix = "arXiv",
	primaryClass = "astro-ph.CO",
	doi = "10.1016/j.dark.2025.101956",
	journal = "Phys. Dark Univ.",
	volume = "48",
	pages = "101956",
	year = "2025"
}

@article{Adil:2023ara,
	author = "Adil, Shahnawaz A. and Mukhopadhyay, Upala and Sen, Anjan A. and Vagnozzi, Sunny",
	title = "{Dark energy in light of the early JWST observations: case for a negative cosmological constant?}",
	eprint = "2307.12763",
	archivePrefix = "arXiv",
	primaryClass = "astro-ph.CO",
	doi = "10.1088/1475-7516/2023/10/072",
	journal = "JCAP",
	volume = "10",
	pages = "072",
	year = "2023"
}

@article{Peebles:2002gy,
	author = "Peebles, P. J. E. and Ratra, Bharat",
	editor = "Hsu, Jong-Ping and Fine, D.",
	title = "{The Cosmological Constant and Dark Energy}",
	eprint = "astro-ph/0207347",
	archivePrefix = "arXiv",
	reportNumber = "KSUPT-02-3",
	doi = "10.1103/RevModPhys.75.559",
	journal = "Rev. Mod. Phys.",
	volume = "75",
	pages = "559--606",
	year = "2003"
}

@article{Hussain:2023kwk,
	author = "Hussain, Saddam and Chatterjee, Anirban and Bhattacharya, Kaushik",
	title = "{Dynamical stability in models where dark matter and dark energy are nonminimally coupled to curvature}",
	eprint = "2305.19062",
	archivePrefix = "arXiv",
	primaryClass = "gr-qc",
	doi = "10.1103/PhysRevD.108.103502",
	journal = "Phys. Rev. D",
	volume = "108",
	number = "10",
	pages = "103502",
	year = "2023"
}

@article{Roy:2022fif,
	author = "Roy, Nandan and Goswami, Sangita and Das, Sudipta",
	title = "{Quintessence or phantom: Study of scalar field dark energy models through a general parametrization of the Hubble parameter}",
	eprint = "2201.09306",
	archivePrefix = "arXiv",
	primaryClass = "astro-ph.CO",
	doi = "10.1016/j.dark.2022.101037",
	journal = "Phys. Dark Univ.",
	volume = "36",
	pages = "101037",
	year = "2022"
}

@article{Setare:2008sf,
	author = "Setare, M. R. and Saridakis, E. N.",
	title = "{Quintom dark energy models with nearly flat potentials}",
	eprint = "0810.4775",
	archivePrefix = "arXiv",
	primaryClass = "astro-ph",
	doi = "10.1103/PhysRevD.79.043005",
	journal = "Phys. Rev. D",
	volume = "79",
	pages = "043005",
	year = "2009"
}

@article{Cai:2025mas,
	author = "Cai, Yifu and Ren, Xin and Qiu, Taotao and Li, Mingzhe and Zhang, Xinmin",
	title = "{The Quintom theory of dark energy after DESI DR2}",
	eprint = "2505.24732",
	archivePrefix = "arXiv",
	primaryClass = "astro-ph.CO",
	month = "5",
	year = "2025",
	doi=" ",
	journal= ""
}

@article{Dutta:2009yb,
	author = "Dutta, Sourish and Saridakis, Emmanuel N. and Scherrer, Robert J.",
	title = "{Dark energy from a quintessence (phantom) field rolling near potential minimum (maximum)}",
	eprint = "0903.3412",
	archivePrefix = "arXiv",
	primaryClass = "astro-ph.CO",
	doi = "10.1103/PhysRevD.79.103005",
	journal = "Phys. Rev. D",
	volume = "79",
	pages = "103005",
	year = "2009"
}

@article{Cicoli:2023opf,
	author = "Cicoli, Michele and Conlon, Joseph P. and Maharana, Anshuman and Parameswaran, Susha and Quevedo, Fernando and Zavala, Ivonne",
	title = "{String cosmology: From the early universe to today}",
	eprint = "2303.04819",
	archivePrefix = "arXiv",
	primaryClass = "hep-th",
	doi = "10.1016/j.physrep.2024.01.002",
	journal = "Phys. Rept.",
	volume = "1059",
	pages = "1--155",
	year = "2024"
}

@article{Hussain:2024yee,
	author = "Hussain, Saddam and Arora, Simran and Rana, Yamuna and Rose, Benjamin and Wang, Anzhong",
	title = "{Interacting models of dark energy and dark matter in Einstein scalar Gauss Bonnet gravity}",
	eprint = "2408.05484",
	archivePrefix = "arXiv",
	primaryClass = "gr-qc",
	doi = "10.1088/1475-7516/2024/11/042",
	journal = "JCAP",
	volume = "11",
	pages = "042",
	year = "2024"
}

@article{Armendariz-Picon:2000nqq,
	author = "Armendariz-Picon, C. and Mukhanov, Viatcheslav F. and Steinhardt, Paul J.",
	title = "{A Dynamical solution to the problem of a small cosmological constant and late time cosmic acceleration}",
	eprint = "astro-ph/0004134",
	archivePrefix = "arXiv",
	doi = "10.1103/PhysRevLett.85.4438",
	journal = "Phys. Rev. Lett.",
	volume = "85",
	pages = "4438--4441",
	year = "2000"
}

@article{Armendariz-Picon:2000ulo,
	author = "Armendariz-Picon, C. and Mukhanov, Viatcheslav F. and Steinhardt, Paul J.",
	title = "{Essentials of k essence}",
	eprint = "astro-ph/0006373",
	archivePrefix = "arXiv",
	doi = "10.1103/PhysRevD.63.103510",
	journal = "Phys. Rev. D",
	volume = "63",
	pages = "103510",
	year = "2001"
}

@article{Chiba:1999ka,
	author = "Chiba, Takeshi and Okabe, Takahiro and Yamaguchi, Masahide",
	title = "{Kinetically driven quintessence}",
	eprint = "astro-ph/9912463",
	archivePrefix = "arXiv",
	reportNumber = "UTAP-352",
	doi = "10.1103/PhysRevD.62.023511",
	journal = "Phys. Rev. D",
	volume = "62",
	pages = "023511",
	year = "2000"
}

@article{Armendariz-Picon:2005oog,
	author = "Armendariz-Picon, C. and Lim, Eugene A.",
	title = "{Haloes of k-essence}",
	eprint = "astro-ph/0505207",
	archivePrefix = "arXiv",
	doi = "10.1088/1475-7516/2005/08/007",
	journal = "JCAP",
	volume = "08",
	pages = "007",
	year = "2005"
}

@article{Arkani-Hamed:2003pdi,
	author = "Arkani-Hamed, Nima and Cheng, Hsin-Chia and Luty, Markus A. and Mukohyama, Shinji",
	title = "{Ghost condensation and a consistent infrared modification of gravity}",
	eprint = "hep-th/0312099",
	archivePrefix = "arXiv",
	reportNumber = "HUTP-03-A081, UMD-PPP-04-012",
	doi = "10.1088/1126-6708/2004/05/074",
	journal = "JHEP",
	volume = "05",
	pages = "074",
	year = "2004"
}

@article{Scherrer:2004au,
	author = "Scherrer, Robert J.",
	title = "{Purely kinetic k-essence as unified dark matter}",
	eprint = "astro-ph/0402316",
	archivePrefix = "arXiv",
	doi = "10.1103/PhysRevLett.93.011301",
	journal = "Phys. Rev. Lett.",
	volume = "93",
	pages = "011301",
	year = "2004"
}

@article{Chatterjee:2021ijw,
	author = "Chatterjee, Anirban and Hussain, Saddam and Bhattacharya, Kaushik",
	title = "{Dynamical stability of the k-essence field interacting nonminimally with a perfect fluid}",
	eprint = "2105.00361",
	archivePrefix = "arXiv",
	primaryClass = "gr-qc",
	doi = "10.1103/PhysRevD.104.103505",
	journal = "Phys. Rev. D",
	volume = "104",
	number = "10",
	pages = "103505",
	year = "2021"
}

@article{Hussain:2022osn,
	author = "Hussain, Saddam and Chatterjee, Anirban and Bhattacharya, Kaushik",
	title = "{Ghost Condensates and Pure Kinetic k-Essence Condensates in the Presence of Field\textendash{}Fluid Non-Minimal Coupling in the Dark Sector}",
	eprint = "2203.10607",
	archivePrefix = "arXiv",
	primaryClass = "gr-qc",
	doi = "10.3390/universe9020065",
	journal = "Universe",
	volume = "9",
	number = "2",
	pages = "65",
	year = "2023"
}

@article{Bhattacharya:2022wzu,
	author = "Bhattacharya, Kaushik and Chatterjee, Anirban and Hussain, Saddam",
	title = "{Dynamical stability in presence of non-minimal derivative dependent coupling of k-essence field with a relativistic fluid}",
	eprint = "2206.12398",
	archivePrefix = "arXiv",
	primaryClass = "gr-qc",
	doi = "10.1140/epjc/s10052-023-11666-w",
	journal = "Eur. Phys. J. C",
	volume = "83",
	number = "6",
	pages = "488",
	year = "2023"
}

@article{Khoeini-Moghaddam:2018znw,
	author = "Khoeini-Moghaddam, Salomeh and Momeni, Farzan and Yousefabadi, Fatemeh",
	title = "{Fermionic tachyons as a source of dark energy}",
	eprint = "1807.05871",
	archivePrefix = "arXiv",
	primaryClass = "gr-qc",
	doi = "10.1016/j.newast.2022.101986",
	journal = "New Astron.",
	volume = "100",
	pages = "101986",
	year = "2023"
}

@article{Bagla:2002yn,
	author = "Bagla, J. S. and Jassal, Harvinder Kaur and Padmanabhan, T.",
	title = "{Cosmology with tachyon field as dark energy}",
	eprint = "astro-ph/0212198",
	archivePrefix = "arXiv",
	doi = "10.1103/PhysRevD.67.063504",
	journal = "Phys. Rev. D",
	volume = "67",
	pages = "063504",
	year = "2003"
}

@article{Liu:2020bmp,
	author = "Liu, Yang",
	title = "{Interacting ghost dark energy in complex quintessence theory}",
	eprint = "2201.00658",
	archivePrefix = "arXiv",
	primaryClass = "hep-th",
	doi = "10.1140/epjc/s10052-020-08786-y",
	journal = "Eur. Phys. J. C",
	volume = "80",
	number = "12",
	pages = "1204",
	year = "2020"
}

@article{Cai:2010uf,
	author = "Cai, Rong-Gen and Tuo, Zhong-Liang and Zhang, Hong-Bo and Su, Qiping",
	title = "{Notes on Ghost Dark Energy}",
	eprint = "1011.3212",
	archivePrefix = "arXiv",
	primaryClass = "astro-ph.CO",
	doi = "10.1103/PhysRevD.84.123501",
	journal = "Phys. Rev. D",
	volume = "84",
	pages = "123501",
	year = "2011"
}

@article{Hussain:2022dhp,
	author = "Hussain, Saddam and Chakraborty, Saikat and Roy, Nandan and Bhattacharya, Kaushik",
	title = "{Dynamical systems analysis of tachyon-dark-energy models from a new perspective}",
	eprint = "2208.10352",
	archivePrefix = "arXiv",
	primaryClass = "gr-qc",
	doi = "10.1103/PhysRevD.107.063515",
	journal = "Phys. Rev. D",
	volume = "107",
	number = "6",
	pages = "063515",
	year = "2023"
}

@article{Wang:2017brl,
	author = "Wang, Anzhong",
	title = "{Ho\v{r}ava gravity at a Lifshitz point: A progress report}",
	eprint = "1701.06087",
	archivePrefix = "arXiv",
	primaryClass = "gr-qc",
	doi = "10.1142/S0218271817300142",
	journal = "Int. J. Mod. Phys. D",
	volume = "26",
	number = "07",
	pages = "1730014",
	year = "2017"
}

@article{Shankaranarayanan:2022wbx,
	author = "Shankaranarayanan, S. and Johnson, Joseph P.",
	title = "{Modified theories of gravity: Why, how and what?}",
	eprint = "2204.06533",
	archivePrefix = "arXiv",
	primaryClass = "gr-qc",
	doi = "10.1007/s10714-022-02927-2",
	journal = "Gen. Rel. Grav.",
	volume = "54",
	number = "5",
	pages = "44",
	year = "2022"
}

@article{Sotiriou:2008rp,
	author = "Sotiriou, Thomas P. and Faraoni, Valerio",
	title = "{f(R) Theories Of Gravity}",
	eprint = "0805.1726",
	archivePrefix = "arXiv",
	primaryClass = "gr-qc",
	doi = "10.1103/RevModPhys.82.451",
	journal = "Rev. Mod. Phys.",
	volume = "82",
	pages = "451--497",
	year = "2010"
}

@article{Nojiri:2006ri,
	author = "Nojiri, Shin'ichi and Odintsov, Sergei D.",
	editor = "Borowiec, Andrzej",
	title = "{Introduction to modified gravity and gravitational alternative for dark energy}",
	eprint = "hep-th/0601213",
	archivePrefix = "arXiv",
	reportNumber = "KARP-2006-06",
	doi = "10.1142/S0219887807001928",
	journal = "eConf",
	volume = "C0602061",
	pages = "06",
	year = "2006"
}

@article{Yang:2010hw,
	author = "Yang, Rong-Jia",
	title = "{New types of $f(T)$ gravity}",
	eprint = "1007.3571",
	archivePrefix = "arXiv",
	primaryClass = "gr-qc",
	doi = "10.1140/epjc/s10052-011-1797-9",
	journal = "Eur. Phys. J. C",
	volume = "71",
	pages = "1797",
	year = "2011"
}

@article{Anagnostopoulos:2021ydo,
	author = "Anagnostopoulos, Fotios K. and Basilakos, Spyros and Saridakis, Emmanuel N.",
	title = "{First evidence that non-metricity f(Q) gravity could challenge \ensuremath{\Lambda}CDM}",
	eprint = "2104.15123",
	archivePrefix = "arXiv",
	primaryClass = "gr-qc",
	doi = "10.1016/j.physletb.2021.136634",
	journal = "Phys. Lett. B",
	volume = "822",
	pages = "136634",
	year = "2021"
}

@article{Sokoliuk:2023ccw,
	author = "Sokoliuk, Oleksii and Arora, Simran and Praharaj, Subhrat and Baransky, Alexander and Sahoo, P. K.",
	title = "{On the impact of f(Q) gravity on the large scale structure}",
	eprint = "2303.17341",
	archivePrefix = "arXiv",
	primaryClass = "astro-ph.CO",
	doi = "10.1093/mnras/stad968",
	journal = "Mon. Not. Roy. Astron. Soc.",
	volume = "522",
	number = "1",
	pages = "252--267",
	year = "2023"
}

@article{Kobayashi:2019hrl,
	author = "Kobayashi, Tsutomu",
	title = "{Horndeski theory and beyond: a review}",
	eprint = "1901.07183",
	archivePrefix = "arXiv",
	primaryClass = "gr-qc",
	reportNumber = "RUP-19-3",
	doi = "10.1088/1361-6633/ab2429",
	journal = "Rept. Prog. Phys.",
	volume = "82",
	number = "8",
	pages = "086901",
	year = "2019"
}

@ARTICLE{new_horn,
	author = {{Horndeski}, Gregory Walter},
	title = "{Second-order scalar-tensor field equations in a four-dimensional space}",
	journal = {International Journal of Theoretical Physics},
	year = 1974,
	month = sep,
	volume = {10},
	number = {6},
	pages = {363-384},
	doi = {10.1007/BF01807638},
	adsurl = {https://ui.adsabs.harvard.edu/abs/1974IJTP...10..363H}
}

@article{Kainulainen:2004vk,
	author = "Kainulainen, Kimmo and Sunhede, Daniel",
	title = "{Dark energy from large extra dimensions}",
	eprint = "astro-ph/0412609",
	archivePrefix = "arXiv",
	doi = "10.1103/PhysRevD.73.083510",
	journal = "Phys. Rev. D",
	volume = "73",
	pages = "083510",
	year = "2006"
}

@article{Saratov:2012ni,
	author = "Saratov, Anton",
	title = "{Effective phantom dark energy in scalar-tensor gravity}",
	eprint = "1204.0369",
	archivePrefix = "arXiv",
	primaryClass = "astro-ph.CO",
	month = "4",
	year = "2012"
}

@article{Langlois:2018dxi,
	author = "Langlois, David",
	title = "{Dark energy and modified gravity in degenerate higher-order scalar\textendash{}tensor (DHOST) theories: A review}",
	eprint = "1811.06271",
	archivePrefix = "arXiv",
	primaryClass = "gr-qc",
	doi = "10.1142/S0218271819420069",
	journal = "Int. J. Mod. Phys. D",
	volume = "28",
	number = "05",
	pages = "1942006",
	year = "2019"
}

@article{Odintsov:2025kyw,
	author = "Odintsov, Sergei D. and Oikonomou, V. K. and Sharov, German S.",
	title = "{Einstein-Gauss-Bonnet cosmology confronted with observations}",
	eprint = "2503.17946",
	archivePrefix = "arXiv",
	primaryClass = "gr-qc",
	doi = "10.1016/j.jheap.2025.100398",
	journal = "JHEAp",
	volume = "47",
	pages = "100398",
	year = "2025"
}

@article{Elizalde:2023rds,
	author = "Elizalde, E. and Nojiri, Shin'ichi and Odintsov, S. D. and Oikonomou, V. K.",
	title = "{Propagation of gravitational waves in a dynamical wormhole background for two-scalar Einstein{\textendash}Gauss{\textendash}Bonnet theory}",
	eprint = "2312.02889",
	archivePrefix = "arXiv",
	primaryClass = "gr-qc",
	doi = "10.1016/j.dark.2024.101536",
	journal = "Phys. Dark Univ.",
	volume = "45",
	pages = "101536",
	year = "2024"
}

@article{Riess:2020fzl,
	author = "Riess, Adam G. and Casertano, Stefano and Yuan, Wenlong and Bowers, J. Bradley and Macri, Lucas and Zinn, Joel C. and Scolnic, Dan",
	title = "{Cosmic Distances Calibrated to 1\% Precision with Gaia EDR3 Parallaxes and Hubble Space Telescope Photometry of 75 Milky Way Cepheids Confirm Tension with $\Lambda$CDM}",
	eprint = "2012.08534",
	archivePrefix = "arXiv",
	primaryClass = "astro-ph.CO",
	doi = "10.3847/2041-8213/abdbaf",
	journal = "Astrophys. J. Lett.",
	volume = "908",
	number = "1",
	pages = "L6",
	year = "2021"
}

@article{Planck:2018vyg,
	author = "Aghanim, N. and others",
	collaboration = "Planck",
	title = "{Planck 2018 results. VI. Cosmological parameters}",
	eprint = "1807.06209",
	archivePrefix = "arXiv",
	primaryClass = "astro-ph.CO",
	doi = "10.1051/0004-6361/201833910",
	journal = "Astron. Astrophys.",
	volume = "641",
	pages = "A6",
	year = "2020",
	note = "[Erratum: Astron.Astrophys. 652, C4 (2021)]"
}

@article{Brout_2022,
	doi = {10.3847/1538-4357/ac8e04},
	url = {https://dx.doi.org/10.3847/1538-4357/ac8e04},
	year = {2022},
	month = {oct},
	publisher = {The American Astronomical Society},
	volume = {938},
	number = {2},
	pages = {110},
	author = "Dillon Brout et al.",
	title = {The Pantheon+ Analysis: Cosmological Constraints},
	journal = {The Astrophysical Journal},
}

@article{Schoneberg:2021qvd,
	author = {Sch{\"o}neberg, Nils and Franco Abell{\'a}n, Guillermo and P{\'e}rez S{\'a}nchez, Andrea and Witte, Samuel J. and Poulin, Vivian and Lesgourgues, Julien},
	title = "{The H0 Olympics: A fair ranking of proposed models}",
	eprint = "2107.10291",
	archivePrefix = "arXiv",
	primaryClass = "astro-ph.CO",
	doi = "10.1016/j.physrep.2022.07.001",
	journal = "Phys. Rept.",
	volume = "984",
	pages = "1--55",
	year = "2022"
}

@article{DiValentino:2020naf,
	author = "Di Valentino, Eleonora and Mukherjee, Ankan and Sen, Anjan A.",
	title = "{Dark Energy with Phantom Crossing and the $H_0$ Tension}",
	eprint = "2005.12587",
	archivePrefix = "arXiv",
	primaryClass = "astro-ph.CO",
	reportNumber = "IPPP/20/89",
	doi = "10.3390/e23040404",
	journal = "Entropy",
	volume = "23",
	number = "4",
	pages = "404",
	year = "2021"
}

@article{Wang:2024vmw,
	author = "Wang, B. and Abdalla, E. and Atrio-Barandela, F. and Pav{\'o}n, D.",
	title = "{Further understanding the interaction between dark energy and dark matter: current status and future directions}",
	eprint = "2402.00819",
	archivePrefix = "arXiv",
	primaryClass = "astro-ph.CO",
	doi = "10.1088/1361-6633/ad2527",
	journal = "Rept. Prog. Phys.",
	volume = "87",
	number = "3",
	pages = "036901",
	year = "2024"
}

@article{Bernal:2016gxb,
	author = "Bernal, Jose Luis and Verde, Licia and Riess, Adam G.",
	title = "{The trouble with $H_0$}",
	eprint = "1607.05617",
	archivePrefix = "arXiv",
	primaryClass = "astro-ph.CO",
	doi = "10.1088/1475-7516/2016/10/019",
	journal = "JCAP",
	volume = "10",
	pages = "019",
	year = "2016"
}

@article{Benisty:2024lmj,
	author = "Benisty, David and Pan, Supriya and Staicova, Denitsa and Di Valentino, Eleonora and Nunes, Rafael C.",
	title = "{Late-time constraints on interacting dark energy: Analysis independent of H0, rd, and MB}",
	eprint = "2403.00056",
	archivePrefix = "arXiv",
	primaryClass = "astro-ph.CO",
	doi = "10.1051/0004-6361/202449883",
	journal = "Astron. Astrophys.",
	volume = "688",
	pages = "A156",
	year = "2024"
}

@article{Pan:2023mie,
	author = "Pan, Supriya and Yang, Weiqiang",
	title = "{On the interacting dark energy scenarios - the case for Hubble constant tension}",
	eprint = "2310.07260",
	archivePrefix = "arXiv",
	primaryClass = "astro-ph.CO",
	doi = "10.1007/978-981-99-0177-7\_29",
	month = "10",
	year = "2023"
}

@article{DiValentino:2021izs,
	author = "Di Valentino, Eleonora and Mena, Olga and Pan, Supriya and Visinelli, Luca and Yang, Weiqiang",
	title = "{In the realm of the Hubble tension review of solutions}",
	eprint = "2103.01183",
	archivePrefix = "arXiv",
	primaryClass = "astro-ph.CO",
	reportNumber = "IPPP/20/108",
	doi = "10.1088/1361-6382/ac086d",
	journal = "Class. Quant. Grav.",
	volume = "38",
	number = "15",
	pages = "153001",
	year = "2021"
}

@article{Huey:2004qv,
	author = "Huey, Greg and Wandelt, Benjamin D.",
	title = "{Interacting quintessence. The Coincidence problem and cosmic acceleration}",
	eprint = "astro-ph/0407196",
	archivePrefix = "arXiv",
	doi = "10.1103/PhysRevD.74.023519",
	journal = "Phys. Rev. D",
	volume = "74",
	pages = "023519",
	year = "2006"
}

@article{Cai:2004dk,
	author = "Cai, Rong-Gen and Wang, Anzhong",
	title = "{Cosmology with interaction between phantom dark energy and dark matter and the coincidence problem}",
	eprint = "hep-th/0411025",
	archivePrefix = "arXiv",
	doi = "10.1088/1475-7516/2005/03/002",
	journal = "JCAP",
	volume = "03",
	pages = "002",
	year = "2005"
}

@article{Montani:2024pou,
	author = "Montani, Giovanni and Carlevaro, Nakia and Escamilla, Luis A. and Di Valentino, Eleonora",
	title = "{Kinetic model for dark energy{\textemdash}dark matter interaction: Scenario for the hubble tension}",
	eprint = "2404.15977",
	archivePrefix = "arXiv",
	primaryClass = "gr-qc",
	doi = "10.1016/j.dark.2025.101848",
	journal = "Phys. Dark Univ.",
	volume = "48",
	pages = "101848",
	year = "2025"
}

@article{Turner:1983he,
	author = "Turner, Michael S.",
	title = "{Coherent Scalar Field Oscillations in an Expanding Universe}",
	reportNumber = "EFI-83-29-CHICAGO",
	doi = "10.1103/PhysRevD.28.1243",
	journal = "Phys. Rev. D",
	volume = "28",
	pages = "1243",
	year = "1983"
}

@article{Hwang:1996xd,
	author = "Hwang, Jai-chan",
	title = "{Roles of a coherent scalar field on the evolution of cosmic structures}",
	eprint = "astro-ph/9610042",
	archivePrefix = "arXiv",
	doi = "10.1016/S0370-2693(97)00400-0",
	journal = "Phys. Lett. B",
	volume = "401",
	pages = "241--246",
	year = "1997"
}

@article{Magana:2012ph,
	author = "Magana, Juan and Matos, Tonatiuh",
	editor = "Barranco, Juan and Contreras, Guillermo and Delepine, David and Napsuciale, Mauro",
	title = "{A brief Review of the Scalar Field Dark Matter model}",
	eprint = "1201.6107",
	archivePrefix = "arXiv",
	primaryClass = "astro-ph.CO",
	doi = "10.1088/1742-6596/378/1/012012",
	journal = "J. Phys. Conf. Ser.",
	volume = "378",
	pages = "012012",
	year = "2012"
}

@article{Urena-Lopez:2019kud,
	author = "Ure\~na-L\'opez, L. Arturo",
	title = "{Brief Review on Scalar Field Dark Matter Models}",
	doi = "10.3389/fspas.2019.00047",
	journal = "Front. Astron. Space Sci.",
	volume = "6",
	pages = "47",
	year = "2019"
}

@article{Gutierrez-Luna:2021tmq,
	author = "Guti\'errez-Luna, Er\'endira and Carvente, Belen and Jaramillo, V\'\i{}ctor and Barranco, Juan and Escamilla-Rivera, Celia and Espinoza, Catalina and Mondrag\'on, Myriam and N\'u\~nez, Dar\'\i{}o",
	title = "{Scalar field dark matter with two components: Combined approach from particle physics and cosmology}",
	eprint = "2110.10258",
	archivePrefix = "arXiv",
	primaryClass = "astro-ph.CO",
	doi = "10.1103/PhysRevD.105.083533",
	journal = "Phys. Rev. D",
	volume = "105",
	number = "8",
	pages = "083533",
	year = "2022"
}

@article{Matos:2008ag,
	author = "Matos, Tonatiuh and Vazquez-Gonzalez, Alberto and Magana, Juan",
	title = "{$\phi^2$ as Dark Matter}",
	eprint = "0806.0683",
	archivePrefix = "arXiv",
	primaryClass = "astro-ph",
	doi = "10.1111/j.1365-2966.2008.13957.x",
	journal = "Mon. Not. Roy. Astron. Soc.",
	volume = "393",
	pages = "1359--1369",
	year = "2009"
}

@article{Solis-Lopez:2019lvz,
	author = "Sol\'\i{}s-L\'opez, Jordi and Guzm\'an, Francisco S. and Matos, Tonatiuh and Robles, Victor H. and Ure\~na-L\'opez, L. Arturo",
	title = "{Scalar field dark matter as an alternative explanation for the anisotropic distribution of satellite galaxies}",
	eprint = "1912.09660",
	archivePrefix = "arXiv",
	primaryClass = "astro-ph.GA",
	doi = "10.1103/PhysRevD.103.083535",
	journal = "Phys. Rev. D",
	volume = "103",
	number = "8",
	pages = "083535",
	year = "2021"
}

@article{Peebles:1998qn,
	author = "Peebles, P. J. E. and Vilenkin, A.",
	title = "{Quintessential inflation}",
	eprint = "astro-ph/9810509",
	archivePrefix = "arXiv",
	doi = "10.1103/PhysRevD.59.063505",
	journal = "Phys. Rev. D",
	volume = "59",
	pages = "063505",
	year = "1999"
}

@article{Matos:2000jx,
	author = "Matos, T. and Guzman, Francisco Siddhartha",
	title = "{Quintessence at galactic level?}",
	eprint = "astro-ph/0002126",
	archivePrefix = "arXiv",
	journal = "Annalen Phys.",
	volume = "9",
	pages = "S1--S133",
	year = "2000"
}

@article{Matos:2000ng,
	author = "Matos, Tonatiuh and Urena-Lopez, L. Arturo",
	title = "{Quintessence and scalar dark matter in the universe}",
	eprint = "astro-ph/0004332",
	archivePrefix = "arXiv",
	doi = "10.1088/0264-9381/17/13/101",
	journal = "Class. Quant. Grav.",
	volume = "17",
	pages = "L75--L81",
	year = "2000"
}

@article{Matos:2000ss,
	author = "Matos, Tonatiuh and Urena-Lopez, L. Arturo",
	title = "{A Further analysis of a cosmological model of quintessence and scalar dark matter}",
	eprint = "astro-ph/0006024",
	archivePrefix = "arXiv",
	doi = "10.1103/PhysRevD.63.063506",
	journal = "Phys. Rev. D",
	volume = "63",
	pages = "063506",
	year = "2001"
}

@article{matos2010accelerated,
	title={Accelerated expansion and structure formation with a single scalar field},
	author={Matos, Tonatiuh and Maga{\~n}a, Juan and Su{\'a}rez, Abril},
	doi = "10.2174/1874381101003010094",
	journal={The Open Astronomy Journal},
	volume={3},
	number={1},
	year={2010}
}

@article{Robles:2018fur,
	author = "Robles, Victor H. and Bullock, James S. and Boylan-Kolchin, Michael",
	title = "{Scalar Field Dark Matter: Helping or Hurting Small-Scale Problems in Cosmology?}",
	eprint = "1807.06018",
	archivePrefix = "arXiv",
	primaryClass = "astro-ph.CO",
	doi = "10.1093/mnras/sty3190",
	journal = "Mon. Not. Roy. Astron. Soc.",
	volume = "483",
	number = "1",
	pages = "289--298",
	year = "2019"
}

@article{Li:2013nal,
	author = "Li, Bohua and Rindler-Daller, Tanja and Shapiro, Paul R.",
	title = "{Cosmological Constraints on Bose-Einstein-Condensed Scalar Field Dark Matter}",
	eprint = "1310.6061",
	archivePrefix = "arXiv",
	primaryClass = "astro-ph.CO",
	doi = "10.1103/PhysRevD.89.083536",
	journal = "Phys. Rev. D",
	volume = "89",
	number = "8",
	pages = "083536",
	year = "2014"
}

@article{Magana:2012xe,
	author = "Magana, Juan and Matos, Tonatiuh and Suarez, Abril and Sanchez-Salcedo, F. J.",
	title = "{Structure formation with scalar field dark matter: the field approach}",
	eprint = "1204.5255",
	archivePrefix = "arXiv",
	primaryClass = "astro-ph.CO",
	doi = "10.1088/1475-7516/2012/10/003",
	journal = "JCAP",
	volume = "10",
	pages = "003",
	year = "2012"
}

@article{Aboubrahim:2024spa,
	author = "Aboubrahim, Amin and Nath, Pran",
	title = "{Interacting ultralight dark matter and dark energy and fits to cosmological data in a field theory approach}",
	eprint = "2406.19284",
	archivePrefix = "arXiv",
	primaryClass = "astro-ph.CO",
	doi = "10.1088/1475-7516/2024/09/076",
	journal = "JCAP",
	volume = "09",
	pages = "076",
	year = "2024"
}

@article{Foidl:2022bpn,
	author = "Foidl, Horst and Rindler-Daller, Tanja",
	title = "{Cosmological structure formation in complex scalar field dark matter versus real ultralight axions: A comparative study using class}",
	eprint = "2203.09396",
	archivePrefix = "arXiv",
	primaryClass = "astro-ph.CO",
	doi = "10.1103/PhysRevD.105.123534",
	journal = "Phys. Rev. D",
	volume = "105",
	number = "12",
	pages = "123534",
	year = "2022"
}

@article{Suarez:2013iw,
	author = "Su\'arez, Abril and Robles, Victor H. and Matos, Tonatiuh",
	editor = "Moreno Gonz\'alez, Claudia and Madriz Aguilar, Jos\'e Edgar and Reyes Barrera, Luz Marina",
	title = "{A Review on the Scalar Field/Bose-Einstein Condensate Dark Matter Model}",
	eprint = "1302.0903",
	archivePrefix = "arXiv",
	primaryClass = "astro-ph.CO",
	doi = "10.1007/978-3-319-02063-1_9",
	journal = "Astrophys. Space Sci. Proc.",
	volume = "38",
	pages = "107--142",
	year = "2014"
}

@article{Suarez:2011yf,
	author = "Suarez, Abril and Matos, Tonatiuh",
	title = "{Structure Formation with Scalar Field Dark Matter: The Fluid Approach}",
	eprint = "1101.4039",
	archivePrefix = "arXiv",
	primaryClass = "gr-qc",
	doi = "10.1111/j.1365-2966.2011.19012.x",
	journal = "Mon. Not. Roy. Astron. Soc.",
	volume = "416",
	pages = "87",
	year = "2011"
}

@article{Poulot:2024sex,
	author = "Poulot, Gaspard and Teixeira, Elsa M. and van de Bruck, Carsten and Nunes, Nelson J.",
	title = "{Scalar field dark matter with time-varying equation of state}",
	eprint = "2404.10524",
	archivePrefix = "arXiv",
	primaryClass = "astro-ph.CO",
	month = "4",
	year = "2024",
	doi =" "
}

@article{Lora:2011yc,
	author = "Lora, V. and Magana, Juan and Bernal, Argelia and Sanchez-Salcedo, F. J. and Grebel, E. K.",
	title = "{On the mass of ultra-light bosonic dark matter from galactic dynamics}",
	eprint = "1110.2684",
	archivePrefix = "arXiv",
	primaryClass = "astro-ph.GA",
	doi = "10.1088/1475-7516/2012/02/011",
	journal = "JCAP",
	volume = "02",
	pages = "011",
	year = "2012"
}

@article{Dev:2016hxv,
	author = "Dev, P. S. Bhupal and Lindner, Manfred and Ohmer, Sebastian",
	title = "{Gravitational waves as a new probe of Bose\textendash{}Einstein condensate Dark Matter}",
	eprint = "1609.03939",
	archivePrefix = "arXiv",
	primaryClass = "hep-ph",
	doi = "10.1016/j.physletb.2017.08.043",
	journal = "Phys. Lett. B",
	volume = "773",
	pages = "219--224",
	year = "2017"
}

@article{Nojiri:2005vv,
	author = "Nojiri, Shin'ichi and Odintsov, Sergei D. and Sasaki, Misao",
	title = "{Gauss-Bonnet dark energy}",
	eprint = "hep-th/0504052",
	archivePrefix = "arXiv",
	reportNumber = "YITP-05-14",
	doi = "10.1103/PhysRevD.71.123509",
	journal = "Phys. Rev. D",
	volume = "71",
	pages = "123509",
	year = "2005"
}

@article{Tsujikawa:2006ph,
	author = "Tsujikawa, Shinji and Sami, M.",
	title = "{String-inspired cosmology: Late time transition from scaling matter era to dark energy universe caused by a Gauss-Bonnet coupling}",
	eprint = "hep-th/0608178",
	archivePrefix = "arXiv",
	doi = "10.1088/1475-7516/2007/01/006",
	journal = "JCAP",
	volume = "01",
	pages = "006",
	year = "2007"
}

@article{DeFelice:2006pg,
	author = "De Felice, Antonio and Hindmarsh, Mark and Trodden, Mark",
	title = "{Ghosts, Instabilities, and Superluminal Propagation in Modified Gravity Models}",
	eprint = "astro-ph/0604154",
	archivePrefix = "arXiv",
	doi = "10.1088/1475-7516/2006/08/005",
	journal = "JCAP",
	volume = "08",
	pages = "005",
	year = "2006",
	
}

@article{Minamitsuji:2024twp,
	author = "Minamitsuji, Masato and Mukohyama, Shinji and Tsujikawa, Shinji",
	title = "{Angular and radial stabilities of spontaneously scalarized black holes in the presence of scalar-Gauss-Bonnet couplings}",
	eprint = "2403.10048",
	archivePrefix = "arXiv",
	primaryClass = "gr-qc",
	reportNumber = "YITP-24-27, IPMU24-0006, WUCG-24-02",
	doi = "10.1103/PhysRevD.109.104057",
	journal = "Phys. Rev. D",
	volume = "109",
	number = "10",
	pages = "104057",
	year = "2024"
}

@article{Tsujikawa:2022aar,
	author = "Tsujikawa, Shinji",
	title = "{Cosmological stability in f(\ensuremath{\phi},G) gravity}",
	eprint = "2212.10022",
	archivePrefix = "arXiv",
	primaryClass = "gr-qc",
	reportNumber = "WUCG-22-12",
	doi = "10.1016/j.physletb.2023.137751",
	journal = "Phys. Lett. B",
	volume = "838",
	pages = "137751",
	year = "2023"
}

@article{DeFelice:2009rw,
	author = "De Felice, Antonio and Mota, David F. and Tsujikawa, Shinji",
	title = "{Matter instabilities in general Gauss-Bonnet gravity}",
	eprint = "0911.1811",
	archivePrefix = "arXiv",
	primaryClass = "gr-qc",
	doi = "10.1103/PhysRevD.81.023532",
	journal = "Phys. Rev. D",
	volume = "81",
	pages = "023532",
	year = "2010"
}

@article{Gross:1986mw,
	author = "Gross, David J. and Sloan, John H.",
	title = "{The Quartic Effective Action for the Heterotic String}",
	reportNumber = "NSF-ITP-87-02",
	doi = "10.1016/0550-3213(87)90465-2",
	journal = "Nucl. Phys. B",
	volume = "291",
	pages = "41--89",
	year = "1987"
}

@article{Bento:1995qc,
	author = "Bento, M. C. and Bertolami, O.",
	title = "{Maximally Symmetric Cosmological Solutions of higher curvature string effective theories with dilatons}",
	eprint = "gr-qc/9503057",
	archivePrefix = "arXiv",
	reportNumber = "CERN-TH-95-63, CERN-TH-95-063, DFTT-19-95",
	doi = "10.1016/0370-2693(95)01519-1",
	journal = "Phys. Lett. B",
	volume = "368",
	pages = "198--201",
	year = "1996"
}

@article{Ferrara:1996hh,
	author = "Ferrara, Sergio and Khuri, Ramzi R. and Minasian, Ruben",
	title = "{M theory on a Calabi-Yau manifold}",
	eprint = "hep-th/9602102",
	archivePrefix = "arXiv",
	reportNumber = "CERN-TH-96-41, UCLA-96-TEP-6, MCGILL-96-05",
	doi = "10.1016/0370-2693(96)00270-5",
	journal = "Phys. Lett. B",
	volume = "375",
	pages = "81--88",
	year = "1996"
}

@article{LIGOScientific:2017zic,
	author = "Abbott, B. P. and others",
	collaboration = "LIGO Scientific, Virgo, Fermi-GBM, INTEGRAL",
	title = "{Gravitational Waves and Gamma-rays from a Binary Neutron Star Merger: GW170817 and GRB 170817A}",
	eprint = "1710.05834",
	archivePrefix = "arXiv",
	primaryClass = "astro-ph.HE",
	reportNumber = "LIGO-P1700308",
	doi = "10.3847/2041-8213/aa920c",
	journal = "Astrophys. J. Lett.",
	volume = "848",
	number = "2",
	pages = "L13",
	year = "2017"
}

@article{Odintsov:2019clh,
	author = "Odintsov, S. D. and Oikonomou, V. K.",
	title = "{Inflationary Phenomenology of Einstein Gauss-Bonnet Gravity Compatible with GW170817}",
	eprint = "1908.07555",
	archivePrefix = "arXiv",
	primaryClass = "gr-qc",
	doi = "10.1016/j.physletb.2019.134874",
	journal = "Phys. Lett. B",
	volume = "797",
	pages = "134874",
	year = "2019"
}

@article{Ezquiaga:2017ekz,
	author = "Ezquiaga, Jose Mar\'\i{}a and Zumalac\'arregui, Miguel",
	title = "{Dark Energy After GW170817: Dead Ends and the Road Ahead}",
	eprint = "1710.05901",
	archivePrefix = "arXiv",
	primaryClass = "astro-ph.CO",
	reportNumber = "IFT-UAM-CSIC-17-096, NORDITA-2017-109",
	doi = "10.1103/PhysRevLett.119.251304",
	journal = "Phys. Rev. Lett.",
	volume = "119",
	number = "25",
	pages = "251304",
	year = "2017"
}

@article{TerenteDiaz:2023iqk,
	author = "Terente D{\'\i}az, Jos{\'e} Jaime and Dimopoulos, Konstantinos and Kar{\v{c}}iauskas, Mindaugas and Racioppi, Antonio",
	title = "{Gauss-Bonnet Dark Energy and the speed of gravitational waves}",
	eprint = "2307.06163",
	archivePrefix = "arXiv",
	primaryClass = "astro-ph.CO",
	doi = "10.1088/1475-7516/2023/10/031",
	journal = "JCAP",
	volume = "10",
	pages = "031",
	year = "2023"
}

@article{Zhang20a,
	author = "Zhang, Shao-Jun and Wang, Bin and Wang, Anzhong and Saavedra, Joel F.",
	title = "{Object picture of scalar field perturbation on Kerr black hole in scalar-Einstein-Gauss-Bonnet theory}",
	eprint = "2010.05092",
	archivePrefix = "arXiv",
	primaryClass = "gr-qc",
	doi = "10.1103/PhysRevD.102.124056",
	journal = "Phys. Rev. D",
	volume = "102",
	number = "12",
	pages = "124056",
	year = "2020"
}

@article{Tomberg:2021ajh,
	author = "Tomberg, Eemeli",
	title = "{Unit conversions and collected numbers in cosmology}",
	eprint = "2110.12251",
	archivePrefix = "arXiv",
	primaryClass = "astro-ph.CO",
	month = "10",
	year = "2021"
}

@article{Ratra:1990me,
	author = "Ratra, Bharat",
	title = "{Expressions for linearized perturbations in a massive scalar field dominated cosmological model}",
	reportNumber = "CALT-68-1665",
	doi = "10.1103/PhysRevD.44.352",
	journal = "Phys. Rev. D",
	volume = "44",
	pages = "352--364",
	year = "1991"
}

@article{Saha:2024irh,
	author = "Saha, Priyanka and Dey, Dipanjan and Bhattacharya, Kaushik",
	title = "{Non-minimal coupling of scalar fields in the dark sector and generalization of the top-hat collapse}",
	eprint = "2410.20736",
	archivePrefix = "arXiv",
	primaryClass = "gr-qc",
	doi = "10.1140/epjc/s10052-025-14080-6",
	journal = "Eur. Phys. J. C",
	volume = "85",
	number = "4",
	pages = "384",
	year = "2025"
}

@article{Urena-Lopez:2015gur,
	author = "Ure\~na-L\'opez, L. Arturo and Gonzalez-Morales, Alma X.",
	title = "{Towards accurate cosmological predictions for rapidly oscillating scalar fields as dark matter}",
	eprint = "1511.08195",
	archivePrefix = "arXiv",
	primaryClass = "astro-ph.CO",
	doi = "10.1088/1475-7516/2016/07/048",
	journal = "JCAP",
	volume = "07",
	pages = "048",
	year = "2016"
}

@article{Urena-Lopez:2023ngt,
	author = "Ure\~na-L\'opez, L. Arturo and Linares Cede\~no, Francisco X.",
	title = "{Cosmological evolution of scalar field dark matter in the class code: Accuracy and precision of numerical solutions}",
	eprint = "2307.05600",
	archivePrefix = "arXiv",
	primaryClass = "astro-ph.CO",
	doi = "10.1103/PhysRevD.109.023512",
	journal = "Phys. Rev. D",
	volume = "109",
	number = "2",
	pages = "023512",
	year = "2024"
}

@article{Bertolami:2012xn,
	author = "Bertolami, Orfeu and Carrilho, Pedro and Paramos, Jorge",
	title = "{Two-scalar-field model for the interaction of dark energy and dark matter}",
	eprint = "1206.2589",
	archivePrefix = "arXiv",
	primaryClass = "gr-qc",
	doi = "10.1103/PhysRevD.86.103522",
	journal = "Phys. Rev. D",
	volume = "86",
	pages = "103522",
	year = "2012"
}

@article{vandeBruck:2022xbk,
	author = "van de Bruck, Carsten and Poulot, Gaspard and Teixeira, Elsa M.",
	title = "{Scalar field dark matter and dark energy: a hybrid model for the dark sector}",
	eprint = "2211.13653",
	archivePrefix = "arXiv",
	primaryClass = "hep-th",
	doi = "10.1088/1475-7516/2023/07/019",
	journal = "JCAP",
	volume = "07",
	pages = "019",
	year = "2023"
}

@article{Lee:2004vm,
	author = "Lee, Seokcheon and Olive, Keith A. and Pospelov, Maxim",
	title = "{Quintessence models and the cosmological evolution of alpha}",
	eprint = "astro-ph/0406039",
	archivePrefix = "arXiv",
	reportNumber = "UMN-TH-2313-04, FTPI-MINN-04-24, UVIC-TH-04-06",
	doi = "10.1103/PhysRevD.70.083503",
	journal = "Phys. Rev. D",
	volume = "70",
	pages = "083503",
	year = "2004"
}

@article{Lee:2006za,
	author = "Lee, Seokcheon and Liu, Guo-Chin and Ng, Kin-Wang",
	title = "{Constraints on the coupled quintessence from cosmic microwave background anisotropy and matter power spectrum}",
	eprint = "astro-ph/0601333",
	archivePrefix = "arXiv",
	doi = "10.1103/PhysRevD.73.083516",
	journal = "Phys. Rev. D",
	volume = "73",
	pages = "083516",
	year = "2006"
}

@article{Damour:1994zq,
	author = "Damour, T. and Polyakov, Alexander M.",
	title = "{The String dilaton and a least coupling principle}",
	eprint = "hep-th/9401069",
	archivePrefix = "arXiv",
	doi = "10.1016/0550-3213(94)90143-0",
	journal = "Nucl. Phys. B",
	volume = "423",
	pages = "532--558",
	year = "1994"
}

@article{Ghosh:2025pbn,
	author = "Ghosh, Subhajit and Boddy, Kimberly K. and Yu, Tien-Tien",
	title = "{Early Universe Constraints on Variations in Fundamental Constants Induced by Ultralight Scalar Dark Matter}",
	eprint = "2511.14532",
	archivePrefix = "arXiv",
	primaryClass = "astro-ph.CO",
	reportNumber = "UT-WI-39-2025",
	month = "11",
	year = "2025"
}

@article{Costa:2025kwt,
	author = "Costa, Marco and Creque-Sarbinowski, Cyril and Simon, Olivier and Weiner, Zachary J.",
	title = "{Dark forces suppress structure growth}",
	eprint = "2510.00098",
	archivePrefix = "arXiv",
	primaryClass = "astro-ph.CO",
	month = "9",
	year = "2025"
}

@article{Brax:2025ahm,
	author = "Brax, Philippe",
	title = "{Weinberg{\textquoteright}s theorem, phantom crossing, and screening}",
	eprint = "2507.16723",
	archivePrefix = "arXiv",
	primaryClass = "astro-ph.CO",
	doi = "10.1103/gcs7-c4c6",
	journal = "Phys. Rev. D",
	volume = "112",
	number = "8",
	pages = "083544",
	year = "2025"
}

@article{He:2011qn,
	author = "He, Jian-Hua and Wang, Bin and Abdalla, Elcio",
	title = "{Deep connection between f(R) gravity and the interacting dark sector model}",
	eprint = "1109.1730",
	archivePrefix = "arXiv",
	primaryClass = "gr-qc",
	doi = "10.1103/PhysRevD.84.123526",
	journal = "Phys. Rev. D",
	volume = "84",
	pages = "123526",
	year = "2011"
}

@article{Amendola:1999er,
	author = "Amendola, Luca",
	title = "{Coupled quintessence}",
	eprint = "astro-ph/9908023",
	archivePrefix = "arXiv",
	doi = "10.1103/PhysRevD.62.043511",
	journal = "Phys. Rev. D",
	volume = "62",
	pages = "043511",
	year = "2000"
}

@article{Das:2019ixt,
	author = "Das, Sudipta and Banerjee, Manisha and Roy, Nandan",
	title = "{Dynamical System Analysis for Steep Potentials}",
	eprint = "1903.02288",
	archivePrefix = "arXiv",
	primaryClass = "gr-qc",
	doi = "10.1088/1475-7516/2019/08/024",
	journal = "JCAP",
	volume = "08",
	pages = "024",
	year = "2019"
}

@article{Bahamonde:2017ize,
	author = {Bahamonde, Sebastian and B{\"o}hmer, Christian G. and Carloni, Sante and Copeland, Edmund J. and Fang, Wei and Tamanini, Nicola},
	title = "{Dynamical systems applied to cosmology: dark energy and modified gravity}",
	eprint = "1712.03107",
	archivePrefix = "arXiv",
	primaryClass = "gr-qc",
	doi = "10.1016/j.physrep.2018.09.001",
	journal = "Phys. Rept.",
	volume = "775-777",
	pages = "1--122",
	year = "2018"
}

@article{Alho:2020cdg,
	author = "Alho, Artur and Uggla, Claes and Wainwright, John",
	title = "{Dynamical systems in perturbative scalar field cosmology}",
	eprint = "2006.00800",
	archivePrefix = "arXiv",
	primaryClass = "gr-qc",
	doi = "10.1088/1361-6382/abb73a",
	journal = "Class. Quant. Grav.",
	volume = "37",
	number = "22",
	pages = "225011",
	year = "2020"
}

@article{Shahalam:2017rit,
	author = "Shahalam, M. and Yang, Weiqiang and Myrzakulov, R. and Wang, Anzhong",
	title = "{Late-time acceleration with steep exponential potentials}",
	eprint = "1802.00326",
	archivePrefix = "arXiv",
	primaryClass = "gr-qc",
	doi = "10.1140/epjc/s10052-017-5468-3",
	journal = "Eur. Phys. J. C",
	volume = "77",
	number = "12",
	pages = "894",
	year = "2017"
}

@article{Chen:2018dbv,
	author = "Chen, Lu and Huang, Qing-Guo and Wang, Ke",
	title = "{Distance Priors from Planck Final Release}",
	eprint = "1808.05724",
	archivePrefix = "arXiv",
	primaryClass = "astro-ph.CO",
	doi = "10.1088/1475-7516/2019/02/028",
	journal = "JCAP",
	volume = "02",
	pages = "028",
	year = "2019"
}

@article{Vagnozzi:2020dfn,
	author = "Vagnozzi, Sunny and Loeb, Abraham and Moresco, Michele",
	title = "{Eppur \`e piatto? The Cosmic Chronometers Take on Spatial Curvature and Cosmic Concordance}",
	eprint = "2011.11645",
	archivePrefix = "arXiv",
	primaryClass = "astro-ph.CO",
	doi = "10.3847/1538-4357/abd4df",
	journal = "Astrophys. J.",
	volume = "908",
	number = "1",
	pages = "84",
	year = "2021"
}

@article{Jimenez:2001gg,
	author = "Jimenez, Raul and Loeb, Abraham",
	title = "{Constraining cosmological parameters based on relative galaxy ages}",
	eprint = "astro-ph/0106145",
	archivePrefix = "arXiv",
	doi = "10.1086/340549",
	journal = "Astrophys. J.",
	volume = "573",
	pages = "37--42",
	year = "2002"
}

@article{Moresco:2012jh,
	author = "Moresco, M. and others",
	title = "{Improved constraints on the expansion rate of the Universe up to z{\textasciitilde}1.1 from the spectroscopic evolution of cosmic chronometers}",
	eprint = "1201.3609",
	archivePrefix = "arXiv",
	primaryClass = "astro-ph.CO",
	doi = "10.1088/1475-7516/2012/08/006",
	journal = "JCAP",
	volume = "08",
	pages = "006",
	year = "2012"
}

@article{Moresco:2015cya,
	author = "Moresco, Michele",
	title = "{Raising the bar: new constraints on the Hubble parameter with cosmic chronometers at z \ensuremath{\sim} 2}",
	eprint = "1503.01116",
	archivePrefix = "arXiv",
	primaryClass = "astro-ph.CO",
	doi = "10.1093/mnrasl/slv037",
	journal = "Mon. Not. Roy. Astron. Soc.",
	volume = "450",
	number = "1",
	pages = "L16--L20",
	year = "2015"
}

@article{Moresco:2016mzx,
	author = "Moresco, Michele and Pozzetti, Lucia and Cimatti, Andrea and Jimenez, Raul and Maraston, Claudia and Verde, Licia and Thomas, Daniel and Citro, Annalisa and Tojeiro, Rita and Wilkinson, David",
	title = "{A 6{\%} measurement of the Hubble parameter at $z\sim0.45$: direct evidence of the epoch of cosmic re-acceleration}",
	eprint = "1601.01701",
	archivePrefix = "arXiv",
	primaryClass = "astro-ph.CO",
	doi = "10.1088/1475-7516/2016/05/014",
	journal = "JCAP",
	volume = "05",
	pages = "014",
	year = "2016"
}

@article{Goliath:2001af,
	author = "Goliath, M. and Amanullah, R. and Astier, P. and Goobar, A. and Pain, R.",
	title = "{Supernovae and the nature of the dark energy}",
	eprint = "astro-ph/0104009",
	archivePrefix = "arXiv",
	doi = "10.1051/0004-6361:20011398",
	journal = "Astron. Astrophys.",
	volume = "380",
	pages = "6--18",
	year = "2001"
}

@article{DES:2025sig,
	author = "Popovic, B. and others",
	collaboration = "DES",
	title = "{The Dark Energy Survey Supernova Program: A Reanalysis Of Cosmology Results And Evidence For Evolving Dark Energy With An Updated Type Ia Supernova Calibration}",
	eprint = "2511.07517",
	archivePrefix = "arXiv",
	primaryClass = "astro-ph.CO",
	reportNumber = "FERMILAB-PUB-25-0842-CSAID-PPD",
	month = "11",
	year = "2025"
}

@article{eBOSS:2020yzd,
	author = "Alam, Shadab and others",
	collaboration = "eBOSS",
	title = "{Completed SDSS-IV extended Baryon Oscillation Spectroscopic Survey: Cosmological implications from two decades of spectroscopic surveys at the Apache Point Observatory}",
	eprint = "2007.08991",
	archivePrefix = "arXiv",
	primaryClass = "astro-ph.CO",
	doi = "10.1103/PhysRevD.103.083533",
	journal = "Phys. Rev. D",
	volume = "103",
	number = "8",
	pages = "083533",
	year = "2021"
}

@article{Elgaroy:2007bv,
	author = "Elgaroy, Oystein and Multamaki, Tuomas",
	title = "{On using the CMB shift parameter in tests of models of dark energy}",
	eprint = "astro-ph/0702343",
	archivePrefix = "arXiv",
	doi = "10.1051/0004-6361:20077292",
	journal = "Astron. Astrophys.",
	volume = "471",
	pages = "65",
	year = "2007"
}

@article{Arendse:2019hev,
	author = "Arendse, Nikki and others",
	title = "{Cosmic dissonance: are new physics or systematics behind a short sound horizon?}",
	eprint = "1909.07986",
	archivePrefix = "arXiv",
	primaryClass = "astro-ph.CO",
	doi = "10.1051/0004-6361/201936720",
	journal = "Astron. Astrophys.",
	volume = "639",
	pages = "A57",
	year = "2020"
}

@article{Hu:1995en,
	author = "Hu, Wayne and Sugiyama, Naoshi",
	title = "{Small scale cosmological perturbations: An Analytic approach}",
	eprint = "astro-ph/9510117",
	archivePrefix = "arXiv",
	reportNumber = "IASSNS-AST-95-42, CFPA-TH-95-18, UTAP-212",
	doi = "10.1086/177989",
	journal = "Astrophys. J.",
	volume = "471",
	pages = "542--570",
	year = "1996"
}

@article{Foreman-Mackey:2012any,
	author = "Foreman-Mackey, Daniel and Hogg, David W. and Lang, Dustin and Goodman, Jonathan",
	title = "{emcee: The MCMC Hammer}",
	eprint = "1202.3665",
	archivePrefix = "arXiv",
	primaryClass = "astro-ph.IM",
	doi = "10.1086/670067",
	journal = "Publ. Astron. Soc. Pac.",
	volume = "125",
	pages = "306--312",
	year = "2013"
}

@article{Lewis:2019xzd,
	author = "Lewis, Antony",
	title = "{GetDist: a Python package for analysing Monte Carlo samples}",
	eprint = "1910.13970",
	archivePrefix = "arXiv",
	primaryClass = "astro-ph.IM",
	month = "10",
	year = "2019"
}

@article{Akaike:1974vps,
	author = "Akaike, H.",
	title = "{A new look at the statistical model identification}",
	doi = "10.1109/TAC.1974.1100705",
	journal = "IEEE Trans. Automatic Control",
	volume = "19",
	number = "6",
	pages = "716--723",
	year = "1974"
}

@article{bic_criterion,
	ISSN = {00905364},
	URL = {http://www.jstor.org/stable/2958889},
	author = {Gideon Schwarz},
	journal = {The Annals of Statistics},
	number = {2},
	pages = {461--464},
	publisher = {Institute of Mathematical Statistics},
	title = {Estimating the Dimension of a Model},
	urldate = {2024-05-18},
	volume = {6},
	year = {1978}
}

@article{Trotta:2008qt,
	author = "Trotta, Roberto",
	title = "{Bayes in the sky: Bayesian inference and model selection in cosmology}",
	eprint = "0803.4089",
	archivePrefix = "arXiv",
	primaryClass = "astro-ph",
	doi = "10.1080/00107510802066753",
	journal = "Contemp. Phys.",
	volume = "49",
	pages = "71--104",
	year = "2008"
}

@article{Fier:2025,
	author = "Fier, Jared and Han, Henry Han and Li, Bowen  and Lin, Kai and   Mukohyama, Shinji and Wang, Anzhong",
	title = "{Gravitational wave cosmology in Einstein-scalar-Gauss-Bonnet gravity}",
	eprint = "2503.01975",
	archivePrefix = "arXiv",
	primaryClass = "gr-qc",
	month = "3",
	year = "2025"
}

@article{Antoniadis:1997eg,
	author = "Antoniadis, Ignatios and Ferrara, S. and Minasian, R. and Narain, K. S.",
	title = "{R**4 couplings in M and type II theories on Calabi-Yau spaces}",
	eprint = "hep-th/9707013",
	archivePrefix = "arXiv",
	reportNumber = "CERN-TH-97-094, CERN-TH-97-94, CPTH-S512-0697",
	doi = "10.1016/S0550-3213(97)00572-5",
	journal = "Nucl. Phys. B",
	volume = "507",
	pages = "571--588",
	year = "1997"
}

@article{Valiviita:2008iv,
	author = "Valiviita, Jussi and Majerotto, Elisabetta and Maartens, Roy",
	title = "{Instability in interacting dark energy and dark matter fluids}",
	eprint = "0804.0232",
	archivePrefix = "arXiv",
	primaryClass = "astro-ph",
	doi = "10.1088/1475-7516/2008/07/020",
	journal = "JCAP",
	volume = "07",
	pages = "020",
	year = "2008"
}

@article{Jackson:2009mz,
	author = "Jackson, Brendan M. and Taylor, Andy and Berera, Arjun",
	title = "{On the large-scale instability in interacting dark energy and dark matter fluids}",
	eprint = "0901.3272",
	archivePrefix = "arXiv",
	primaryClass = "astro-ph.CO",
	doi = "10.1103/PhysRevD.79.043526",
	journal = "Phys. Rev. D",
	volume = "79",
	pages = "043526",
	year = "2009"
}

@article{Wang:2016lxa,
	author = "Wang, B. and Abdalla, E. and Atrio-Barandela, F. and Pavon, D.",
	title = "{Dark Matter and Dark Energy Interactions: Theoretical Challenges, Cosmological Implications and Observational Signatures}",
	eprint = "1603.08299",
	archivePrefix = "arXiv",
	primaryClass = "astro-ph.CO",
	doi = "10.1088/0034-4885/79/9/096901",
	journal = "Rept. Prog. Phys.",
	volume = "79",
	number = "9",
	pages = "096901",
	year = "2016"
}

@article{Kase:2019veo,
	author = "Kase, Ryotaro and Tsujikawa, Shinji",
	title = "{Scalar-Field Dark Energy Nonminimally and Kinetically Coupled to Dark Matter}",
	eprint = "1910.02699",
	archivePrefix = "arXiv",
	primaryClass = "gr-qc",
	doi = "10.1103/PhysRevD.101.063511",
	journal = "Phys. Rev. D",
	volume = "101",
	number = "6",
	pages = "063511",
	year = "2020"
}

@article{Tamanini:2015iia,
	author = "Tamanini, Nicola",
	title = "{Phenomenological models of dark energy interacting with dark matter}",
	eprint = "1504.07397",
	archivePrefix = "arXiv",
	primaryClass = "gr-qc",
	doi = "10.1103/PhysRevD.92.043524",
	journal = "Phys. Rev. D",
	volume = "92",
	number = "4",
	pages = "043524",
	year = "2015"
}

@article{Lesgourgues:2013qba,
	author = "Lesgourgues, Julien",
	title = "{Cosmological Perturbations}",
	booktitle = "{Theoretical Advanced Study Institute in Elementary Particle Physics: Searching for New Physics at Small and Large Scales}",
	eprint = "1302.4640",
	archivePrefix = "arXiv",
	primaryClass = "astro-ph.CO",
	reportNumber = "CERN-PH-TH-2013-031, LAPTH-CONF-012-13",
	doi = "10.1142/9789814525220_0002",
	pages = "29--97",
	year = "2013"
}

@article{Amendola:2007ni,
	author = "Amendola, Luca and Charmousis, Christos and Davis, Stephen C.",
	title = "{Solar System Constraints on Gauss-Bonnet Mediated Dark Energy}",
	eprint = "0704.0175",
	archivePrefix = "arXiv",
	primaryClass = "astro-ph",
	doi = "10.1088/1475-7516/2007/10/004",
	journal = "JCAP",
	volume = "10",
	pages = "004",
	year = "2007"
}

@article{Fier:2025huc,
	author = "Fier, Jared and Han, Henry and Li, Bowen and Lin, Kai and Mukohyama, Shinji and Wang, Anzhong",
	title = "{Gravitational wave cosmology in Einstein-scalar-Gauss-Bonnet gravity}",
	eprint = "2503.01975",
	archivePrefix = "arXiv",
	primaryClass = "gr-qc",
	reportNumber = "YITP-25-29, IPMU25-0008",
	month = "3",
	year = "2025"
}

	\end{document}